\documentclass[twocolumn,fleqn,preprint]{article}    

\usepackage{amssymb}
\usepackage{amscd}
\usepackage{amsmath}
\usepackage{amsthm}
\usepackage{amsfonts}
\usepackage{graphicx}
\usepackage{balance}
\usepackage[all]{xy}
\usepackage{latexsym}
\usepackage{natbib}
\usepackage{dirtytalk} 
\usepackage[dvipsnames,usenames]{color}
\usepackage[colorlinks=true,
pdfstartview=FitV, linkcolor=cyan,
citecolor=cyan, urlcolor=cyan]{hyperref}
\usepackage{flushend} 
\usepackage[makeroom]{cancel} 

\newtheorem{definition}{Definition}[section]
\newtheorem{remark}{Remark}[section]
\newtheorem{lemma}{Lemma}[section]
\newtheorem{proposition}{Proposition}[section]
\newtheorem{theorem}{Theorem}[section]

\usepackage{titlesec}
\titleformat*{\section}{\bfseries}
\titleformat*{\subsection}{\bfseries}
\titleformat*{\subsubsection}{\it}
\providecommand{\keywords}[1]{
\textrm{\textit{Keywords:}} #1}

\begin{document}


\def\mattre#1#2#3#4#5#6#7#8#9{\left[
\begin{array}{ccc}
#1\,\,&\,\,#2\,\,&\,\,#3
\\[4pt] #4\,\,&\,\,#5\,\,&\,\,#6
\\[4pt] #7\,\,&\,\,#8\,\,&\,\,#9
\end{array}\right]}
\def\vectre#1#2#3{\left[
\begin{array}{c}
#1
\\[4pt] #2
\\[4pt] #3
\end{array}\right]}
\def\matdue#1#2#3#4{\left[
\begin{array}{cr}
\,\,#1\,\,&\,\,#2\,\,
\\[4pt] 
\,\,#3\,\,&\,\,#4\,\,
\end{array}\right]}
\def\vecdue#1#2{\left[
\begin{array}{c}
#1
\\[4pt] #2
\end{array}\right]}

\newcommand{\immers}{\Bi}
\newcommand{\extens}{\Bj}

\newcommand{\intprod}{\rfloor} 
\newcommand{\msec}{$\mathsection$}
\newcommand{\MIL}{\textsc{M.I.L.}}
\newcommand{\EIL}{\textsc{E.I.L.}}
\newcommand{\timeform}{{\Btheta}}
\newcommand{\relfactor}{\relfac_\traslpar}
\newcommand{\lightvel}{c}
\newcommand{\relfac}{\gamma}
\newcommand{\relspeed}{\ww}
\newcommand{\relvel}{\Bw}
\newcommand{\fourax}{\BX}
\newcommand{\threeax}{\Bx}

\newcommand{\map}{\zeta}
\newcommand{\mapL}{\map_{\textsc{L}}}
\newcommand{\mapG}{\map_{\textsc{G}}}
\newcommand{\mapS}{\map_\EU}

\newcommand{\Eul}{\textsc{Eul}}
\newcommand{\Lieder}{\cL}
\newcommand{\LiederS}{\cL^\EU}
\newcommand{\lapse}{\alpha}
\newcommand{\blapse}{\beta}
\newcommand{\extder}{\mathtt{d}}

\newcommand{\rotor}{\mathrm{rot}}
\newcommand{\identvec}{\Br}

\newcommand{\lininv}{\JJ_1}
\newcommand{\Vect}{\BV}
\newcommand{\vect}{\Bv}

\newcommand{\threevel}{{\vect_{\moto}}}
\newcommand{\fourvel}{{\Vect_{\moto}}}
\newcommand{\vel}{{\vect_{\moto}}}
\newcommand{\fourvelVE}{{\Vect_{\moto}^\VEVE}}
\newcommand{\fourvelHE}{{\Vect_{\moto}^\HEVE}}

\newcommand{\Exter}{\BOmega^k}
\newcommand{\exter}{\Bomega^k}
\newcommand{\splitZ}{{\BR}}
\newcommand{\splitS}{{\BP}}

\newcommand{\FORMS}{\Lambda}

\newcommand{\linea}{\BGamma}
\newcommand{\surf}{\Sigma}
\newcommand{\surfconf}{{\Sigma\di\conf}}
\newcommand{\bulk}{\cV}

\newcommand{\Tsurf}{\TT\surf}

\newcommand{\emf}{\textsc{emf}}
\newcommand{\mmf}{\textsc{mmf}}
\newcommand{\lineinn}{\linea_\inne}
\newcommand{\lineout}{\linea_\oute}
\newcommand{\surfinn}{\surf_\inne}
\newcommand{\surfout}{\surf_\oute}
\newcommand{\bulkinn}{\bulk_\inne}
\newcommand{\bulkout}{\bulk_\oute}

\newcommand{\Current}{\BOmega^3_{\BA}}
\newcommand{\Magnet}{\BOmega^3_{\BF}}
\newcommand{\Faradpot}{\BOmega^1_{\BF}}
\newcommand{\Farad}{\BOmega^2_{\BF}}
\newcommand{\Amper}{\BOmega^2_{\BA}}

\newcommand{\image}{\mathbf{Im}}
\newcommand{\nucleo}{\mathbf{Ker}}


\newcommand{\elecurr}{\BJ}
\newcommand{\magind}{\BB}
\newcommand{\eledisp}{\BD}
\newcommand{\magfield}{\BH}
\newcommand{\elefield}{\BE}
\newcommand{\elepot}{\PP_\elefield}
\newcommand{\magpot}{\PP_\magfield}
\newcommand{\magvecpot}{\BA}
\newcommand{\elecurrvecpot}{\BY}
\newcommand{\eledisppot}{\BN}

\newcommand{\eleform}{\Boo^1_\elefield}
\newcommand{\Eleform}{\BOO^1_\elefield}
\newcommand{\magmom}{\Boo^1_\BM}
\newcommand{\Magmom}{\BOO^1_\BM}
\newcommand{\elefluxpot}{\Boo^1_\eledisp}
\newcommand{\eleflux}{\Boo^2_\eledisp}
\newcommand{\Eleflux}{\BOO^2_\eledisp}
\newcommand{\magwind}{\Boo^1_\magfield}
\newcommand{\Magwind}{\BOO^1_\magfield}
\newcommand{\magcurl}{\Boo^2_\BM}
\newcommand{\Magcurl}{\BOO^2_{\BM}}
\newcommand{\elecurrflux}{\Boo^2_\elecurr}
\newcommand{\Elecurrflux}{\BOO^2_{\elecurr}}
\newcommand{\elecurrpot}{\Boo^1_\elecurr}

\newcommand{\chargescalar}{\rho}

\newcommand{\elestatform}{\Boo^0_\elefield}
\newcommand{\Elestatform}{\BOO^0_\elefield}
\newcommand{\elechargeform}{\Boo_{\elecharge}}
\newcommand{\Elechargeform}{\BOO_{\elecharge}}

\newcommand{\elecharge}{\rho}
\newcommand{\elecurrsource}{\Boo^3_\elecurr}
\newcommand{\Elecurrsource}{\BOO^3_\elecurr}

\newcommand{\elecurrscalar}{\sss_\elecurr}


\newcommand{\EU}{\EUCLID}
\newcommand{\EUL}{\textsc{Eul}}
\newcommand{\EUC}{\EU_\conf}

\newcommand{\curvela}{\chi}
\newcommand{\bending}{\MM}
\newcommand{\reazione}{r}

\newcommand{\euler}{e}
\renewcommand{\aa}{a}
\newcommand{\bb}{b}
\newcommand{\dd}{d}
\newcommand{\jj}{j}

\newcommand{\Strain}{\cD}
\newcommand{\respstress}{\response_\Sigma}
\newcommand{\respstrain}{\response_\Strain}

\newcommand{\elaop}{\EE}
\newcommand{\invelaop}{{\inv\elaop}}
\newcommand{\vincop}{\KK}
\newcommand{\constop}{\CC}
\newcommand{\constopsr}{\sqrt\CC}
\newcommand{\Hilb}{\cH}
\newcommand{\response}{\cR}
\newcommand{\meno}{\backslash}
\newcommand{\zeroset}{\set{\zerovec}}
\newcommand{\Replus}{[0,\pinfty[}

\newcommand{\spz}{\,;\,}

\newcommand{\ave}{\cA}

\newcommand{\elcurvup}{\chi^{el}}
\newcommand{\distr}{\mathbb{T}}
\newcommand{\distrinf}{\mathbb{T}^\infty}
\newcommand{\zerofun}{{\bf0}}
\newcommand{\unofun}{{\bf1}}
\newcommand{\idnfun}{{\bf{x}}}
\newcommand{\source}{\sss}
\newcommand{\mixture}{\mm}
\newcommand{\coeff}{\beta}
\newcommand{\spost}{\uu}
\newcommand{\abb}{\vv}
\newcommand{\test}{\cD}

\newcommand{\paramone}{\alpha}
\newcommand{\paramtwo}{\beta}

\renewcommand{\AA}{A}
\newcommand{\FFeu}{\mathfrak{F}}
\newcommand{\MMeu}{\mathfrak{M}}
\newcommand{\elcurvnlplus}{{\chi^+_{nl}}}
\newcommand{\elcurvnlminus}{{\chi^-_{nl}}}
\newcommand{\geocurv}{\chi}

\newcommand{\geoext}{\varepsilon}

\newcommand{\barra}[2]{ \bigg\vert _{\hbox{$\scriptstyle#1$}}^{\hbox{$\scriptstyle#2$}}  \, }
\newcommand{\enlarge}[1][2]{\rule[-.8#1\baselineskip]{0pt}{#1\baselineskip}}
\newcommand{\Enlarge}[1][3]{\rule[-1.2#1\baselineskip]{0pt}{#1\baselineskip}}
\newcommand{\normboxed}{\setlength\fboxsep{0.25cm}\boxed}
\newcommand{\svert}{\\ \vspace{5ex}}
\newcommand{\CTRLmap}{\Bxi}
\newcommand{\tube}{\cT_\Path}
\newcommand{\equidef}{\stackrel{\mathrm{def}}\equi}
\newcommand{\choord}{\phi}
\newcommand{\percorso}{{\Bgamma}}
\newcommand{\bvec}{\Bd}
\newcommand{\stdvec}{\Ba}
\newcommand{\lag}{\textit{Lag}}
\newcommand{\BXs}{\BX^*}
\newcommand{\LEM}{\persone{L.E.M.}}
\newcommand{\LHP}{\persone{L.H.P.}}
\newcommand{\loc}{{\textsc{loc}}}
\newcommand{\noloc}{\textsc{nloc}}
\newcommand{\nl}{\textsc{nl}}
\newcommand{\res}{{\textsc{res}}}

\newcommand{\TTs}{{\TT^*}}
\newcommand{\canform}{\BTheta}
\newcommand{\twoform}{\Boo^2_\LAG}

\newcommand{\curv}{\textsc{curv}}
\newcommand{\tors}{\textsc{tors}}
\newcommand{\TORS}{\textsc{Tors}}
\newcommand{\CLRV}{\textsc{Curv}}

\newcommand{\abs}[1]{|#1|}
\newcommand{\curvtot}{\chi}
\newcommand{\elcurv}{\curvtot_{el}}
\newcommand{\elcurvnl}{{\overline\curvtot_{el}}}
\newcommand{\elcurvnlp}{{\overline\curvtot{\,}'_{el}}}
\newcommand{\elcurvnlpp}{{\overline\curvtot{\,}''_{el}}}
\newcommand{\elcurvstd}{{\chi_{el}}}
\newcommand{\bendmom}{M}
\newcommand{\axnorm}{N}
\newcommand{\elext}{{\varepsilon_{el}}}
\newcommand{\bendvar}{\delta\bendmom}
\newcommand{\load}{\ell}

\newcommand{\distor}{\curvtot_0}
\newcommand{\nonloc}{\lambda}
\newcommand{\LLc}{\LL_{c}}
\newcommand{\param}{\pp}
\newcommand{\Bparam}{\Bp}
\newcommand{\mixpar}{\xx}
\newcommand{\fourvar}{\omega}

\newcommand{\INT}{\textsc{Int}}
\newcommand{\DIF}{\textsc{Dif}}
\newcommand{\AUX}{\textsc{Aux}}
\newcommand{\BND}{\textsc{Bnd}}
\newcommand{\CBND}{\textsc{Bnd}}
\newcommand{\EQL}{\textsc{Equil}}

\newcommand{\LAG}{L}
\newcommand{\LAGloc}{L_\loc}
\newcommand{\LAGT}{{\LAG_\TT}}

\newcommand{\HAM}{H}
\newcommand{\HAMloc}{H_\loc}
\newcommand{\HAMT}{\HAM_\TT}

\newcommand{\ENE}{E}
\newcommand{\ENEloc}{E_\loc}
\newcommand{\ENET}{\ENE_\TT}

\newcommand{\KIN}{K}
\newcommand{\KINloc}{K_\loc}
\newcommand{\KINT}{\KIN_\TT}

\newcommand{\ACT}{A}
\newcommand{\ACTloc}{A_\loc}

\newcommand{\ACTs}{B}
\newcommand{\BOH}{J}

\newcommand{\POT}{\Pi}
\newcommand{\LAGact}{\Boo^1_\LAG}
\newcommand{\HAMact}{\Boo^1_\HAM}

\newcommand{\PC}{{\textsc{PC}}}
\newcommand{\HP}{{\textsc{HP}}}
\newcommand{\PCact}{\Boo^1_\PC}
\newcommand{\HPact}{\Boo^1_\HP}

\newcommand{\LEGE}{\textsc{Leg}}
\newcommand{\LEGT}{\textsc{Leg}_\TT}

\newcommand{\CTRL}{\BM}
\newcommand{\TC}{\TT\CTRL}
\newcommand{\TCs}{(\TT\CTRL)^*}
\newcommand{\TCP}{(\TC)_\Path}
\newcommand{\TCPs}{\TCP^*}

\newcommand{\VC}{\VERT\CTRL}
\newcommand{\VCs}{(\VERT\CTRL)^*}
\newcommand{\VCP}{(\VC)_\Path}
\newcommand{\VCPs}{\VCP^*}

\newcommand{\TTC}{\TT\TT\CTRL}
\newcommand{\VTC}{\VERT\TC}

\newcommand{\projC}{\Btau_\CTRL}
\newcommand{\projTC}{\Btau_{\TC}}
\newcommand{\projCs}{\Btau^*_\CTRL}
\newcommand{\cansold}{\BJ}
\newcommand{\Liouv}{\BTheta}
\newcommand{\BVT}{{\BV}}
\newcommand{\Bus}{\Bu^*}
\newcommand{\Bws}{\Bw^*}
\newcommand{\Bvs}{\Bv^*}
\newcommand{\LAGTform}{\Boo^1_\LAGT}
\newcommand{\HAMTform}{\Boo^1_\TT}

\newcommand{\morphism}{{\Bchi}}
\newcommand{\morphismE}{\morphism_\EVE}
\newcommand{\basemorphism}{\Bphi}

\newcommand{\mapE}{\map_\EVE}

\newcommand{\control}{\CC}
\newcommand{\pcontrol}{\partial\control}
\newcommand{\controltout}{\control_\oute}
\newcommand{\mapcontr}{\Bxi^\EVE}
\newcommand{\TjC}{{\Tj_\control}}
\newcommand{\BuS}{{\Bu_{\Sigma}}}
\newcommand{\BvS}{{\Bv_\Sigma}}
\newcommand{\BwS}{{\Bw_{\Sigma}}}
\newcommand{\BvE}{{\Bv_\EVE}}

\newcommand{\tT}{\ttt_{\TT}}
\newcommand{\tTs}{\ttt_{\TT}^*}
\newcommand{\dtT}{\derT\tT}
\newcommand{\dtTs}{\derT\tT^*}
\newcommand{\flip}[1]{\Bk_{#1}}
\newcommand{\Fform}{\Boo^1_\BF}
\newcommand{\force}{\Boo^1_\Bf}
\newcommand{\secfield}{Sect}
\newcommand{\DIFOPG}{\BD_\Lpath}
\newcommand{\torsformT}{\torsform_\TT}
\newcommand{\BgT}{\Bg_\TT}
\newcommand{\Bip}{\partial\Bi}

\newcommand{\sourceT}{\source_\TT}
\newcommand{\BooT}{\Boo_\TT}
\newcommand{\oneT}{\Boo^1_\TT}
\newcommand{\BXT}{{\BX_\TT}}
\newcommand{\BYT}{{\BY_\TT}}
\newcommand{\LAGformT}{\Btheta_\LAGT}
\newcommand{\LAGform}{\Btheta_\LAG}
\newcommand{\HAMform}{\Btheta_\HAM}
\newcommand{\LC}{\Btheta}
\newcommand{\Path}{{\BGamma}}
\newcommand{\Paths}{{\BGamma^*}}
\newcommand{\PathT}{{\BGamma_\TT}}
\newcommand{\PathTs}{{\BGamma_\TT^*}}
\newcommand{\RIG}{\textsc{RIG}}
\newcommand{\rig}{\textsc{rig}}
\newcommand{\rigT}{\rig_\TT}
\newcommand{\syn}{\textsc{syn}}
\newcommand{\synT}{\syn_\TT}
\newcommand{\powT}{\Boo^1_{\BSigma}}
\newcommand{\equil}{\Boo^1_{\bf eq}}
\newcommand{\equilT}{\Boo^1_{\bf Eq}}

\newcommand{\place}{\cP}
\newcommand{\Hil}{\cH}
\newcommand{\HilC}{\cH_\CTRL}
\newcommand{\HilE}{\cH_\EVE}
\newcommand{\HilT}{\cH_\TT}
\newcommand{\HilM}{\cH_\MMM}
\newcommand{\HilCs}{\cH_\CTRL^*}
\newcommand{\HilTs}{\cH_\TT^*}
\newcommand{\derE}{\der_{\EVE}}
\newcommand{\derS}{\der}
\newcommand{\derT}{\der_\TT}
\newcommand{\derC}{\der_\CTRL}

\newcommand{\flu}{\textsc{flu}}
\newcommand{\sol}{\textsc{sol}}
\newcommand{\rel}{\textsc{rel}}
\newcommand{\ske}{\textsc{ske}}
\newcommand{\oute}{\textsc{out}}
\newcommand{\inne}{\textsc{inn}}
\newcommand{\inout}{\BSigma^\inne_\oute}
\newcommand{\skel}{\conf_\ske}
\newcommand{\ZEIT}{\cZ}
\newcommand{\Zeit}{\timearrow}
\newcommand{\zeit}{\ttt}

\newcommand{\coord}{\Bchi}

\newcommand{\extrusion}[1]{\Tj_{#1}}
\newcommand{\immersTj}{\Bi_{\Tj}}
\newcommand{\immersS}{\Bi_\EU}
\newcommand{\Bproj}{\Bp}
\newcommand{\SHIFT}[1]{\textsc{shift}_{#1}}
\newcommand{\shift}{\theta}
\newcommand{\dshift}{\delta\ttt}
\newcommand{\dvZ}{\delta\vv_\ZEIT}
\newcommand{\qvel}{\vect}
\newcommand{\dqvel}{{\delta\qvel}}

\newcommand{\qmov}{\Bp}
\newcommand{\dqmov}{{\delta\qmov}}
\newcommand{\qmovS}{\qmov_S}
\newcommand{\dqmovS}{\delta\qmov_S}

\newcommand{\qvec}{{\Bu}}
\newcommand{\dqvec}{{\delta\qvec}}

\newcommand{\dfourvel}{{\delta\fourvel}}
\newcommand{\dtheta}{{\delta\theta}}
\newcommand{\dthetap}{{\delta\theta'}}

\newcommand{\fourvelst}{{\overline\fourvel}}
\newcommand{\fourvels}{\Bv^*}
\newcommand{\fourvelS}{{\fourvel_\EU}}
\newcommand{\fourvelZ}{{\fourvel_\ZEIT}}
\newcommand{\fourvelT}{{\fourvel_\TT}}

\newcommand{\dvel}{{\delta\fourvel}}
\newcommand{\dvelS}{{\delta\fourvelS}}
\newcommand{\fourvelSs}{{\qvel^*_\EU}}
\newcommand{\dvels}{{\delta\fourvel^*}}
\newcommand{\fourvelC}{\Bv_\CTRL}
\newcommand{\vT}{{\qvel_\TT}}
\newcommand{\vTs}{{\qvel_\TT^*}}
\newcommand{\dvT}{{\dqvel_\TT}}
\newcommand{\dvTs}{{\dqvel_{\TT^*}}}
\newcommand{\TZ}{{\timearrow_\TT}}
\newcommand{\dBY}{{\delta\BY}}
\newcommand{\dBV}{{\delta\BV}}
\newcommand{\simpform}{\Boo^2}
\newcommand{\LsimpformT}{\Boo^2_\LAGT}
\newcommand{\Lsimpform}{\Boo^2_\LAG}
\newcommand{\sMMM}{\BGamma}

\newcommand{\LLL}{\textrm{L}}
\newcommand{\CTRLM}{\cC}
\newcommand{\TRAJM}{\cT_{\cC}}
\newcommand{\pMAP}{\textsc{pos}}
\newcommand{\BiET}{\Bi_{\EVE,\Tj}}
\newcommand{\Dt}{{\Delta\ttt}}

\newcommand{\liftlab}{{\textsc{lift}}}
\newcommand{\unliftlab}{{\textsc{proj}}}
\newcommand{\coliftlab}{{\textsc{co-lift}}}

\newcommand{\lift}{\,\stackrel{\liftlab\ }\longrightarrow\,}
\newcommand{\unlift}{\,\stackrel{\unliftlab\ \,}\longrightarrow\,}
\newcommand{\colift}{\,\stackrel{\coliftlab\ }\longrightarrow\,}

\newcommand{\fliparrow}{\,\stackrel{\flip\CTRL}\longrightarrow\,}
\newcommand{\TTarrow}{\,\stackrel{\TT}\longrightarrow\,}

\newcommand{\EVE}{\cE}
\newcommand{\SEVE}{\EU\EVE}
\newcommand{\EVES}{\EVE_\EU}
\newcommand{\EVEZ}{\EVE_\ZEIT}
\newcommand{\EVEr}{\EVE_\morph}
\newcommand{\Flip}[1]{\mathbf{Flip}_{#1}}
\newcommand{\torsform}{\BT}
\newcommand{\torsformS}{\BT^\EU}
\newcommand{\torsformE}{\BT^\EVE}
\newcommand{\torsformC}{\BT^\CTRL}

\newcommand{\curvform}{\BR}
\newcommand{\nablators}{\nabla\textsc{T}}
\newcommand{\nablaS}{\nabla^\EU}
\newcommand{\nablaE}{\nabla^\EVE}
\newcommand{\nablaC}{\nabla^\CTRL}

\newcommand{\FEM}{\textsc{F.E.M.}}

\newcommand{\qmovT}{(\qmov)_\TT}

\newcommand{\push}{{\uparrow}}
\newcommand{\pull}{{\downarrow}}
\newcommand{\forw}{{\Uparrow}}
\newcommand{\back}{{\Downarrow}}
\newcommand{\forwE}{{\Uparrow^\EVE}}
\newcommand{\forwC}{{\Uparrow^\CTRL}}
\newcommand{\backE}{{\Downarrow^\EVE}}
\newcommand{\backEs}{{\Downarrow^\EVEs}}
\newcommand{\forwS}{{\Uparrow^\EU}}
\newcommand{\backS}{\Downarrow^{\EU}}
\newcommand{\forwZ}{{\Uparrow^\ZEIT}}
\newcommand{\backZ}{{\Downarrow^\II}}
\newcommand{\forwES}{{\Uparrow^{\EVE,\EU}}}
\newcommand{\backES}{{\Downarrow^{\EVE,\EU}}}
\newcommand{\forwEZ}{{\Uparrow^{\EVE,\II}}}
\newcommand{\backEZ}{{\Downarrow^{\EVE,\II}}}
\newcommand{\diag}{\textsc{diag}}
\newcommand{\VPROJ}{\textsc{P}_V}
\newcommand{\HPROJ}{\textsc{P}_H}
\newcommand{\Hlift}{\BH}
\newcommand{\Tder}[1]{\TT_{#1}}
\newcommand{\Null}[1]{\textbf{ker}\di{#1}}
\newcommand{\Range}[1]{\textbf{im}\di{#1}}
\newcommand{\mapl}{\FL\fourvel\nonloc}
\newcommand{\vv}{v}
\newcommand{\vvS}{v_S}
\newcommand{\dv}{\delta{v}}
\newcommand{\defequal}{=:}
\newcommand{\epsdvp}{{\dot{\Bee}}_{vp}}
\newcommand{\INF}{\mathop{\inf}}
\newcommand{\epi}{\mathrm{epi}\,}
\newcommand{\diverg}{\mathrm{div}}

\newcommand{\indconc}[1]{\sqcap_{#1}}
\newcommand{\indconv}[1]{\sqcup_{\,#1}}
\newcommand{\suppconv}[1]{\conj{\indconv#1}}
\newcommand{\conj}[1]{#1^*}

\newcommand{\zerovec}{\mathbf{0}}
\newcommand{\dt}{\der\ttt}

\newcommand{\strain}{\Bee}
\newcommand{\BTT}{\BT}
\newcommand{\BsM}{\Bs}

\newcommand{\fourvelTj}{\fourvel_{\Tj}}
\newcommand{\refe}{\textsc{ref}}
\newcommand{\derplusF}{\der^+_F}
\newcommand{\trZ}{\Btheta}

\newcommand{\basis}{\Bd}
\newcommand{\cLo}{\cL_o}
\newcommand{\acc}{\Ba_\EU}
\newcommand{\bkernel}{{\Bphi}}
\newcommand{\pkernel}{{\Bpsi}}
\newcommand{\kernel}{\varphi}
\newcommand{\kernelmod}{\psi}
\newcommand{\kernelH}{h}
\newcommand{\volphi}{\mathrm{vol}_{\phi_\nonloc}}
\newcommand{\kernelbar}{\omega}
\newcommand{\kernelerr}{\kernel^{\textsc{err}}}


\newcommand{\BVE}{\BV^\EVE}
\newcommand{\fourvelE}{{\Bv_\EVE}}
\newcommand{\BvEs}{{\Bv_{\EVEs}}}
\newcommand{\traslpar}{\ww}
\newcommand{\traslvel}{\Bw}
\newcommand{\tensor}{\Bs}
\newcommand{\Tensor}{\BS}
\newcommand{\Dl}{\Delta\nonloc}
\newcommand{\mat}{\textsc{mat}}
\newcommand{\spa}{\textsc{spa}}

\newcommand{\VERT}{V}
\newcommand{\HORI}{H}
\newcommand{\TANG}{T}

\newcommand{\dVERT}{\delta{\BX}}
\newcommand{\TEVE}{{\TANG\EVE}}
\newcommand{\VEVE}{{\VERT\EVE}}
\newcommand{\HEVE}{{\HORI\EVE}}
\newcommand{\TTEVE}{{\TANG^2\EVE}}
\newcommand{\CTEVE}{{\CTANG}\EVE}
\newcommand{\TEVEs}{{(\TEVE)^*}}
\newcommand{\VEVEs}{{(\VEVE)^*}}
\newcommand{\Tj}{\cT}
\newcommand{\TTj}{{\TANG\Tj}}
\newcommand{\VTj}{{\VERT\Tj}}
\newcommand{\derTj}{\der_\Tj}

\newcommand{\TjE}{{\Tj_\EVE}}
\newcommand{\VTjE}{{\VTj_\EVE}}
\newcommand{\VETjE}{{(\VEVE)_\TjE}}
\newcommand{\VETjEs}{{(\VEVE)^*_\TjE}}
\newcommand{\TTjE}{{\TT\Tj_\EVE}}
\newcommand{\TTTjE}{{\TT^2\Tj_\EVE}}

\newcommand{\tE}{\ttt_\EVE}
\newcommand{\tC}{\ttt_\CTRL}
\newcommand{\tTj}{\ttt_\Tj}

\newcommand{\tTE}{\ttt_\TEVE}
\newcommand{\TtE}{\TT\ttt_\EVE}
\newcommand{\TtTE}{\TT\ttt_\TEVE}

\newcommand{\EVEs}{{\EVE_\map}}
\newcommand{\Tjs}{{\Tj_\map}}
\newcommand{\tEr}{\tbar_\EVE}

\newcommand{\tEs}{\ttt_{\EVE_\map}}

\newcommand{\dtE}{\der\tE}
\newcommand{\dtTj}{\der\tTj}
\newcommand{\dtTE}{\der\tTE}

\newcommand{\dtC}{\der\tC}

\newcommand{\stmap}{\Bxi}

\newcommand{\conf}{{\BOmega}}
\newcommand{\pconf}{\partial\conf}

\newcommand{\mani}{{\cM}}
\newcommand{\pmani}{\partial\mani}
\newcommand{\manin}{{\cN}}
\newcommand{\pmanin}{\partial\manin}

\newcommand{\Tconf}{{\TANG\conf}}
\newcommand{\conft}{\conf_\ttt}
\newcommand{\confo}{{\BOmega_{o}}}
\newcommand{\confref}{\conf_\ref}

\newcommand{\Fthrust}{\Bf_{\textsc{thr}}}
\newcommand{\fluido}{\conf_\flu}
\newcommand{\solido}{\conf_\sol}
\newcommand{\rocket}{\textsc{rocket}}
\newcommand{\sprinkler}{\textsc{sprinkler}}
\newcommand{\pipe}{\textsc{pipe}}

\newcommand{\dhvelS}{\delta\hat\fourvel_\EU}

\newcommand{\sinistra}{l}
\newcommand{\centro}{c}
\newcommand{\destra}{r}
\newcommand{\timearrow}{\BZ}

\newcommand{\ext}{\textsc{ext}}
\newcommand{\inn}{\textsc{int}}
\newcommand{\dyn}{\textsc{dyn}}
\newcommand{\eq}{\textsc{eq}}
\newcommand{\sing}{\textsc{sing}}
\newcommand{\fext}{\Bf_\ext}
\newcommand{\finn}{\Bf_\inn}
\newcommand{\fdyn}{\Bf_\dyn}
\newcommand{\feq}{\Bf_\eq}
\newcommand{\Fdyn}{\BF_\dyn}
\newcommand{\fsing}{\Bf_\sing}
\newcommand{\fpot}{\PP}

\newcommand{\GNT}{\persone{GNT}}
\newcommand{\DIN}{\persone{DIN}}
\newcommand{\DG}{\persone{DG}}
\newcommand{\vol}{\textsc{vol}}
\newcommand{\isohomo}{\BL^\ISO}
\newcommand{\ciclo}{\Bo}
\newcommand{\stressref}{{\stress_{\refe}}}

\newcommand{\IDN}{\mathbb{I}}



\newcommand{\BuE}{\Bu_{\EVE}}
\newcommand{\BsE}{\Bs_{\EVE}}

\newcommand{\omissis}{...omissis...}

\newcommand{\TENS}{\textsc{Tens}}

\newcommand{\partialF}{\partial_F}
\newcommand{\derF}{\der_F}
\newcommand{\derdueF}{\der^2_F}

\newcommand{\stress}{\Bss}
\newcommand{\stretching}{\Beps}
\newcommand{\stressing}{\dot\stress}
\newcommand{\vstress}{\delta\stress}

\newcommand{\Celast}{\BPsi}
\newcommand{\dCelast}{\dot\Celast}
\newcommand{\dCelasts}{\dot\Celast^*}
\newcommand{\Gelast}{\Xi}
\newcommand{\dGelast}{\dot\Gelast}
\newcommand{\Gelasts}{\Gelast^*}
\newcommand{\dGelasts}{\dot\Gelast^*}

\newcommand{\pneg}[1]{#1^{-}}
\newcommand{\ppos}[1]{#1^{+}}
\newcommand{\pinfty}{+\infty}

\newcommand{\es}{\Be\Bs}
\newcommand{\des}{\dot\Be\Bs}

\newcommand{\elstate}{\Be}
\newcommand{\elstrain}{\Bee}
\newcommand{\ela}{\textsc{el}}
\newcommand{\pla}{\textsc{pl}}

\newcommand{\locstrain}{\stretching_\loc}
\newcommand{\selfstress}{\stress_\zerofun}
\newcommand{\Stress}{\BSigma}
\newcommand{\Selfstress}{\BSigma_o}
\newcommand{\stressvar}{\delta\stress}

\newcommand{\metric}{\Bg}
\newcommand{\metricS}{\Bg_\EU}
\newcommand{\metriciso}{\BJ}
\newcommand{\selfmom}{\MM_o}
\newcommand{\impdis}{\ww}
\newcommand{\dvv}{{\delta\vv}}


\newcommand{\BUN}{{\textsc{Bun}}}
\newcommand{\FUN}{{\textsc{Fun}}}
\newcommand{\VEC}{{\textsc{Vec}}}
\newcommand{\SYM}{{\textsc{Sym}}}
\newcommand{\MIX}{{\textsc{Mix}}}
\newcommand{\MAX}{{\textsc{Max}}}
\newcommand{\COV}{{\textsc{Cov}}}
\newcommand{\CON}{{\textsc{Con}}}
\newcommand{\ALT}{{\textsc{Alt}}}
\newcommand{\VOL}{\textsc{Vol}}
\newcommand{\QUA}{\textsc{Qua}}
\newcommand{\VBUN}{{\textsc{VBUN}}}
\newcommand{\VCOV}{{\textsc{VCOV}}}
\newcommand{\VSYM}{{\textsc{VSYM}}}
\newcommand{\BIL}{\textsc{BIL}}

\newcommand{\dual}[1]{#1^*}
\newcommand{\di}[1]{(#1)}
\newcommand{\disq}[1]{[#1]}
\newcommand{\pair}[2]{\{#1\,,#2\}}
\newcommand{\coppia}[2]{(#1\,,#2)}
\newcommand{\terna}[3]{(#1,#2,#3)} 
\newcommand{\Lbrack}[2]{[#1\,,#2]}
\newcommand{\LbrackT}[2]{[#1\,,#2]_\TT}

\newcommand{\CT}{\TT^*}

\newcommand{\ID}{\textsc{id}}

\newcommand{\massform}{\Bm}
\newcommand{\massformT}{\massform}
\newcommand{\massformE}{\massform_\EVE}
\newcommand{\volform}{\Bmu}
\newcommand{\volformg}{\volform_\metric}
\newcommand{\volformT}{\volform_\Tj}
\newcommand{\volformE}{\volform_\EVE}
\newcommand{\massdens}{\rho}

\newcommand{\unmezzo}{{\scriptstyle\onehalf}}
\newcommand{\unmezzotext}{\tfrac{1}{2}}
\newcommand{\unterzotext}{\tfrac{1}{3}}
\newcommand{\unquarto}{{\scriptstyle\frac{1}{4}}}
\newcommand{\jump}[1]{[[#1]]}
\newcommand{\Jump}[1]{\biggl[\biggl[#1\biggr]\biggr]}

\newcommand{\bigdi}[1]{\big(#1\big)}
\newcommand{\Bigdi}[1]{\Big(#1\Big)}
\newcommand{\biggdi}[1]{\bigg(#1\bigg)}

\newcommand{\cont}{\mathrm{C}}
\newcommand{\parder}[2]{\partial_{#1=#2}\,}
\newcommand{\FIBRE}{\FFF}
\newcommand{\VF}{\VERT\FFF}
\newcommand{\TF}{\TANG\FFF}

\newcommand{\MMM}{\mathbf{M}}
\newcommand{\NNN}{\mathbf{N}}
\newcommand{\HHH}{\mathbf{H}}
\newcommand{\FFF}{\mathbf{F}}
\newcommand{\SSS}{\mathbb{S}}
\newcommand{\CCC}{\mathbb{C}}
\newcommand{\DDD}{\mathbb{D}}
\newcommand{\RRR}{\mathbb{R}}
\newcommand{\ZZZ}{\mathbb{Z}}

\newcommand{\TN}{\TANG\NNN}
\newcommand{\CTN}{\TANG^*\NNN}
\newcommand{\TM}{\TANG\MMM}
\newcommand{\CTM}{\TANG^*\MMM}
\newcommand{\TTM}{\TANG\TM}
\newcommand{\VTM}{\VERT\TANG\MMM}

\newcommand{\rrr}{r}
\newcommand{\xx}{x}
\newcommand{\zz}{z}
\newcommand{\ii}{i}
\newcommand{\nn}{n}
\newcommand{\gi}{g}

\newcommand{\sym}{\textrm{sym}}

\newcommand{\cov}{\textsc{cov}}
\newcommand{\con}{\textsc{con}}
\newcommand{\mix}{\textsc{mix}}

\newcommand{\sfield}{\psi}
\newcommand{\sfieldN}{\phi}

\newcommand{\BpiE}{\Bpi_\EU}
\newcommand{\BpiT}{\Bpi_\ZEIT}
\newcommand{\referspace}{\conf\times\ZEIT}

\newcommand{\linspan}[1]{\mathbf{Span}{(#1)}} 

\newcommand{\Bee}{\boldsymbol{\varepsilon}}
\newcommand{\BDelta}{\boldsymbol{\Delta}}
\newcommand{\BPhi}{\boldsymbol{\phi}}
\newcommand{\Baa}{\boldsymbol{\alpha}}
\newcommand{\Bomega}{\boldsymbol{\omega}}
\newcommand{\Boo}{\Bomega}
\newcommand{\BOmega}{\boldsymbol{\Omega}}
\newcommand{\BOO}{\BOmega}
\newcommand{\BGamma}{\boldsymbol{\Gamma}}
\newcommand{\BTheta}{\boldsymbol{\Theta}}
\newcommand{\BSigma}{\boldsymbol{\Sigma}}
\newcommand{\Btau}{\boldsymbol{\tau}}
\newcommand{\Bchi}{\boldsymbol{\chi}}
\newcommand{\BPi}{\boldsymbol{\Pi}}
\newcommand{\Bpi}{\boldsymbol{\pi}}
\newcommand{\Bpsi}{\boldsymbol{\psi}}
\newcommand{\Bss}{\boldsymbol{\sigma}}
\newcommand{\Bmu}{\boldsymbol{\mu}}
\newcommand{\Bgamma}{\boldsymbol{\gamma}}
\newcommand{\Balpha}{\boldsymbol{\alpha}}
\newcommand{\Bvarphi}{\boldsymbol{\varphi}}
\newcommand{\Bphi}{\boldsymbol{\phi}}
\newcommand{\Bzeta}{{\boldsymbol{\zeta}}}
\newcommand{\Beta}{\boldsymbol{\eta}}
\newcommand{\Brho}{\boldsymbol{\rho}}
\newcommand{\Blambda}{\boldsymbol{\lambda}}
\newcommand{\Btheta}{\boldsymbol{\theta}}
\newcommand{\Biota}{\boldsymbol{\iota}}
\newcommand{\Bxi}{\boldsymbol{\xi}}
\newcommand{\BXi}{\boldsymbol{\Xi}}
\newcommand{\Bbeta}{\boldsymbol{\beta}}
\newcommand{\BPsi}{\boldsymbol{\Psi}}
\newcommand{\Beps}{\boldsymbol{\epsilon}}

\newcommand{\Ba}{\mathbf{a}}
\newcommand{\Bb}{\mathbf{b}}
\newcommand{\Bc}{\mathbf{c}}
\newcommand{\Bd}{\mathbf{d}}
\newcommand{\Be}{\mathbf{e}}
\newcommand{\Bf}{\mathbf{f}}
\newcommand{\Bg}{\mathbf{g}}
\newcommand{\Bh}{\mathbf{h}}
\newcommand{\Bi}{\mathbf{i}}
\newcommand{\Bj}{\mathbf{j}}
\newcommand{\Bk}{\mathbf{k}}
\newcommand{\Bl}{\mathbf{l}}
\newcommand{\Bm}{\mathbf{m}}
\newcommand{\Bn}{\mathbf{n}}
\newcommand{\Bo}{\mathbf{o}}
\newcommand{\Bp}{\mathbf{p}}
\newcommand{\Bq}{\mathbf{q}}
\newcommand{\Br}{\mathbf{r}}
\newcommand{\Bs}{\mathbf{s}}
\newcommand{\Bt}{\mathbf{t}}
\newcommand{\Bu}{\mathbf{u}}
\newcommand{\Bv}{\mathbf{v}}
\newcommand{\Bw}{\mathbf{w}}
\newcommand{\Bx}{\mathbf{x}}
\newcommand{\By}{\mathbf{y}}
\newcommand{\Bz}{\mathbf{z}}
\newcommand{\BB}{\mathbf{B}}
\newcommand{\BC}{\mathbf{C}}
\newcommand{\BP}{\mathbf{P}}
\newcommand{\BR}{\mathbf{R}}
\newcommand{\BS}{\mathbf{S}}
\newcommand{\BK}{\mathbf{K}}
\newcommand{\BG}{\mathbf{G}}
\newcommand{\BA}{\mathbf{A}}
\newcommand{\BE}{\mathbf{E}}
\newcommand{\BF}{\mathbf{F}}
\newcommand{\BM}{\mathbf{M}}
\newcommand{\BN}{\mathbf{N}}
\newcommand{\BD}{\mathbf{D}}
\newcommand{\BI}{\mathbf{I}}
\newcommand{\BL}{\mathbf{L}}
\newcommand{\BX}{\mathbf{X}}
\newcommand{\BY}{\mathbf{Y}}
\newcommand{\BZ}{\mathbf{Z}}
\newcommand{\BW}{\mathbf{W}}
\newcommand{\BQ}{\mathbf{Q}}
\newcommand{\BT}{\mathbf{T}}
\newcommand{\BJ}{\mathbf{J}}
\newcommand{\BH}{\mathbf{H}}
\newcommand{\BV}{\mathbf{V}}
\newcommand{\BO}{\mathbf{O}}
\newcommand{\BU}{\mathbf{U}}
\newcommand{\cP}{\mathcal{P}}
\newcommand{\cM}{\mathcal{M}}
\newcommand{\cA}{\mathcal{A}}
\newcommand{\cE}{\mathcal{E}}
\newcommand{\cB}{\mathcal{B}}
\newcommand{\cC}{\mathcal{C}}
\newcommand{\cD}{\mathcal{D}}
\newcommand{\cR}{\mathcal{R}}
\newcommand{\cL}{\mathcal{L}}
\newcommand{\cN}{\mathcal{N}}
\newcommand{\cV}{\mathcal{V}}
\newcommand{\cT}{\mathcal{T}}
\newcommand{\cI}{\mathcal{I}}
\newcommand{\cO}{\mathcal{O}}
\newcommand{\cK}{\mathcal{K}}
\newcommand{\cF}{\mathcal{F}}
\newcommand{\cH}{\mathcal{H}}
\newcommand{\cS}{\mathcal{S}}
\newcommand{\cX}{\mathcal{X}}
\newcommand{\cY}{\mathcal{Y}}
\newcommand{\cU}{\mathcal{U}}
\newcommand{\cW}{\mathcal{W}}
\newcommand{\cZ}{\mathcal{Z}}

\newcommand{\VV}{V}
\newcommand{\XX}{X}
\newcommand{\YY}{Y}
\newcommand{\FF}{F}
\newcommand{\PP}{P}
\newcommand{\QQ}{Q}
\newcommand{\RR}{R}
\newcommand{\UU}{U}
\newcommand{\EE}{E}
\newcommand{\GG}{G}
\newcommand{\KK}{K}
\newcommand{\LL}{L}
\newcommand{\JJ}{J}
\newcommand{\TT}{T}
\newcommand{\DD}{D}
\newcommand{\CC}{C}
\newcommand{\HH}{H}
\newcommand{\MM}{M}
\newcommand{\ZZ}{Z}
\newcommand{\WW}{W}
\newcommand{\mm}{m}
\newcommand{\pp}{p}
\newcommand{\qq}{q}
\newcommand{\ff}{f}
\newcommand{\cc}{c}
\newcommand{\hh}{h}
\newcommand{\kk}{k}
\newcommand{\II}{I}
\newcommand{\yy}{y}
\newcommand{\ww}{w}
\newcommand{\uu}{u}

\newcommand{\field}{\ff}
\newcommand{\testfun}{\hh}
\newcommand{\dom}{\textrm{dom }}

\newcommand{\scalar}[2]{{\langle}\kern.1em#1,#2\kern.1em{\rangle}}
\newcommand{\scalarC}[2]{{\langle}\kern.1em#1,#2\kern.1em{\rangle}_\conf}
\newcommand{\scalarT}[2]{{\langle}\kern.1em#1,#2\kern.1em{\rangle}_\TT}
\newcommand{\scalarTC}[2]{{\langle}\kern.1em#1,#2\kern.1em{\rangle}_{\TANG\conf}}

\newcommand{\inde}{\,\mathit{d}}
\newcommand{\integrale}[2]{\int_{#1}^{#2}}
\newcommand{\ointegrale}[2]{\oint_{#1}^{#2}}
\newcommand{\Kernel}{\mathrm{Ker}}
\newcommand{\Image}{\mathrm{Im}}
\newcommand{\domain}{\textrm{dom }}
\newcommand{\anti}{\textrm{skew }}

\newcommand{\medvectre[3]}{\left|\begin{array}{ccc}#1\\ #2 \\ #3\end{array}\right|}

\newcommand{\Bendspace}{\Sigma}
\newcommand{\Bendeq}{\Sigma}
\newcommand{\Bendself}{\Sigma_o}
\newcommand{\Bendperm}{\Xi}

\newcommand{\ee}{\varepsilon}
\newcommand{\ttt}{t}
\newcommand{\sss}{s}
\newcommand{\mathscr}{\mathcal}
\newcommand{\Real}{\mathscr{R}}
\newcommand{\tr}{\mathrm{tr}}
\newcommand{\onehalf}{\frac{1}{2}}

\newcommand{\Linmap}[1]{\textrm{\textit{L}}\,\di{#1}}
\newcommand{\spazio}{\,;}

\newcommand{\equi}{\;\Longleftrightarrow\;}
\newcommand{\implica}{\;\Longrightarrow\;}

\newcommand{\sub}[1]{{}_{\lower2pt\hbox{$\scriptstyle#1$}}}
\newcommand{\inv}[1]{#1^{-1}}
\newcommand{\invmezzo}[1]{#1^{-1/2}}
\newcommand{\punto}{\cdot}
\newcommand{\duepunti}{\,:\,}
\newcommand{\suchthat}{\mid}
\newcommand{\equaldef}{:=}
\newcommand{\perogni}{\quad\forall\,}
\newcommand{\medset}[1]{\{\,#1\,\}}
\newcommand{\set}[1]{\{#1\}}
\newcommand{\der}{d}
\newcommand{\lineare}[1]{[#1]}
\newcommand{\talechesia}{\, : \,}
\newcommand{\norma}[1]{\|#1\|}
\newcommand{\enorma}[2]{\norma{#1}_{#2}}
\newcommand{\sqnorma}[1]{\norma{#1}^2}
\newcommand{\aster}{*}
\newcommand{\bil}[3]{#1\,(#2,#3)}
\newcommand{\quadrato}{\blacksquare}
\newcommand{\tonde}[2]{(#1,#2)}
\newcommand{\duetonde}[2]{((#1,#2))}

\def\meas{\textsc{meas}}
\def\signum{\textsc{signum}}
\def\sign{\textbf{sgn}}


\newcommand{\persone}[1]{\textcolor{cyan}{\textsc{#1}}}

\newcommand{\pow}{\textsc{power}}
\newcommand{\umov}{\textsc{umov}}
\newcommand{\powerEH}{\Boo^3_\pow}
\newcommand{\powerflux}{\Boo^2_\umov}

\newcommand{\Romano}{\persone{G. Romano}}
\newcommand{\Newton}{\persone{Newton}}
\newcommand{\Maclaurin}{\persone{Maclaurin}}
\newcommand{\Hodge}{\persone{Hodge}}
\newcommand{\Hilbert}{\persone{Hilbert}}
\newcommand{\Noether}{\persone{Noether}} 
\newcommand{\Hill}{\persone{Hill}}
\newcommand{\Haar}{\persone{Haar}}
\newcommand{\Gram}{\persone{Gram}} 
\newcommand{\Truesdell}{\persone{Truesdell}} 
\newcommand{\Noll}{\persone{Noll}} 
\newcommand{\Thomas}{\persone{Thomas}}
\newcommand{\Zaremba}{\persone{Zaremba}} 
\newcommand{\Jaumann}{\persone{Jaumann}}
\newcommand{\Oersted}{\persone{{\O}rsted}}
\newcommand{\Oldroyd}{\persone{Oldroyd}}
\newcommand{\Bernstein}{\persone{Bernstein}}
\newcommand{\Tricomi}{\persone{Tricomi}}
\newcommand{\Green}{\persone{Green}}
\newcommand{\Gauss}{\persone{Gauss}}
\newcommand{\Umov}{\persone{Umov}}
\newcommand{\Winkler}{\persone{Winkler}}
\newcommand{\Zimmermann}{\persone{Zimmermann}}
\newcommand{\Wieghardt}{\persone{Wieghardt}}
\newcommand{\Gateaux}{\persone{G{\^a}teaux}}
\newcommand{\Fubini}{\persone{Fubini}}
\newcommand{\Riemann}{\persone{Riemann}}
\newcommand{\Leibniz}{\persone{Leibniz}}
\newcommand{\Lie}{\persone{Lie}}
\newcommand{\LeviCivita}{\persone{Levi-Civita}}
\newcommand{\Koszul}{\persone{Koszul}}
\newcommand{\Cauchy}{\persone{Cauchy}}
\newcommand{\Eringen}{\persone{Eringen}}
\newcommand{\Rogula}{\persone{Rogula}}
\newcommand{\Dominik}{\persone{Dominik}}
\newcommand{\Capurso}{\persone{Capurso}}

\newcommand{\Kirchhoff}{\persone{Kirchhoff}}
\newcommand{\Euler}{\persone{Euler}}
\newcommand{\Piola}{\persone{Piola}}
\newcommand{\Eshelby}{\persone{Eshelby}}
\newcommand{\Almansi}{\persone{Almansi}}
\newcommand{\Finger}{\persone{Finger}}
\newcommand{\Schwarz}{\persone{Schwarz}}
\newcommand{\Pythagoras}{\persone{Pythagoras}}
\newcommand{\Lame}{\persone{Lam}}
\newcommand{\Poisson}{\persone{Poisson}}
\newcommand{\Young}{\persone{Young}}
\newcommand{\Poincare}{\persone{Poincar{\'e}}}
\newcommand{\Dieudonne}{\persone{Dieudonn{\'e}}}
\newcommand{\Cartan}{\persone{Cartan}}
\newcommand{\Banach}{\persone{Banach}}
\newcommand{\Whitney}{\persone{Whitney}}
\newcommand{\Legendre}{\persone{Legendre}}
\newcommand{\Euclid}{\persone{Euclid}}
\newcommand{\Leonhard}{\persone{Leonhard}}
\newcommand{\Herman}{\persone{Herman}}
\newcommand{\Navier}{\persone{Navier}}
\newcommand{\Stokes}{\persone{Stokes}}
\newcommand{\StVenant}{\persone{St.Venant}}
\newcommand{\Lagrange}{\persone{Lagrange}}
\newcommand{\Hamilton}{\persone{Hamilton}}
\newcommand{\Maupertuis}{\persone{Maupertuis}}
\newcommand{\Galilei}{\persone{Galilei}}
\newcommand{\Galileo}{\persone{Galileo}}
\newcommand{\Dalembert}{\persone{d'Alembert}}
\newcommand{\Bernoulli}{\persone{Bernoulli}}
\newcommand{\desCartes}{\persone{des Cartes}}
\newcommand{\Coriolis}{\persone{Coriolis}}
\newcommand{\Kronecker}{\persone{Kronecker}}
\newcommand{\Simo}{\persone{Sim\'o}}
\newcommand{\Ortiz}{\persone{Ortiz}}
\newcommand{\Pinsky}{\persone{Pinsky}}
\newcommand{\Pister}{\persone{Pister}}
\newcommand{\Reynolds}{\persone{Reynolds}}
\newcommand{\Liouville}{\persone{Liouville}}
\newcommand{\Koenig}{\persone{K{\"o}nig}}
\newcommand{\Holder}{\persone{H{\"o}lder}}
\newcommand{\Jacobi}{\persone{Jacobi}}
\newcommand{\Klein}{\persone{Klein}}
\newcommand{\Hermann}{\persone{Hermann}}
\newcommand{\Fermat}{\persone{Fermat}}
\newcommand{\Huygens}{\persone{Huygens}}
\newcommand{\Hankel}{\persone{Hankel}}
\newcommand{\Kelvin}{\persone{Kelvin}}
\newcommand{\Thomson}{\persone{Thomson}}
\newcommand{\Ampere}{\persone{Amp{\`e}re}}
\newcommand{\Snell}{\persone{Snell}}
\newcommand{\Pontryagin}{\persone{Pontryagin}}
\newcommand{\Dirac}{\persone{Dirac}}
\newcommand{\Heaviside}{\persone{Heaviside}}
\newcommand{\Poynting}{\persone{Poynting}}
\newcommand{\Fredholm}{\persone{Fredholm}}
\newcommand{\Helmholtz}{\persone{Helmholtz}}
\newcommand{\Pfaff}{\persone{Pfaff}}
\newcommand{\Russel}{\persone{Russel}}
\newcommand{\Marsden}{\persone{Marsden}}
\newcommand{\Hughes}{\persone{Hughes}}
\newcommand{\Naghdi}{\persone{Naghdi}}
\newcommand{\Bruhns}{\persone{Bruhns}}
\newcommand{\Xiao}{\persone{Xiao}}
\newcommand{\Meyers}{\persone{Meyers}}
\newcommand{\Korn}{\persone{Korn}} 
\newcommand{\Hencky}{\persone{Hencky}}
\newcommand{\Liu}{\persone{Liu}}

\newcommand{\Galerkin}{\persone{Galerkin}}
\newcommand{\Riesz}{\persone{Riesz}}
\newcommand{\Frechet}{\persone{Fr{\'e}chet}}

\newcommand{\Polizzotto}{\persone{Polizzotto}}
\newcommand{\Borino}{\persone{Borino}}
\newcommand{\Bazant}{\persone{Bazant}}
\newcommand{\Jirasek}{\persone{Jirasek}}
\newcommand{\Peddieson}{\persone{Peddieson}}
\newcommand{\PijaudierCabot}{\persone{Pijaudier-Cabot}}
\newcommand{\JNReddy}{\persone{J.N. Reddy}}
 \newcommand{\ARSrinivasa}{\persone{A.R. Srinivasa}}
\newcommand{\Khodabakhshi}{\persone{Khodabakhshi}}
\newcommand{\Fernandezsaez}{\persone{Fern{\'a}ndez S{\'a}ez}}
\newcommand{\Barretta}{\persone{R. Barretta}}
\newcommand{\Silling}{\persone{Silling}}

\newcommand{\Langendonck}{\persone{van Langendonck}}
\newcommand{\Sollazzo}{\persone{Sollazzo}}
\newcommand{\Ylinen}{\persone{Ylinen}}
\newcommand{\Mikkola}{\persone{Mikkola}}
\newcommand{\Aifantis}{\persone{Aifantis}}
\newcommand{\Mindlin}{\persone{Mindlin}}
\newcommand{\Timoshenko}{\persone{Timoshenko}}
\newcommand{\Foeppl}{\persone{F{\"o}ppl}}

\newcommand{\Sedov}{\persone{Sedov}}
\newcommand{\Lee}{\persone{Lee}}
\newcommand{\Palais}{\persone{Palais}}
\newcommand{\Donkin}{\persone{Donkin}}
\newcommand{\Godbillon}{\persone{Godbillon}}
\newcommand{\Lichnerowicz}{\persone{Lichnrowicz}}
\newcommand{\Bertrand}{\persone{Bertrand}}
\newcommand{\Volterra}{\persone{Volterra}}
\newcommand{\Deahna}{\persone{Deahna}}
\newcommand{\Frobenius}{\persone{Frobenius}}
\newcommand{\deDonder}{\persone{de Donder}}
\newcommand{\Sommerfeld}{\persone{Sommerfeld}}
\newcommand{\Meshchersky}{\persone{Meshchersky}}
\newcommand{\Cayley}{\persone{Cayley}}
\newcommand{\Buquoy}{\persone{von Buquoy}}
\newcommand{\Delambre}{\persone{Delambre}}
\newcommand{\Fourier}{\persone{Fourier}}
\newcommand{\Arago}{\persone{Arago}}

\newcommand{\Laplace}{\persone{Laplace}}
\newcommand{\Carnot}{\persone{Carnot}}
\newcommand{\Irschik}{\persone{Irschik}}
\newcommand{\Neumann}{\persone{Neumann}}
\newcommand{\Lorentz}{\persone{Lorentz}}
\newcommand{\Maxwell}{\persone{Maxwell}}
\newcommand{\Clerk}{\persone{Clerk}}
\newcommand{\ClerkMaxwell}{\persone{Clerk-Maxwell}}
\newcommand{\ThomsonJJ}{\persone{J.J. Thomson}}

\newcommand{\Clifford}{\persone{Clifford}}
\newcommand{\Hertz}{\persone{Hertz}}
\newcommand{\Voigt}{\persone{Voigt}}
\newcommand{\Coulomb}{\persone{Coulomb}}
\newcommand{\Feynman}{\persone{Feynman}}
\newcommand{\Faraday}{\persone{Faraday}}
\newcommand{\Trautman}{\persone{Trautman}}
\newcommand{\Minkowski}{\persone{Minkowski}}
\newcommand{\Einstein}{\persone{Einstein}}
\newcommand{\Lenz}{\persone{Lenz}}
\newcommand{\Henry}{\persone{Henry}}

\newcommand{\BgS}{{\Bg_\spa}}
\newcommand{\BtauE}{{\Btau_\EVE}}
\newcommand{\Vlift}{\textsc{vlift}}

\newcommand{\FL}[2]{\BF\Bl^{#1}_{#2}}

\newcommand{\TETj}{(\TEVE)_\TjE}
\newcommand{\TTETj}{\TANG\TETj}
\newcommand{\CTANG}{T^*}

\newcommand{\BPs}{\BP^*}

\newcommand{\moto}{\Bphi}
\newcommand{\motoT}{\moto^\Tj}
\newcommand{\motoE}{\moto^\EVE}
\newcommand{\motoVE}{\moto^\VEVE}
\newcommand{\motoHE}{\moto^\HEVE}
\newcommand{\motoC}{\moto^\CTRL}
\newcommand{\dmoto}{\delta\moto}
\newcommand{\dmotoC}{\dmoto^\CTRL}
\newcommand{\dmotoT}{\delta\motoT}
\newcommand{\motoS}{\moto^\EU}
\newcommand{\motoZ}{\moto^\ZEIT}

\newcommand{\vmoto}{\delta\moto}

\newcommand{\motorel}{\moto^\rel}
\newcommand{\motosol}{\moto^\sol}
\newcommand{\motoflu}{\moto^\flu}
\newcommand{\motoske}{\moto^\ske}

\newcommand{\mapmoto}{(\map\push\moto)}
\newcommand{\fourvelmap}{{\Vect_{\mapmoto}}}

\newcommand{\fourvelrel}{{\Vect_\rel}}
\newcommand{\velrel}{{\vect_\rel}}

\newcommand{\massformflu}{\massform_\flu}
\newcommand{\massformsol}{\massform_\sol}

\newcommand{\fourvelSflu}{{\fourvel^\flu_\EU}}
\newcommand{\fourvelSsol}{{\fourvel^\sol_\EU}}
\newcommand{\fourvelSske}{{\fourvel^\ske_\EU}}
\newcommand{\accsol}{{\acc^\sol}}
\newcommand{\accske}{{\acc^\ske}}
\newcommand{\fourvelSrel}{{\fourvel^\rel_\EU}}

\newcommand{\fourvelEsol}{{\fourvel^\sol_\EVE}}
\newcommand{\fourvelEflu}{{\fourvel^\flu_\EVE}}
\newcommand{\fourvelEske}{{\fourvel^\ske_\EVE}}

\newcommand{\EUCLID}{S}
\newcommand{\TEU}{\TANG\EUCLID}
\newcommand{\CTEU}{\CTANG\EUCLID}
\newcommand{\TEUconf}{\TEU_\conf}
\newcommand{\CTEUconf}{\CTEU_\conf}

\newcommand{\deform}{\mathbf{\BB}}
\newcommand\bdeform{\bar\BB}

\newcommand{\data}{\Bd}
\newcommand{\NAT}{\mathbb{N}}
\newcommand{\sqenorma}[2]{\| #1 \|_{#2}^2}
\newcommand{\Dloadres}{{\Delta\Brho_\res}}
\newcommand{\loadres}{{\Brho_\res}}
\newcommand{\Rload}{\Br}
\newcommand{\Dstrainres}{{\Delta\Be_\res}}
\newcommand{\strainres}{{\Be_\res}}

\newcommand{\soluz}{\Bt}



\title{\bf Electrodynamics without Lorentz force}


\author{\bf Giovanni Romano\\\\
\small
University of Naples Federico II\\
\small
Naples, Italy}

\date{}

\maketitle

\begin{abstract}
This communication is devoted to a brief historical framework
and to a comprehensive critical discussion concerning 
foundational issues of Electrodynamics.
Attention is especially focused on the events which, 
about the end of XIX century, 
led to the notion of \Lorentz\ force,
still today ubiquitous
in literature on Electrodynamics.
Is this a noteworthy instance of a rule
which, generated by an improper simplification
of \Maxwell-\ThomsonJJ\ formulation,
is in fact physically untenable but,
this notwithstanding, highly successful.
Modelling of electromagnetic fields and fluxes in spacetime 
respectively as \emph{even} and \emph{odd} spatial differential forms and 
the formulation of induction laws by means of
exterior and \Lie\ derivatives,
make their covariance manifest under any smooth spacetime transformations,
contrary to the usual affirmation in literature
which confines this property to relativistic frame-changes.
A remarkable consequence is that
there is no entanglement between electric and magnetic fields and fluxes 
under special relativity transformations.
In particular, relativistic support to \Lorentz\ force rule is thus deactivated.
For translational motions
of charged bodies immersed in a uniform and constant magnetic field,
the induced electric field in such a frame, 
is equal to one half the 
\Lorentz\ force term.
The qualitative successful application of the
\Lorentz\ force rule to experimental evidence
of special observers
is therefore explained.

\medskip

\keywords{
Electromagnetic induction;%
Lineal rule;
Vortex rule;
Frame covariance;
Lorentz force;
Relativistic entanglement.}
\end{abstract}

\section{Introduction}
\label{sec: Introduction}

The story I am going to tell you,
provides a sound confirmation of a sentence 
by Samuel Langhorne \persone{Clemens}
(1835--1910)
best known by pen name, \persone{Mark Twain}:

\smallskip

\emph{When even the brightest mind in our world has been trained up from childhood 
in a superstition of any kind, 
it will never be possible for that mind, in its maturity, 
to examine sincerely, dispassionately, and conscientiously 
any evidence or any circumstance which shall seem to cast a doubt 
upon the validity of that superstition.}
\par\noindent
Autobiography (1959).

\smallskip

The difficulty evidenced in the sentence
is even greater for those who might have 
actively contributed in disseminating that superstition.%
\footnote{\label{fn: auth}
The author himself graduated in electronic engineering
in 1965, an epoch where a training in differential geometry was not
included in educational plans, even at postgraduate level.
}

Physically biased readers would deem the present treatment
to be rather mathematical than physical in style.
In this respect the famous words by \cite{GalileoGalilei1623} 
about the essential role of Mathematics in modelling physical phenomena,
might however be remembered.%
\footnote{\label{fn: Gal}
\Galileo\ \Galilei, \emph{ad litteram}, in Italian:
\say{La filosofia natu\-rale è scritta in questo grandissimo libro che continuamente ci sta aperto innanzi agli occhi, 
io dico l’universo, ma non si può intendere se prima non s’impara a intender la lingua e conoscer i caratteri nei quali è scritto. 
Egli è scritto in lingua matematica, e i caratteri son triangoli, cerchi ed altre figure geometriche,
senza i quali mezzi è impossibile a intenderne umanamente parola; 
senza questi è un aggirarsi vanamente per un oscuro labirinto.}
}

In this respect, we will show that it is precisely the
improper mathematical treatment of 
basic laws of electrodynamics
which is responsible for long standing issues and vain debates
concerning tentative and flawed interpretations of improper formulations.

For instance, a common misdeed is the confusion 
between convective and parallel derivatives,
with consequent geometric misstatements about frame covariance
of electromagnetic induction laws.

A related source of improper statements
is the usual representation of fields and fluxes 
in terms of vector fields and partial derivatives
rather than in terms of 
differential forms and exterior derivatives.
These differential geometric
notions are in fact the ones directly stemming from
integral formulations of physical laws
and the ones naturally susceptible of a clean geometric treatment.

An early temptative title for this paper was conceived as
\emph{The Tragicomical History of Lorentz Force},
boldly borrowed from 
\emph{The Tragicomical History of Thermodynamics} \citep{Truesdell1980},
authored by Clifford Ambrose \Truesdell\ III,
a master of style and a passionate scholar and historian of Science,
credited with an exceptional range of knowledge, as witnessed by
treatises 
\citep{TruesdellToupin1960,TruesdellNoll1965}
and by widely ranging interests and publications.

This title was eventually dismissed since somebody was
deceptively brought to think that the paper was an historical essay
rather than an original contribution to foundational aspects of Electrodynamics.

The \Lorentz\ force was so
named after the influential dutch physicist 
Hendrik Antoon \Lorentz\ (1853--1928),
who displayed it in \citep{Lorentz1892,Lorentz1899,Lorentz1904}.

The relevant mathematical expression was however formulated 
much before by James \cite{Maxwell1855} 
in his wonderful completion of the pioneering 
formulations of electromagnetism by 
André-Marie \cite{Ampere1826},
Michael \cite{Faraday1838}
and Franz Ernst \cite{FENeumann1846}.%
\footnote{\label{fn: Max}
The celebrated Scottish scientist James \ClerkMaxwell\ (1831--1879),
commonly abridged to \Maxwell\ in literature since the end of XIXth century,
is too often improperly cited as C. \Maxwell, or \Maxwell, J.C.,
even in historical essays \citep{Katz1979,OlivierDarrigol2000,Bucci2014}.}

Still nowadays the role of \Lorentz\ force
is universally considered to be so basic in Electrodynamics
that anyone casting a doubt upon the validity of this notion
may seriously run the risk of being deemed a heretic.

In this respect we must keep in mind an often verifiable
aspect of human belief effectively commented by Bertrand \Russel\ 
by the crude words \citep{Russell1929}:

\emph{The fact that an opinion has been widely held 
is no evidence whatever that it is not utterly absurd; 
indeed in view of the silliness of the majority of mankind, 
a widespread belief is more likely to be foolish than sensible.}

Occasional criticisms in the past years had to face harshly against
a widely spread resort to the notion of  \Lorentz\ force,
spanning from research papers and treatises
on Electrodynamics, to high-school and university textbooks
and presently repeated also in a multitude of web sites \citep{Munley2004}.

Yet difficulties concerning the 
notion of \Lorentz\ force have been early raised
in literature, by authoritative scholars, see fn.\ref{fn: lorforce}.

Vain relativistic arguments
have also been brought in support of \Lorentz\ force
\citep[II.13-6]{Feynman1964}, \citep[Ch.5]{Purcell1965}.

These relativistic interpretations are in manifest contrast with
the evidence of tests concerning the action of magnetic
fields upon a beam of electric point charges
(the cathodic tube rays of Joseph John \Thomson).%
\footnote{\label{fn: Thom}
Successor of \Maxwell\ as Cavendish Professor of Physics
at Cambridge. 
His name is usually abridged to J.J. \Thomson\
to avoid confusion with William \Thomson\ (Lord \Kelvin).}

There in fact speeds far below the limit value 
pertaining to light \emph{in vacuo} are involved.

Accordingly, it is quite reasonable to sustain that
relativistic arguments should not
play any role in this matter.

This is the conclusion of the present analysis,
as we shall eventually see.

To bring a contribution aimed at clarifying the issue,
we will deal with a spacetime formulation of Electrodynamics
and with the 
relativistic phenomenon of electromagnetic entanglement
taken as well-founded in literature.


In special relativity,
the transformation of electromagnetic fields
under \Lorentz\ frame-changes
\citep{Lorentz1904,Poincare1905,Einstein1905a}
were conceived in the wake of early
treatments by Oliver \cite{Heaviside1885,Heaviside1892}
and Heinrich \cite{Hertz1892}.%
\footnote{\label{fn: Min}
According to \citep{Minkowski1908}
the transformations proposed by
\cite{Lorentz1904} and reformulated as a mathematical group by 
Henri \cite{Poincare1905},
were first conceived by Woldemar \cite{VoigtWoldemar1887a,VoigtWoldemar1887b}.}

These pioneering analyses were later reproduced in literature
without significant modifications
\citep{Sommerfeld1952,PanofskiPhillips1962,Feynman1964,Purcell1965,%
MisnerThorneWheeler1973,StephenParrott1987,
LandauLifshits1987,Jackson1999}.

The involved
entanglements of electric and magnetic fields were accordingly
assumed to persist in the classical limit,
that is for a vanishing ratio between boost speed 
and speed of light \emph{in vacuo}.


Inspection of early treatments
\citep{Lorentz1892,Lorentz1904,Poincare1905,Einstein1905a,Einstein1905b}
reveals however that entanglements of electric and magnetic fields
were stipula\-ted under the assumptions of an alleged
\emph{form-invariance} of electromagnetic induction laws 
and of conservation of electric charge under
\Lorentz\ transformation rule,
both contrasted by geometric evidence,
as will be here shown in {\S}\ref{sec: framechange}.

In particular, transformations of \Maxwell-\Hertz\
equations for empty space due to action of a \Lorentz\
frame-change,
were stipulated without explicit proof in \citep[Part II, {\S}4]{Einstein1905a} 
as outcome of a simple substitution, by appealing to \emph{form-invariance}.%
\footnote{\label{fn: Ein}
\Einstein\ statement was literally:
''If we apply to these (\Maxwell-\Hertz) equations 
the (\Lorentz) transformation developed in §3, 
by referring the electromagnetic processes to the system of co-ordinates 
there introduced, moving with the velocity v, we obtain the equations....''.}
With any evidence the procedure there sketched
was a partial rephrasing
of seminal treatments pioneered at the same time
by \cite{Lorentz1904} and \cite{Poincare1905}.
The differential geometric analysis developed 
in {\S}\ref{sec: framechange} of the present contribution,
leads to the outstanding conclusion that the laws of electromagnetic induction,
when properly formulated in terms of integrals of differential forms,
are fully covariant under any frame-change and that
no electromagnetic entanglement does occur.

More precisely, the resulting transformation rule 
may be enunciated in the  
following neat simple terms.

The components of the
transformed electromagnetic fields,
represented by time-vertical exterior forms,
when evaluated in the original framing,
obey the following rule.

Either they undergo an amplification
by the scalar relativistic factor or otherwise remain invariant.

The alternative depending on wether,
in evaluating the component under investigation,
the direction of \Lorentz\ boost 
is included in the list of argument vectors or does not.

The relativistic factor goes to infinity 
when the boost speed tends to 
the limit value of light speed \emph{in vacuo},
and to unity in the classical limit,
for a boost speed smaller and smaller  than
the light speed in vacuo.

Consequently, in the classical limit
the transformed electromagnetic fields tend rapidly to coincide 
with those evaluated under the action of the standard \Galilei\ transformation group.
This is exactly what was to be expected on a physico-mathematical ground,
due to the continuous dependence of \Lorentz\ transformations upon
the relativistic factor.
The contributed amendment to the alleged transformation rules
of special relativity
deactivates any relevance of 
relativistic effects in support of \Lorentz\ force.

Basic features of electrodynamics up to the sixties 
are comprehensively illustrated in the monumental treatise 
\citep[Ch. F]{TruesdellToupin1960}.
Formulations in terms of differential forms were carried out in 
\citep{MisnerThorneWheeler1973} and in
\citep{StephenParrott1987}.

In the recent book \citep{HehlObukhov2003} 
on foundations of classical electrodynamics,
the \Lorentz\ force law is introduced as an additional axiom.
Accurate overviews on formulations of Electrodynamics 
and historical spotlights
have also been contributed
in \citep{HehlObukhov2000,Hehl2010}.

The present treatment starts by observing that the law of electrical induction,
usually stated by means of the well-known \emph{flux-rule},
is rather formulated in the most general and clearest way in terms of path integrals
of electric and magnetic one-forms along arbitrary piecewise smooth paths.

The \emph{flux-rule},
restated here as \emph{vorticity rule},
being expressed in terms of \emph{inner} oriented surfaces
and \emph{even} exterior forms, 
is applicable only to circuits 
which are boundaries of surfaces undergoing regular motions.

Consequently the \emph{vorticity rule}
can be adopted only in simple model cases
of scant applicative interest,
or at best as a convenient approximation
in describing the functioning of technical devices such as 
\emph{solenoids}.

The approach in terms of 
magnetic vector potential 
was the one originally undertaken by
\cite{Ampere1826},
\cite{Maxwell1855,Maxwell1861} and later adopted also by 
Hermann von \cite{Helmholtz1870,Helmholtz1873,Helmholtz1874} 
and by \cite{ThomsonJJ1881,ThomsonJJ1893}.


Simplifying modifications, introduced soon later
by \cite{Heaviside1885,Heaviside1892}, 
\cite{Hertz1892} and \cite{Lorentz1892},
were highly successful in the engineering
community due to the computationally convenient
substitution of the magnetic vector potential field
(named
\emph{magnetic momentum} by \Maxwell)
with its curl, the magnetic induction vector field.%
\footnote{\label{fn: Heav}
\Heaviside\ deemed potentials to be 
treacherous and useless, due to intrinsic lack of uniqueness
\citep{Deschamps1981}.}

This convenience may be na{\"i}vely illustrated by
considering the analogy with the kinematics of
an act of rigid rotation, characterised by a non-uniform
velocity field with a uniform field of curls
(the skew symmetric part of the derivative). 

In introducing the simplification,
a velocity dependent scalar potential in the expression of the electric vector field
was consequently and correctly
ignored as not influential for the computation of the curl.

The resulting induction law was however
improperly still applied
to evaluate the electric field itself and not just its curl,
so that the whole affair went unexpectedly along a wrong way.

As a consequence the expression of the electric vector field 
became velocity dependent,
even under the action of \Galilei\ frame transformations.

\Galilei\ relativity principle of Classical Dynamics, 
stating covariance of the law of motion 
when passing from one inertial frame to another still inertial,
was thus violated.

\goodbreak

To find a way out of this embarrassing situation,
the answer sometimes given
to naturally arising questions about who is measuring the velocity of
a charged particle moving in the magnetic vortex field,
is that
\say{measurements ought to be made in the laboratory frame}!
But which lab?

When the full spacetime expression of the electric vector field
induced by a magnetic momentum is adopted,
the resulting scenario becomes by far different and full covariance of 
electric induction laws, under any 
change of spacetime frame, is attained.

A differential geometric treatment reveals in fact that 
the expression of the electric field induced by a magnetic field is given by 
the negative%
\footnote{\label{fn: lenz}
This is Emil \cite{LenzEmil1834} law.}
of the \Lie\ derivative of magnetic momentum along 
the spacetime motion detected in the observer frame.

This convective derivative behaves in a natural way, 
being covariant under any spacetime frame-change transformation.
When the involved vector fields and the motion
are transformed by covariance according to frame-change,
the spacetime velocity
and the convective derivative along it
also transform by covariance
so that the law of electric induction is still fulfilled
\citep{Electro2013}.

The general formula for the electric field, in terms of 
\Lie\ derivative along the spacetime  motion,
may be split into three additive terms.

The first term, given by the partial time-derivative of the magnetic momentum vortex,
is covariant under any frame-change.
The remaining two terms are both dependent on the spatial velocity 
which is not covariant in the general group of diffeomorphic transformations.

None of these two terms is separately covariant under any frame-change
but their sum is such.

One of them is what in literature
has been abusively labeled \say{\Lorentz\ force}.%
\footnote{\label{fn: lor}
Attribution to \Lorentz\ is historically unfounded.
The term
$\,\vel\times\magind\,$
was in fact introduced, without giving it the meaning of force,
by James \Clerk-\Maxwell\
in \citep{Maxwell1855} 
when he was twenty-four
and Hendrik \Lorentz\ was only two years old.}

The other term is the differential of
a functional given by the inner product between
magnetic momentum and spatial velocity.

In \Maxwell's original treatment \citep[p.485]{Maxwell1865} 
this last term was merged with the functional 
expressing the electrostatic potential,
to simplify the resulting formula
\citep[Eq. D. table 5.2]{Bucci2014}.
The undesired serious collateral effect was however 
that to most scholars this velocity dependent term remained hidden 
behind a symbolic curtain.

Here is located the very beginning of our story.

However, when the attention turned to evaluation of 
the curl of the electric field, the last term disappeared
being the curl of a gradient 
(or, more in general, because the exterior derivative
is nilpotent, which means that
iterated exterior derivatives do
vanish).

It is to be ascribed to merits of \cite{ThomsonJJ1893},
the discoverer of the electron,
to point out the importance of bringing this 
velocity dependent functional back to full visibility.

The independent analysis performed by the author
in \citep{InductionLaws2012}, in the context of a 
differential geometric formulation in spacetime,
and reproduced for convenience below,
fully confirmed the expression
contributed in \cite[Ch VII, Eq(1) p.538]{ThomsonJJ1893}.

This early findings were still unknown to me at the time when
the related theoretical developments were independently carried out
by me relying on the tools of differential geometry.

A practical advantage of the simple \Lorentz\ force formula 
is that it yields the electric
field acting on a moving charge just as cross product of
magnetic momentum and velocity at the point of evaluation.

The further neglected term requires in addition to compute the spatial differential
of the inner product between
magnetic momentum and spatial velocity.

This differential does not vanish even in case of a uniform magnetic vortex
since the magnetic momentum,
being its potential form, is not spatially uniform.

Therefore knowledge of the involved fields in a neighbourhood
of that point is needed.

Electrical engineers certainly
will not be glad of having lost the possibility of adopting 
the convenient and simple, but untenable formula, 
provided by the \Lorentz\ force rule, for their computations.

An effective, even if partial, remedy to their disappointment,
and a motivation of the many qualitative successful applications of
\Lorentz\ force rule to experimental evidence will be given
in {\S}\ref{sec: unifmag}.

Indeed the explicit calculation of the 
force acting on a beam of charged particles, 
in motion with uniform velocity
under the action of a time-constant and uniform magnetic momentum,
will be carried out therein.


When the correct local expression of 
the induction law is applied by an observer who
evaluates a time-constant and uniform magnetic momentum 
acting on a beam of charged particles
in motion with uniform velocity,
a simple result is shown to hold true.

In fact in this special case 
the partial time-derivative of magnetic momentum vanishes
and the sum of the remaining two velocity dependent terms
give a result equal to one-half the expression of \Lorentz\ force.

This correction factor of one-half is in accord
with findings in \citep{ThomsonJJ1881},
by a different procedure,
as discussed in \citep[p.430]{OlivierDarrigol2000}.

The limited validity of halved \Lorentz\ force rule, 
which is confined just to special observers, 
deprives the term of the physical meaning of force,
a caveat about \Lorentz\ force rule
already clearly expressed in \citep[XVI-2, p.248]{Hertz1892},
but completely ignored ever since.%
\footnote{{\ }\label{fn: lorforce}%
In \citep[XVI-2, p.248]{Hertz1892} 
the statement concerning the \Lorentz\ \emph{force} was:%
\say{Now the resultant of $\,\terna{\XX_1}{\XX_2}{\XX_3}\,$
is an electric force which arises as soon as a body moves in the magnetic field.
It is that force which in a narrower sense we are accustomed 
to denote as the electromotive force induced through the motion.
But it should be observed that, according to our views, the separation of this from 
the total force can have no physical meaning.}
}
The relevance of the factor one-half
is evident since the unit of measure
for the magnetic vortex is still presently fixed on the basis of the 
\Lorentz\ force rule.


\section{Exterior derivatives}
\label{sec: extder}

A multilinear function, 
mapping a list of vector fields on a manifold
to a target linear space,
is said to be \emph{tensorial} 
if the value at a point depends only on the values of the argument vector fields
at that point, viz. the map \say{lives at points} \citep{Spivak1970}.

Differential forms (or simply \emph{forms})
are fields of piecewise smooth alternating $\kk$-tensors
on a manifold $\,\mani\,$ with finite
geometric dimension $\,\dim\di\mani=\mm\,$.

The linear space of $\,\kk\,$-forms on $\,\mani\,$
will be denoted by $\,\FORMS^\kk\di{\TT\mani}\,$,
with $\,\TT\,$ tangent functor.

All $\,\kk$-forms $\,\Boo^{\kk}\,$ on $\,\mani\,$,
due to the alternating property
(their value change sign when to arguments are swapped)
vanish if the argument list vectors are linear dependent.

Therefore forms with $\,\kk>\mm\,$ vanish identically.

Forms of maximal degree $\,\Boo^{\nn}\,$ 
on a manifold $\,\manin\,$ ($\,\dim\di\manin=\nn\,$)
(volume forms)
are proportional one another
and are the geometric objects that can be integrated on a
$n$D compact manifold $\,\manin\,$.

\goodbreak

The exterior derivative  
of a $\,(\nn-1)$-form $\,\Boo^{\nn-1}\,$ on $\,\manin\,$
is the $\,\nn$-form $\,\extder\Boo^{\nn-1}\,$ on $\,\manin\,$
fulfilling
\Kelvin-\Stokes-\Volterra\ 
integral formula:%
\footnote{\label{fn: vito}
The general theorem is due to Vito
\cite{Volterra1889a,Volterra1889b}
with subsequent reformulations by Henri \Poincare\ and {\'E}lie \Cartan\
\citep{Poincare1887,CartanE1899}, see
Victor Joseph \cite{Katz1979}
and Hans \cite{Samelson2001}.
By extending a 2D formula due to Andr{\'e}-Marie \Ampere,
William \Thomson\ (lord \Kelvin) communicated 
the 3D result 
in a letter on July 1850
to Gabriel \Stokes\ who lectured on it in Cambridge.
It is commonly referred to as \Stokes' formula,
even sometimes with awful typo, as \persone{Stoke}'s formula.}
\begin{equation}
\integrale{\manin}{}\extder\Boo^{\nn-1}=\ointegrale{\pmanin}{}\Boo^{\nn-1}\,.
\label{fm: VS}
\end{equation}
for any compact $\,\nn$-submanifold $\,\manin\subset\mani\,$ 
with
\begin{equation}
\dim\di\manin=\nn\le\mm\,,
\label{fm: }
\end{equation}
with boundary $\,\pmanin\,$ ($\,\dim\di\pmanin=\nn-1\,$),
and any differentiable form $\,\Boo^{\nn-1}\,$ on $\,\manin\,$.

Iterated boundary operator of any manifold generates a null manifold,
and the iterated exterior derivative of any differential form generates a zero form:%
\begin{equation}
\partial\partial=\zerovec\equi\extder\extder=\zerovec\,.
\label{fm: nihilpotent}
\end{equation}

This equivalence may be deduced by rewriting Eq.\eqref{fm: VS} 
 as a duality relation between 
 exterior derivative $\,\extder\,$ and boundary operator $\,\partial\,$:
\begin{equation}
\scalar{\extder\Boo^{\nn-1}}{\manin}=\scalar{\Boo^{\nn-1}}{\pmanin}\,.
\label{fm: duality}
\end{equation}
Then:%
\begin{equation}
\setlength{\jot}{8pt}
\begin{aligned}
\scalar{\Boo^{\nn-2}}{\partial\pmanin}
&\,=\scalar{\extder\Boo^{\nn-2}}{\pmanin}
\\
&\,=\scalar{\extder\extder\Boo^{\nn-2}}{\manin}=0\,.
\end{aligned}
\label{fm: }
\end{equation}

\section{Lie and covariant derivatives}
\label{sec: liecovder}

The flow:
\begin{equation}
\FL{\vect}{\lambda}:\mani\mapsto\mani\,,
\label{fm: }
\end{equation}
of a tangent vector field 
$\,\vect:\mani\mapsto\TT\mani\,$ is composed of integral 
envelopes parametrised so that:%
\begin{equation}
\vect=\parder\lambda0\FL{\vect}{\lambda}\,.
\label{fm: flow}
\end{equation}

\medskip\goodbreak

The \Lie\ or \emph{convective} derivative%
\footnote{\label{fn: conv}
\emph{Convective} derivatives where first considered by \cite{Maxwell1855} 
and \cite{Helmholtz1858}. 
The \Lie\ derivatives of general tensor fields were introduced by
\cite{SlebodinskiWladyslaw1931}.
The naming after \Lie\ is due to David \cite{Dantzig1932}.
}
of a vector field $\,\Bu:\mani\mapsto\TT\mani\,$
along a vector field $\,\vect:\mani\mapsto\TT\mani\,$
is the $\,\lambda\,$-derivative of the pull-back
along the flow $\,\FL{\vect}{\lambda}\,$:%
\begin{equation}
\setlength{\jot}{8pt}
\begin{aligned}
\Lieder_{\vect}\di\Bu
\equaldef&\,\parder\lambda0\di{\FL{\vect}{\lambda}\pull\Bu}
\\
=&\,\parder\lambda0
\Bigdi{\TT\FL{\vect}{-\lambda}\punto\di{\Bu\circ\FL{\vect}{\lambda}}}\,.
\end{aligned}
\label{fm: lieder}
\end{equation}

The letter $\,\TT\,$ denotes the tangent functor 
which to a smooth map between two manifolds associates the 
corresponding differential map relating the relevant tangent bundles,
in a fiberwise linear manner.

The symbols $\,\push,\pull\,$ are push, pull operations
on tensors induced by tangent maps.

The \Lie\ derivative $\,\Lieder_{\vect}\,$
and the \emph{parallel} (also named \emph{covariant}) derivative $\,\nabla_{\vect}\,$ 
along a vector field $\,\vect:\mani\mapsto\TT\mani\,$ 
differs in the way backward evaluation is performed.

In \Lie\ derivatives the evaluation tool is a pull-back $\,\pull\,$ 
along the flow $\,\FL{\vect}{\lambda}:\mani\mapsto\mani\,$
so that the values of the field $\,\vect:\mani\mapsto\TT\mani\,$  
in a neighbourhood of the evaluation point are involved.

Dependence on $\,\vect:\mani\mapsto\TT\mani\,$ is therefore not tensorial.
In \emph{parallel} derivatives the evaluation tool is 
a backward parallel transport $\,\back\,$ along the curve
$\,\FL{\vect}{\lambda}:\mani\mapsto\mani\,$
and the only restriction to that curve of the field to be differentiated 
is significant.

Moreover the result at a point $\,\Bx\in\mani\,$
depends linearly on the sole vector $\,\vect_\Bx\in\TT_\Bx\mani\,$,
so that the parallel derivative $\,\nabla_{\vect}\,$ is tensorial in $\,\vect\,$.

\Lie\ derivative $\,\Lieder\,$,
covariant derivative $\,\nabla\,$  
and exterior derivative $\,\extder\,$
are coincident for scalar fields.

The \Lie\ derivative and the covariant derivative of tensor fields 
are defined by a formal application of \Leibniz\ rule, 
taking into account invariance
of scalar fields under pull-back by a flow 
and under backward parallel transport along a curve
\citep{Spivak1970}.

\section{Spacetime framings}
\label{sec: spacetime}

The ambient of a proper electromagnetic analysis
is the $4$D spacetime manifold $\,\EVE\,$
without boundary
and its tangent bundle $\,\TEVE\,$.

Each observer endows the tangent bundle $\,\TEVE\,$
with two geometric fields:
\begin{enumerate}
\itemsep=5pt
\item
A nowhere vanishing
field of tangent time-arrows
$\,\timearrow:\EVE\mapsto\TEVE\,$,
pointing towards the future and
named \emph{rigging} \citep{FriedmanMichael1983}
or \emph{observer field} \citep{Fecko1997}
according to the suggestive language of physicists.
\item
A \emph{clock} one-form %
\footnote{\label{fn: dual}
As customary, a superscript $\,*\,$ denotes duality.}
$\,\timeform:\EVE\mapsto\TEVEs\,$ which is closed, i.e. such that:
\begin{equation}
\extder\timeform=\zerovec\,.
\label{fm: timeformvan}
\end{equation}
\end{enumerate}

It is convenient to stipulate,
between the \emph{clock} and the future pointing \emph{observer field},
fulfilment of the \emph{tuning} relation: 
\begin{equation}
\scalar{\timeform}{\timearrow}=1\,.
\label{fm: tuning}
\end{equation}

\Volterra's theorem (\Poincare\ Lemma)
states that in star shaped manifolds
closed forms are exact.%
\footnote{\label{fn: volter}
This theorem, first proved by Vito \cite{Volterra1889a,Volterra1889b},
and subsequently quoted by Henri \cite{Poincare1899}, 
is known in literature
as \Poincare\ Lemma \citep{Samelson2001}.}
The potential $\,\tE:\EVE\mapsto\ZEIT\,$ is 
 defined to within an additive constant by
 the requirement:
\begin{equation}
\timeform=\dtE\,.
\label{fm: timediff}
\end{equation}
The map $\,\tE:\EVE\mapsto\ZEIT\,$
defines time-projection (surjective submersion)
onto an oriented $1$D \emph{time-axis} $\,\ZEIT\,$.%
\footnote{\label{fn: subm}
A submersion is a map with a surjective differential at any point.
$\,\ZEIT\,$ is initial of the German word \emph{Zeit} meaning \emph{time}.}

\medskip\goodbreak

\Deahna-\Frobenius\ theorem,%
\footnote{\label{fn: deahna}
First proved by Feodor \cite{DeahnaFeodor1840},
and later investigated upon and simplified by 
Ferdinand Georg \cite{FrobeniusGeorg1875},
the result is commonly known as \Frobenius\ Theorem \citep{Samelson2001}.}
provides the condition for	
integrability of the time-vertical tangent distribution,
composed of tangent vector fields $\,\Vect:\EVE\mapsto\TEVE\,$
fulfilling the \Pfaff\ condition 
$\,\scalar\timeform\Vect=\zerovec\,$,
in the form of
vanishing of the exterior derivative in Eq.\eqref{fm: timeformvan}.
For a proof see \citep{KMS1993,MarsdenRatiuAbraham2003}.

\medskip\goodbreak

The spacetime manifold $\,\EVE\,$ is doubly foliated into:
\begin{enumerate}
\itemsep=5pt
\item
Leaves of \emph{isochronous} events ($\,3$D \emph{spatial slices}), 
integral manifolds of the kernel distribution of $\,\dtE\,$.
\item
Lines of \emph{isotopic} events ($1$D spatial positions).
\end{enumerate}
They are mutually transversal due to \emph{tuning} Eq.\eqref{fm: tuning}.

By item 1 it is licit to consider the
\emph{time-vertical} subbundle 
$\,\VEVE\,$ of the tangent bundle $\,\TEVE\,$
whose fibers are slices of \emph{isochronous} events (\emph{spatial slices}).

Spacetime tensor fields of degree greater than zero
are \emph{time-vertical} if they vanish
when any of their arguments is time-horizontal, i.e. tangent to a time-line,
and are \emph{time-horizontal}
if they vanish
when any of their arguments is time-vertical, i.e. tangent to a spatial slice.

\begin{definition}[Framing]\label{lm: framing}
An observer is described in geometrical terms 
by a field of rank-one linear projectors on the time-\emph{rigging}
$\,\timearrow:\EVE\mapsto\TEVE\,$,
named a \emph{framing}:%
\begin{equation}
\splitZ\equaldef\timeform\otimes\timearrow\,.
\label{fm: framing}
\end{equation}
Then, for all $\,\BX\in\TANG\EVE\,$:
\begin{equation}
\splitZ\BX
=(\timeform\otimes\timearrow)\BX=\scalar{\dtE}{\BX}\,\timearrow\,.
\label{fm: }
\end{equation}

The \emph{idempotency} property,
characteristic of linear projectors,
is equivalent to \emph{tuning}:
\begin{equation}
\splitZ\splitZ=\splitZ\equi\scalar\timeform\timearrow=1_\ZEIT\circ\tE\,.
\label{fm: tuningequiv}
\end{equation}
The \emph{time-vertical} complementary projector defined by
$\,\splitS=\BI-\splitZ\,$, is then \emph{idempotent} too:
\begin{equation}
\setlength{\jot}{8pt}
\begin{aligned}
\splitS\splitS&\,=(\BI-\splitZ)(\BI-\splitZ)
=\BI-\splitZ-\splitZ+\splitZ\splitZ
\\
&\,=\BI-\splitZ=\splitS\,.
\end{aligned}
\label{fm: name}
\end{equation}
Then also:
\begin{equation}
\setlength{\jot}{8pt}
\left\{
\begin{aligned}
&\,\splitS\splitZ=\splitZ\splitS=\zerovec\,,
\\
&\,\splitZ\BZ=\BZ\,,
\\
&\,\splitS\BZ=\zerovec\,,
\\
&\,\nucleo\di\timeform=\image\di\splitS\,.
\end{aligned}
\right.
\label{fm: name}
\end{equation}
\end{definition}

\section{Trajectory and motion}
\label{sec: difvector}

A \emph{motion} along a trajectory submanifold
$\,\TjE\subset\EVE\,$
is a one-parameter 
($\lapse=$ time-lapse)
commutative group of automorphic \emph{movements} 
$\,\moto_\lapse:\TjE\mapsto\TjE\,$,
with $\,\moto_0\,$ the identity and the group composition rule:
\begin{equation}
\moto_\lapse\circ\moto_\blapse
=\moto_\blapse\circ\moto_\lapse
=\moto_{(\lapse+\blapse)}\,.
\label{fm: group}
\end{equation}
A \emph{movement} $\,\moto_\lapse\,$ is a trajectory automorphism covering the
time translation $\,\shift_\lapse\di{\ttt}\equaldef\ttt+\lapse\,$ so that:
\begin{equation}
\tE\circ\moto_\lapse=\shift_\lapse\circ\tE\,.
\label{fm: timetrasl}
\end{equation}
This means isochronous events at time $\,\ttt\,$
are mapped into isochronous events at time $\,\ttt+\lapse\,$.

Taking the derivative $\,\parder\lapse0\,$
of Eq.\eqref{fm: timetrasl},
the spacetime velocity:
\begin{equation}
\fourvel=\parder\lapse0\moto_\lapse:\EVE\mapsto\TEVE\,,
\label{fm: Vel}
\end{equation}
fulfils the property:
\begin{equation}
\setlength{\jot}{8pt}
\begin{aligned}
\scalar{\dtE}{\fourvel}
&\,=\parder\lapse0\di{\tE\circ\moto_\lapse}
\\
&\,=\parder\lapse0\di{\shift_\lapse\circ\tE}
=1_\ZEIT\circ\tE
\,.
\end{aligned}
\label{fm: velspacetime}
\end{equation}
From Eq.\eqref{fm: framing} and Eq.\eqref{fm: velspacetime} we get:
\begin{equation}
\splitZ\fourvel=\scalar{\dtE}{\fourvel}\punto\timearrow=\timearrow\,.
\label{fm: }
\end{equation}
The spacetime velocity then splits into time-vertical 
$\,\vel=\splitS\fourvel\,$
and time-horizontal
$\,\timearrow\,$
components, according to the formula:
\begin{equation}
\fourvel
=\splitS\fourvel+\timearrow
=\vel+\timearrow\,.
\label{fm: spatimecomp}
\end{equation}
The tangent bundle $\,\TEVE\,$ is accordingly split as direct sum of
a time-vertical bundle $\,\VEVE\,$ and a time-horizontal bundle $\,\HEVE\,$,
with $\,\VEVE=\image\di\splitS\,$ and $\,\HEVE=\image\di\splitZ\,$.

\section{Splitting the motion}
\label{sec: motionsplit}

The spacetime motion can be split into commutative chain compositions
of time-vertical and time-horizontal flows:
\begin{equation}
\moto_\lapse=\motoVE_\lapse\circ\motoHE_\lapse
=\motoHE_\lapse\circ\motoVE_\lapse\,.
\label{fm: motosplit}
\end{equation}
Taking the derivative $\,\parder{\lapse}{0}\,$
we infer:
\begin{equation}
\fourvel=\fourvelVE+\fourvelHE\,,
\label{fm: splitVel}
\end{equation}
where
\begin{equation}
\setlength{\jot}{8pt}
\left\{
\begin{aligned}
\fourvelVE=\parder{\lapse}{0}\motoVE_{\lapse}\,,
\\
\fourvelHE=\parder{\lapse}{0}\motoHE_{\lapse}\,.
\end{aligned}
\right.
\label{fm: defsplitVel}
\end{equation}
Being $\,\fourvelHE=\timearrow\,$,
by uniqueness of the additive decomposition Eq.\eqref{fm: spatimecomp}
we infer:
\begin{equation}
\fourvelVE=\splitS\fourvel=\vel\,.
\label{fm: }
\end{equation}

\section{Spacetime and spatial homotopy formulae}
\label{sec: Spatimeext}

To a spacetime 1-form $\,\BOO^1\in\FORMS^1\di\TEVE\,$
there corresponds a spatial 1-form 
$\,\Boo^1=\splitS\pull\BOO^{1}\in\FORMS^1\di\VEVE\,$
defined by:
\begin{equation}
\setlength{\jot}{8pt}
\begin{aligned}
\Boo^1\di{\Vect}=&\,\di{\splitS\pull\BOO^1}\di{\Vect}
\\
\equaldef&\,\BOO^1\di{\splitS\Vect}\,,
\perogni\Vect\in\TEVE\,,
\end{aligned}
\label{fm: }
\end{equation}
which is time-vertical since:
\begin{equation}
\Boo^{1}\di{\timearrow}=0\,.
\label{fm: vertcomp}
\end{equation}
Similarly for any \kk-form.

We will need some basic results
concerning convective and exterior derivatives in spacetime,
posted below in 
Eq.\eqref{fm: transport}, Eq.\eqref{fm: extrusion}, Eq.\eqref{fm: STdiffhomo} and Eq.\eqref{fm: diffhomo}.%
\footnote{\label{fn: state}
For one-forms the homotopy formula is due to James
\cite{Maxwell1861,Maxwell1873}
and for
two-forms and three-forms to Hermann von
\cite{Helmholtz1874}.
According to Andrzej \cite{Trautman2008}, 
the proof for differential forms of any degree
is due to \'Elie \cite{CartanE1922}.
The statement in terms of spatial forms was
introduced in \citep{Manifolds2017}.
}
\begin{lemma}[Extrusion in spacetime]
\label{lem: STtraspextr}
Let us consider a spacetime motion 
$\,\moto_\lapse:\EVE\mapsto\EVE\,$
with velocity 
$\,\fourvel\equaldef\parder\lapse0\moto_\lapse:\EVE\mapsto\TEVE\,$.
The time-rate of variation of the integral of a spacetime form 
$\,\BOO^k\,$ with $\kk\le\dim\di\EVE\,$, 
over the moving image of a $\,\kk$D compact submanifold $\,\surf\,$,
is expressed in terms of \Lie-derivative by the transport formula:
\begin{equation}
\setlength{\jot}{8pt}
\begin{aligned}
\parder\lapse0\integrale{\moto_\lapse\di\surf}{}\BOO^\kk
&\,=\parder\lapse0\integrale{\surf}{}\moto_\lapse\pull\BOO^\kk
\\
&\,=\integrale{\surf}{}\Lieder_{\fourvel}\di{\BOO^\kk}
\,.
\end{aligned}
\label{fm: transport}
\end{equation}
In terms of spacetime exterior derivatives
we get the extrusion formula:
\begin{equation}
\parder\lapse0\integrale{\moto_\lapse\di\surf}{}\BOO^\kk
=\integrale{\surf}{}\di{\extder\BOO^\kk}\punto\fourvel
+\integrale{\surf}{}\extder\di{\BOO^\kk\punto\fourvel}\,.
\label{fm: extrusion}
\end{equation}
\end{lemma}

\begin{lemma}[Spacetime homotopy formula]
\label{lem: SThomo}
Substituting the transport formula Eq.\eqref{fm: transport} 
into the extrusion formula Eq.\eqref{fm: extrusion}
and localising we get:
\begin{equation}
\setlength{\jot}{8pt}
\begin{aligned}
\Lieder_{\fourvel}\di{\BOO^\kk}
&\,=\di{\extder\BOO^\kk}\punto\fourvel
+\extder\di{\BOO^\kk\punto\fourvel}\,.
\end{aligned}
\label{fm: STdiffhomo}
\end{equation}
This gives the expression of the 
\Lie\ derivative of a spacetime form 
$\,\BOO^k\,$ with $\kk\le\nn+1=\dim\di{\TT_\Be\EVE}\,$
for all $\,\Be\in\EVE\,$,
along the motion velocity field 
$\,\fourvel:\EVE\mapsto\TEVE\,$,
in terms of spacetime exterior derivatives.
\end{lemma}
\begin{proposition}[Spatial homotopy formula]
\label{lem: SPhomo}
The \Lie\ derivative, along the spatial motion velocity field $\,\vel:\EVE\mapsto\VEVE\,$,
of a spatial form $\,\Boo^k\,$ with $\kk\le\nn=\dim\di{\VV_\Be\EVE}\,$,
for all $\,\Be\in\EVE\,$,
is expressed in terms of exterior derivatives by:
\begin{equation}
\Lieder_\vel\di{\Boo^\kk}
=\di{\extder\Boo^\kk}\punto\vel+\extder\di{\Boo^\kk\punto\vel}\,.
\label{fm: diffhomo}
\end{equation}
\end{proposition}
\begin{proof}
Setting $\,\fourvel=\timearrow\,$ and $\,\BOO^\kk=\Boo^\kk\,$
in Eq.\eqref{fm: STdiffhomo},
by time-verticality of $\,\Boo^\kk\,$ Eq.\eqref{fm: vertcomp} we get:
\begin{equation}
\setlength{\jot}{8pt}
\begin{aligned}
\Lieder_\timearrow\di{\Boo^\kk}
&\,=\di{\extder\Boo^\kk}\punto\timearrow
+\extder\di{\xcancel{\Boo^\kk\punto\timearrow}}
\,.
\end{aligned}
\label{fm: diffhomospe}
\end{equation}
Splitting
$\,\fourvel=\splitS\fourvel+\timearrow\,$
as in Eq.\eqref{fm: spatimecomp}
and setting $\,\BOO^\kk=\Boo^\kk\,$
in Eq.\eqref{fm: STdiffhomo},
by linearity of exterior and \Lie\ derivatives,
recalling Eq.\eqref{fm: diffhomospe} we get:
\begin{equation}
\setlength{\jot}{8pt}
\begin{aligned}
\Lieder_{(\splitS\fourvel)}\di{\Boo^\kk}
=\di{\extder\Boo^\kk}\punto\splitS\fourvel
+\extder\di{\Boo^\kk\punto\splitS\fourvel}
\,.
\end{aligned}
\label{fm: prefinhomo}
\end{equation}
Consequently Eq.\eqref{fm: diffhomo} follows from Eq.\eqref{fm: prefinhomo} 
by setting $\,\vel\equaldef\splitS\fourvel\,$.
\end{proof}

Denoting by $\,\intprod\,$ the contraction operator,
Eq.\eqref{fm: diffhomo} can be written 
as a relation between graded derivatives
$\,\Lieder,\extder,\intprod\,$
respectively of degree $\,0,1,-1\,$, \citep{KMS1993}:
\begin{equation}
\Lieder=\intprod\circ\extder+\extder\,\circ\intprod\,.
\label{fm: gradder}
\end{equation}

\section{Differential forms versus vector fields}
\label{sec: difvector}

Denoting by $\,\metric:\VEVE\mapsto\dual{(\VEVE)}\,$ the spatial metric tensor field,
the electric field one-form $\,\eleform\,$
and the magnetic momentum one-form $\,\magmom\,$
may be expressed by:%
\footnote{\label{fn: upper}
Uppercase letters here adopted are standard for 
vector fields in Electrodynamics.}
\begin{equation}
\setlength{\jot}{8pt}
\left\{
\begin{aligned}
\eleform&=\metric\punto\elefield\,,
\\
\magmom&=\metric\punto\magvecpot\,.
\end{aligned}
\right.
\label{fm: elemag}
\end{equation}
with $\,\elefield\,$ electric vector field and
$\,\magvecpot\,$ magnetic vector potential.

\goodbreak

Denoting by $\,\volformg\,$ the volume-form compatible with the metric,%
\footnote{\label{fn: unit}
This means the unit cube has unitary volume.}
the magnetic vortex vector field $\,\magind\,$ is related to
the magnetic vortex two-form $\,\magcurl\,$ by:
\begin{equation}
\magcurl=\volformg\punto\magind\,.
\label{fm: magcurl}
\end{equation}
Then
\begin{equation}
\magcurl\punto\vel
=\volformg\punto\magind\punto\vel
=\metric\punto\Bigdi{\magind\times\vel}\,.
\label{fm: vectprod}
\end{equation}

Exterior derivatives of forms and 
differential operators on vector fields are related by:
\begin{equation}
\setlength{\jot}{8pt}
\left\{
\begin{aligned}
&\,\extder\eleform=\extder\di{\metric\punto\elefield}
=\volformg\punto\Bigdi{\rotor\di\elefield}\,,
\\
&\,\extder\magmom=\extder\di{\metric\punto\magvecpot}
=\volformg\punto\Bigdi{\rotor\di\magvecpot}\,,
\\
&\,\extder\magcurl
=\extder\Bigdi{\volformg\punto\magind}
=\Bigdi{\diverg\di\magind}\punto\volformg\,,
\,.
\end{aligned}
\right.
\label{fm: extdermagcurl}
\end{equation}
The magnetic vortex $2$-form $\,\magcurl\,$ and
the magnetic momentum $1$-form $\,\magmom\,$
are spatial fields related by
\begin{equation}
\magcurl=\extder\magmom\,.
\label{fm: mompot}
\end{equation}
In terms of vector fields:
\begin{equation}
\magind=\rotor\di\magvecpot\,.
\label{fm: mompotvec}
\end{equation}
The
Eq.\eqref{fm: mompot}-\eqref{fm: mompotvec}
are expressions of \Gauss\ principle
stating nonexistence of magnetic charges:
\begin{equation}
\extder\magcurl=\zerovec
\equi
\diverg\di\magind=\zerovec\,.
\label{fm: }
\end{equation}

In \Euclid\ spacetime, push along the flow generated by
the time-arrows field
and parallel transport along time-lines are coincident
so that time-independence of the time-vertical metric tensor field
is expressed by
\begin{equation}
{\Lieder_\timearrow}\di\metric={\nabla_\timearrow}\di\metric=\zerovec\,.
\label{fm: notimemetric}
\end{equation}

\section{Electric induction}
\label{sec: Einduction}

\begin{definition}
The overall electromotive force $\,\emf\di\lineinn\,$
along an inner oriented spatial path $\,\lineinn\,$
is the integral:%
\begin{equation}
\vcenter{\halign{
\hfil$\displaystyle#$&$#$\hfil&$#$\hfil&$#$\hfil\cr
\emf\di\lineinn
\equaldef
\integrale{\lineinn}{}\eleform\,.
\cr}}
\label{fm: FBLdis}
\end{equation}
The electric field $\,\eleform\,$ is an \emph{even} one-form.%
\footnote{\label{fn: inner}
\emph{Inner} and \emph{outer} oriented manifolds and 
\emph{even} and \emph{odd} (or \emph{twisted}) forms
are treated in 
\citep{Schouten1951,deRham1955,Tonti1995,Marmo2005}.
\emph{Even} forms are simply exterior forms
to be integrated on \emph{inner} oriented manifolds.
\emph{Odd} forms are to be integrated on \emph{outer} oriented manifolds.
Their sign changes by changing orientation of ambient manifold.
\emph{Odd} forms are best described by sets made of two opposite pairs, 
each one made of an exterior form and of an orientation.
\emph{Even} forms represent circulations and vortices, 
\emph{odd} forms have the meaning of 
sources, winding around and flux through.
A thorough discussion is offered in \citep{Bossavit1998}.
}
\end{definition}
\begin{proposition}[Lineal electric induction]
Along any spatial inner oriented path $\,\lineinn\,$
dragged by a piecewise regular spacetime motion
$\,\moto_\lapse:\TjE\mapsto\TjE\,$,
the induced electromotive force $\,\emf\di\lineinn\,$
is given by the negative time-rate of the magnetic momentum
along the motion:
\begin{equation}
\normboxed{\,
\vcenter{\halign{
\hfil$\displaystyle#$&$\displaystyle#$\hfil&$#$\hfil&$#$\hfil\cr
\integrale{\lineinn}{}\eleform
=&\,-\,\parder\lapse0
\integrale{\moto_\lapse\di{\lineinn}}{}\magmom
\,.\cr}}
\,}
\label{fm: EIL}
\end{equation}
\end{proposition}

Applying \Lie-\Reynolds\ transport formula:%
\footnote{\label{fn: most}
In most treatments of electromagnetics
a parallel derivative along the motion appears 
in place of the \Lie\ derivative.
An instance is provided by the treatment in \citep[{\S}1.3.4, p.12--14]{ThideBo2012}
which is consequently erroneous.
The formulation of \Faraday's law of induction
in \citep[{\S}9-3, p.160]{PanofskiPhillips1962} 
is emblematic of the physicists' way of deriving
the homotopy formula for the \Lie\ derivative.
}
\begin{equation}
\parder\lapse0
\integrale{\moto_\lapse\di{\lineinn}}{}\magmom
=\integrale{\lineinn}{}{\Lieder_{\fourvel}}\di\magmom\,,
\label{fm: }
\end{equation}
and localising the integral Eq.\eqref{fm: EIL}, we get the rule:
\begin{equation}
\normboxed{\,
-\eleform
=\Lieder_{\fourvel}\di\magmom\,.
\,}
\label{fm: DiffEIL}
\end{equation}
The \Lie\ derivative of the spatial  metric tensor field $\,\metric\,$
along spacetime motion,
taking into account the time-independence 
in Eq.\eqref{fm: notimemetric}, is given by:%
\begin{equation}
\setlength{\jot}{8pt}
\begin{aligned}
\Lieder_{\fourvel}\di\metric
=\Lieder_{\vel}\di\metric
=\metric\punto2\,\Eul\di\vel\,,
\end{aligned}
\label{fm: EUL}
\end{equation}
The \Euler\ stretching tensor is given by:
\begin{equation}
\Eul\di\vel\equaldef\inv\metric\punto\unmezzotext\,\Lieder_{\vel}\di\metric=\sym\nabla\di\vel\,,
\label{fm: Eul}
\end{equation}
with $\,\nabla\,$ the connection in \Euclid\ spacetime.

\begin{proposition}[Spacetime split]
Splitting the spacetime velocity 
into the sum of space and time components
the law of electric induction Eq.\eqref{fm: DiffEIL}
becomes:%
\begin{equation}
\kern-5pt
\normboxed{\,
\begin{aligned}
-\eleform={\Lieder_\timearrow}\di\magmom
+\magcurl\punto\vel
+\extder(\magmom\punto\vel)
\,,
\label{fm: DiffEILsplit}
\end{aligned}
\,}
\end{equation}
and in vectorial terms, by time-independence Eq.\eqref{fm: notimemetric}:%
\begin{equation}
\normboxed{\,
-\,\elefield={\Lieder_\timearrow}\di\magvecpot
+\magind\times\vel
+\nabla\bigdi{\metric\coppia\magvecpot\vel}\,.
\label{fm: DiffEILvectsplit}
\,}
\end{equation}
The convective derivative of the field $\,\magvecpot\,$
at r.h.s. is due to \cite{Helmholtz1892} and
Eq.\eqref{fm: DiffEILvectsplit}
is the electric induction law exposed in
\citep[Eq(1) p.534]{ThomsonJJ1893}.
\end{proposition}
\begin{proof}
By additivity of \Lie\ derivative,
splitting the spacetime velocity into time and spatial components gives:%
\begin{equation}
\Lieder_{\fourvel}\di{\magmom}
=\Lieder_\timearrow\di{\magmom}
+\Lieder_{\vel}\di\magmom\,.
\label{fm: diffsplit}
\end{equation}
Setting $\,\Boo^{\kk}=\magmom\,$,
the spatial homotopy formula Eq.\eqref{fm: diffhomo}
yields Eq.\eqref{fm: DiffEILsplit}.

By Eq.$\eqref{fm: elemag}_2$ 
$\,\magmom=\metric\punto\magvecpot\,$
and by time-independence Eq.\eqref{fm: notimemetric}
we may rewrite in vectorial terms:
\begin{equation}
\setlength{\jot}{8pt}
\left\{
\begin{aligned}
\Lieder_\timearrow\di{\metric\punto\magvecpot}
&\,=\metric\punto\Lieder_\timearrow\di{\magvecpot}\,,
\\
\Lieder_{\vel}\di{\metric\punto\magvecpot}
&\,=\extder\di{\metric\punto\magvecpot}\punto\vel
+\extder(\metric\punto\magvecpot\punto\vel)\,.
\end{aligned}
\right.
\label{fm: EILsplitformsvect}
\end{equation}
Observing that:
\begin{equation}
\setlength{\jot}{8pt}
\left\{
\begin{aligned}
\extder\di{\metric\punto\magvecpot}\punto\vel
&\,=\volformg\punto\magind\punto\vel
=\metric\punto\bigdi{\magind\times\vel}\,,
\\
\extder(\metric\punto\magvecpot\punto\vel)
&\,=\metric\punto\nabla\bigdi{\metric\di{\magvecpot\punto\vel}}
\,,
\end{aligned}
\right.
\label{fm: name}
\end{equation}
and that by Eq.$\eqref{fm: elemag}_1$
$\,\eleform=\metric\punto\elefield\,$,
we get Eq.\eqref{fm: DiffEILvectsplit}.
\end{proof}

\begin{proposition}[Vorticity rule]\label{prop: fluxrule}
In the special case when the path $\,\lineinn\,$ 
is the boundary of an inner oriented spatial surface $\,\surfinn\,$
undergoing a regular motion, 
we have:
\begin{equation}
\setlength{\jot}{8pt}
\left\{
\begin{aligned}
&\,\lineinn=\partial\surfinn\,,
\\
&\,\partial\lineinn=\partial\partial\surfinn=\zerovec\,.
\end{aligned}
\right.
\label{fm: name}
\end{equation}
Applying \Stokes-\Volterra\ formula
to the integral at r.h.s. of Eq.\eqref{fm: EIL}
we get the \Lenz-\Faraday\ \emph{rule}:%
\footnote{\label{fn: vorticityvsflux}
The denomination \emph{flux rule}
was changed to \emph{vorticity rule} to conform with the
assumption of an inner oriented surface and boundary path,
with a clearer physical meaning and in accord with
\Maxwell\ point of view.}
\begin{equation}
\normboxed{\,
\vcenter{\halign{
\hfil$\displaystyle#$&$#$\hfil&$#$\hfil&$#$\hfil\cr
-\ointegrale{\lineinn}{}\eleform
=\parder\lapse0\integrale{\moto_\lapse\di{\surfinn}}{}\magcurl\,.
\cr}}
\,}
\label{fm: FVR}
\end{equation}
\end{proposition}
\begin{proof}
Being $\,\partial\lineinn=\partial\partial\surfinn=0\,$, we get
\begin{equation}
\vcenter{\halign{
\hfil$\displaystyle#$&$\displaystyle#$\hfil&$#$\hfil&$#$\hfil\cr
\ointegrale{\moto_\lapse\di{\partial\surfinn}}{}\magmom
&\,=\ointegrale{\partial\di{\moto_\lapse\di{\surfinn}}}{}\magmom
\vspace{8pt}\cr
&\,=\integrale{\moto_\lapse\di{\surfinn}}{}\extder\magmom
\,.\cr}}
\label{fm: toSFL}
\end{equation}
Then Eq.\eqref{fm: mompot} and the \EIL\  Eq.\eqref{fm: EIL}  
give Eq.\eqref{fm: FVR}.
From \Gauss\ law
$\,\extder\magcurl=\zerovec\,$,
stating absence of magnetic monopoles,
we have for any inner oriented $3$D domain $\,\bulkinn\,$:
\begin{equation}
\ointegrale{\partial\bulkinn}{}\magcurl
=\integrale{\bulkinn}{}\extder\magcurl=0
\,.
\label{fm: nomagcharge}
\end{equation}
so that independence of the choice of an inner oriented surface $\,\surfinn\,$
such that $\,\lineinn=\partial\surfinn\,$ is inferred.
\end{proof}

Localisation of the integrals in Eq.\eqref{fm: FVR}
and recalling Eq.\eqref{fm: magcurl},
we get the differential law:
\begin{equation}
-\extder\eleform
={\Lieder_{\fourvel}}\di{\magcurl}
={\Lieder_{\fourvel}}\di{\volformg\punto\magind}
\,.
\label{fm: Diffvort}
\end{equation}
From Eq.\eqref{fm: extdermagcurl}
applying \Leibniz\ rule,
Eq.\eqref{fm: Diffvort} 
may be expressed, 
in terms of the magnetic induction vector field $\,\magind\,$,
as:
\begin{equation}
-\volformg\punto\rotor\di\elefield
=\volformg\punto{\Lieder_{\fourvel}}\di\magind
+\Bigdi{ {\Lieder_{\fourvel}}\di\volformg }\punto\magind
\label{fm: Diffvortvect}
\end{equation}
Denoting by $\,\lininv\,$ the linear invariant,
and setting $\,\vect=\splitS\Vect\,$
for any spacetime tangent vector field $\,\Vect:\EVE\mapsto\TEVE\,$,
we have \citep{Manifolds2017}:
\begin{equation}
\setlength{\jot}{8pt}
\begin{aligned}
\Lieder_{\Vect}\di{\volformg}
&\,=\lininv\Bigdi{
\inv\Bg\punto\unmezzotext\,\Lieder_{\vect}\di{\Bg}
}
\punto\volformg
\\
&\,=\lininv\di{\EUL\di\vect}\punto\volformg\,.
\end{aligned}
\label{fm: gfinalorth}
\end{equation}
Then by definition of divergence:
\begin{equation}
\Lieder_\vel\di\volformg=\diverg\di\vel\punto\volformg\,,
\label{fm: }
\end{equation}
from Eq.\eqref{fm: notimemetric} we infer
\begin{equation}
\setlength{\jot}{8pt}
\begin{aligned}
\Lieder_{\fourvel}\di\volformg
&\,=\Lieder_\vel\di\volformg+\xcancel{{\Lieder_\timearrow}\di\volformg}
\\
&\,={\diverg\di\vel}\punto\volformg\,,
\end{aligned}
\label{fm: }
\end{equation}
with $\,\diverg\di\vel=\lininv\di{\EUL\di\vel}\,$
volumetric stretching.
The vorticity rule Eq.\eqref{fm: Diffvortvect} then writes:
\begin{equation}
\normboxed{\,
-\rotor\di\elefield
={\Lieder_{\fourvel}}\di\magind
+{\diverg\di\threevel}\punto\magind\,.
\,}
\label{fm: Diffvortvectexpl}
\end{equation}

The vorticity rule 
Eq.\eqref{fm: FVR}
is independent of the choice of a surface $\,\surfinn\,$
such that $\,\lineinn=\partial\surfinn\,$.

The proof is readily got by appealing to Eq.\eqref{fm: mompot} and
to \Stokes\ formula:
\begin{equation}
\integrale{\moto_\lapse\di{\surfinn}}{}\extder\magmom
=\ointegrale{\moto_\lapse\di{\partial\surfinn}}{}\magmom
=\ointegrale{\moto_\lapse\di{\lineinn}}{}\magmom
\,.
\label{fm: surfin}
\end{equation}

\begin{remark}
Performing the splitting:
\begin{equation}
\Lieder_{\fourvel}\di\magcurl
=\Lieder_{\timearrow}\di\magcurl
+\Lieder_{\vel}\di\magcurl\,,
\label{fm: diffsplitcurl}
\end{equation}
and applying Eq.\eqref{fm: magcurl} and Eq.\eqref{fm: diffhomo}
an alternative expression for Eq.\eqref{fm: Diffvortvect} is got.

Indeed, from
\begin{equation}
-\volformg\punto\rotor\di\elefield
={\Lieder_\timearrow}\di{\volformg\punto\magind}
+{\Lieder_\vel}\di{\volformg\punto\magind}\,,
\label{fm: Diffvortvectdue}
\end{equation}
recalling that $\,{\Lieder_\timearrow}\di\volformg=\zerovec\,$,
we get
\begin{equation}
-\volformg\punto\rotor\di\elefield
=\volformg\punto{\Lieder_\timearrow}\di{\magind}
+{\Lieder_\vel}\di{\volformg\punto\magind}\,,
\label{fm: Diffvortvectdue}
\end{equation}
and from the homotopy formula Eq.\eqref{fm: diffhomo}:
\begin{equation}
{\Lieder_\vel}\di{\volformg\punto\magind}
=\extder\di{\volformg\punto\magind}\punto\vel
+\extder\di{\volformg\punto\magind\punto\vel}\,.
\label{fm: }
\end{equation}
Being
\begin{equation}
\extder\di{\volformg\punto\magind}
=\diverg\di{\magind}\punto\volformg\,,
\label{fm: name}
\end{equation}
and
\begin{equation}
\setlength{\jot}{8pt}
\left\{
\begin{aligned}
\extder\di{\volformg\punto\magind\punto\vel}
&\,=\extder\bigdi{\metric\punto\di{\magind\times\vel}}
\\
&\,=\volformg\punto\bigdi{\rotor\di{\magind\times\vel}}\,,
\end{aligned}
\right.
\label{fm: name}
\end{equation}
we eventually get:
\begin{equation}
-\,\rotor\di\elefield={\Lieder_\timearrow}\di\magind
+\rotor\di{\magind\times\vel}
+\xcancel{\diverg\di\magind}\punto\vel\,.
\label{fm: rotDiffEILvectsplit}
\end{equation}
The divergence in the last term of 
Eq.\eqref{fm: rotDiffEILvectsplit}
vanishes due to \Gauss\ principle expressed by
Eqs.\eqref{fm: mompot}, \eqref{fm: mompotvec}.
The r.h.s of Eq.\eqref{fm: rotDiffEILvectsplit}
is the expression of the convective derivative due to 
Hermann \cite{Helmholtz1858} and Kazimierz \cite{Zorawski1900}.
\end{remark}

\subsection{Motion in a uniform magnetic field}
\label{sec: unifmag}

The split lineal electric induction law Eq.\eqref{fm: DiffEILvectsplit}
reveals that, if the magnetic vortex $\,\magind\,$
is time-independent ($\,\Lieder_\timearrow\di\magind=\zerovec\,$)
and spatially uniform ($\,\nabla\di\magind=\zerovec\,$)
then the induced electric field is expressed by:
\begin{equation}
\normboxed{\,
\elefield=\onehalf\Bigdi{\vel\times\magind}\,.
\label{fm: }
\,}
\end{equation}
The proof of this result 
is made of two steps aimed at establishing the evaluation:
\begin{equation}
\extder(\magmom\punto\vel)
=-\onehalf\Bigdi{\magcurl\punto\vel}\,.
\label{fm: }
\end{equation}

\medskip\goodbreak

To this end, 
we observe that, by an application of \Leibniz\ rule,
the \Lie\ derivative
of a spatial one-form $\,\Boo^1\,$ or two-form $\,\Boo^2\,$,
along the flow of a time-vertical vector field $\,\vect\in\VEVE\,$,
is expressed, in terms of parallel derivatives $\,\nabla\,$ 
according to a connection with vanishing torsion,
as:
\begin{subequations}
\setlength{\jot}{8pt}
\begin{align}
&\,(\Lieder_\Bv-\nabla_\Bv)\di{\Boo^1}
=\,\dual{\nabla\di\Bv}\punto\Boo^1\,,
\label{fm: Liecov3}
\\
&\,(\Lieder_\Bv-\nabla_\Bv)\di{\Boo^2}
=\Boo^2\punto\nabla\di\Bv
+\,\dual{\nabla\di\Bv}\punto\Boo^2\,,
\label{fm: Liecov4}
\end{align}
\end{subequations}
where the star $\,{}^*\,$ denotes duality.

\begin{theorem}[Linear Faraday potential]\label{lm: potlin}
A spatially uniform magnetic vortex two-form
$\,\nabla\di{\magcurl}=\zerovec\,$,
admits a magnetic momentum potential one-form
$\,\magmom\,$, that is
$\,\magcurl=\extder\magmom\,$,
having the linear distribution:
\begin{equation}
\normboxed{\,
\magmom
\equaldef\unmezzotext\extder\magmom\punto\identvec
=\unmezzotext\magcurl\punto\identvec
=\unmezzotext\volformg\punto\magind\punto\identvec\,.
\,}
\label{fm: spaconst}
\end{equation}
to within the differential of a scalar potential.
The position vector field $\,\identvec\,$ is defined by 
\begin{equation}
\identvec\di\Bp\equaldef\Bx\,,
\label{fm: }
\end{equation}
for all $\,\Bx=\Bp-\Bo\,$.
\end{theorem}
\begin{proof}
For any increment of position $\,\Bh\,$ we have
\begin{equation}
\setlength{\jot}{8pt}
\begin{aligned}
\nabla_\Bh\di\identvec
&\,=\lim_{\epsilon\to0}\inv\epsilon(\identvec\di{\Bp+\epsilon\Bh}-\identvec\di\Bp)
\\
&\,=\lim_{\epsilon\to0}\inv\epsilon(\Bx+\epsilon\Bh-\Bx)
=\Bh\,,
\end{aligned}
\label{fm: identder}
\end{equation}
so that, denoting by $\,\BI\,$ identity map
and by $\,\dual\BI\,$ the dual identity map:
\begin{equation}
\setlength{\jot}{8pt}
\left\{
\begin{aligned}
\nabla\di\identvec&\,=\BI\,,
\\
(\nabla\identvec)^*&\,=\dual\BI\,.
\end{aligned}
\right.
\label{fm: identanddual}
\end{equation}

Under the assumption 
$\,\nabla\di{\magcurl}=\zerovec\,$,
the homotopy formula and the expression in 
Eq.\eqref{fm: Liecov4}
of \Lie\ derivative in terms of parallel derivative, 
recalling \Gauss\ law Eq.\eqref{fm: mompot} 
and Eq.\eqref{fm: identanddual}, give:
\begin{equation}
\setlength{\jot}{8pt}
\begin{aligned}
\extder(\magcurl\punto\identvec)
&\,=\Lieder_{\identvec}\di{\magcurl}
-\xcancel{(\extder\magcurl)\punto\identvec}
\\
&\,\kern-60pt=\xcancel{\nabla_{\identvec}\di{\magcurl}}
+\magcurl\punto\nabla\identvec
+(\nabla\identvec)^*\punto\magcurl
=2\,\magcurl\,,
\end{aligned}
\label{fm: finfor}
\end{equation}
which was to be proved.
\end{proof}
\begin{theorem}[Electric field on translating charges]
\label{prop: unmezzo}
A charged body in translational motion, across a region of spatially uniform 
magnetic vortex $\,\nabla\magcurl=0\,$,
experiences an electric field given by
\begin{equation}
\setlength{\jot}{8pt}
\begin{aligned}
-\eleform&\,=\Lieder_{\timearrow}\di{\magmom}
+{\unmezzotext\magcurl\punto\vel}\,,
\\
&\,=\Lieder_{\timearrow}\di{\magmom}
-{\extder\di{\magmom\punto\vel}}\,.
\end{aligned}
\label{fm: trasl}
\end{equation}
In vector terms
\begin{equation}
\setlength{\jot}{8pt}
\begin{aligned}
-\elefield&\,={\Lieder_\timearrow}\di\magvecpot-\unmezzotext\,(\vel\times\magind)\,,
\\
&\,={\Lieder_\timearrow}\di\magvecpot-\extder\,\metric\coppia{\magvecpot}{\vel}\,.
\end{aligned}
\label{fm: traslvec}
\end{equation}
\end{theorem}
\begin{proof}
Let an observer be detecting a translational motion 
$\,\moto_\lapse\in\cont^1\di{\TjE\spz\TjE}\,$ 
and measuring the spacetime velocity 
$\,\fourvel\equaldef\parder\lapse0\moto_{\lapse}=\vel+\timearrow\,$,
whose spatial component is uniform, i.e. $\,\nabla\vel=\zerovec\,$.

From the expression of \Lie\ derivative of a one-form
in terms of parallel derivatives
Eq.\eqref{fm: Liecov3},
we get
\begin{equation}
\setlength{\jot}{8pt}
\begin{aligned}                             
\Lieder_{\fourvel}\di{\magmom}
&\,=\nabla_{\fourvel}\di{\magmom}
+\xcancel{(\nabla\fourvel)^*\punto\di{\magmom}}\,.
\end{aligned}      
\label{fm: finforuno}
\end{equation}
Being $\,\nabla\di{\magcurl}=\zerovec\,$, 
Lemma \ref{lm: potlin} gives
\begin{equation}
\magmom=\unmezzotext\magcurl\punto\identvec\,.
\label{fm: }
\end{equation}
Then
\begin{equation}
\Lieder_{\fourvel}\di{\magmom}
=\nabla_{\fourvel}\di{\magmom}
=\unmezzotext\di{\magcurl}\punto\vel\,.
\label{fm: }
\end{equation}

By assumption $\,\extder\elestatform=\zerovec\,$
so that from $\,\eqref{fm: DiffEIL}\,$:
\begin{equation}
\setlength{\jot}{8pt}
\begin{aligned}
-\di{\eleform}
&\,=\Lieder_{\fourvel}\di\magmom
\\
&\,=\Lieder_\timearrow\di\magmom
\Lieder_\vel\di\magmom\,,
\\
&\,=\Lieder_\timearrow\di\magmom
+\unmezzotext\magcurl\punto\vel\,.
\end{aligned}
\label{fm: name}
\end{equation}
Finally the computation
\begin{equation}
\setlength{\jot}{8pt}
\begin{aligned}
\extder(\magmom\punto\vel)
&\,=\unmezzotext\extder\Bigdi{\extder\magmom\punto\identvec\punto\vel}
\\
&\,=-\unmezzotext\extder\Bigdi{\di{\extder\magmom}\punto\vel\punto\identvec}
\\
&\,=-\unmezzotext\di{\extder\magmom}\punto\vel\,,
\end{aligned}
\label{fm: elepot}
\end{equation}
shows electric field has a velocity linearly dependent potential.
\end{proof}

\section{Magnetic induction}
\label{sec: Minduction}

Let us consider an outer oriented surface $\,\surf\,$
with boundary $\,\partial\surf\,$
and define the magnetomotive force $\,\mmf\di{\partial\surf}\,$
along an outer oriented spatial path $\,\lineout\,$ by:
\begin{equation}
\mmf\di{\lineout}\equaldef\integrale{\lineout}{}\magwind\,,
\label{fm: }
\end{equation}
and adopt the expressions:
\begin{equation}
\setlength{\jot}{8pt}
\left\{
\begin{aligned}
\eleflux&\,=\volformg\punto\eledisp\,,
\\
\magwind&\,=\metric\punto\magfield\,,
\\
\elecurrflux&\,=\volformg\punto\elecurr\,,
\\
\elechargeform&\,=\rho\punto\volformg\,.
\end{aligned}
\right.
\label{fm: elevect}
\end{equation}
The spatial vector fields are:
$\,\eledisp\,$ electric displacement,
$\,\magfield\,$ magnetic winding,
$\,\elecurr\,$ electric current. 
The scalar field $\,\rho\,$ is the electric charge density per unit volume.
\begin{proposition}[Magnetic induction law]
When the spatial outer oriented circuit $\,\lineout\,$
is dragged by a piecewise regular spacetime motion
$\,\moto_\lapse:\TjE\mapsto\TjE\,$,
a magnetomotive force $\,\mmf\di{\lineout}\,$ is induced.

In regions where there are no electric charges and no sources of
electric currents, so that:
\begin{equation}
\setlength{\jot}{8pt}
\left\{
\begin{aligned}
&\,\extder\eleflux=\zerovec\equi\eleflux=\extder\elefluxpot\,,
\\
&\,\;\extder\elecurrflux=\zerovec\equi\;\elecurrflux=\extder\elecurrpot\,,
\end{aligned}
\right.
\label{fm: nocharge}
\end{equation}
the magnetomotive force is given by the
time-rate of the global electric flux potential
plus the global electric current potential,
along the path $\,\lineout\,$:
\begin{equation}
\normboxed{\,
\vcenter{\halign{
\hfil$\displaystyle#$&$\displaystyle#$\hfil&$#$\hfil&$#$\hfil\cr
\integrale{\lineout}{}\magwind
=&\,\parder\lapse0
\integrale{\moto_\lapse\di{\lineout}}{}\elefluxpot
+\integrale{\lineout}{}\elecurrpot
\,.\cr}}
\,}
\label{fm: MILopen}
\end{equation}
Out of these regions,
the magnetic force induced
along the boundary 
$\,\lineout=\partial\surfout\,$
of a spatial surface $\,\surfout\,$
can be expressed as
time-rate of the global electric displacement flux
plus the global electric current flux,
by the formula:
\begin{equation}
\normboxed{\,
\vcenter{\halign{
\hfil$\displaystyle#$&$\displaystyle#$\hfil&$#$\hfil&$#$\hfil\cr
\ointegrale{\partial\surfout}{}\magwind
=&\,\parder\lapse0
\integrale{\moto_\lapse\di{\surfout}}{}\eleflux
+\integrale{\surfout}{}\elecurrflux
\,.\cr}} 
\,}
\label{fm: MIL}
\end{equation}
\end{proposition}

Independence of the choice of a spatial surface $\,\surfout\,$ 
such that $\,\partial\surfout=\lineout\,$,
is inferred from balance of electric charge along the motion:

\begin{proposition}[Balance of charges and currents]
Through the boundary $\,\partial\bulkout\,$
of an outer oriented spatial domain $\,\bulkout\,$
the total outward flux of currents (due to electric displacement and free charges)
is vanishing:
\begin{equation}
\normboxed{\,
\parder\lapse0
\ointegrale{\moto_\lapse\di{\partial\bulkout}}{}\eleflux
+\ointegrale{\partial\bulkout}{}\elecurrflux=0\,.
\,}
\label{fm: currbal}
\end{equation}
\end{proposition}
\begin{proof}
Being
$\,\moto_\lapse\di{\partial\bulkout}=\partial\di{\moto_\lapse\di{\bulkout}}\,$,
and setting:
\begin{equation}
\extder\eleflux=\elechargeform=\rho\punto\volformg\,,
\label{fm: coulombbal}
\end{equation}
equivalence between Eq.\eqref{fm: currbal} and
\Coulomb's balance of electric charge along the motion:
\begin{equation}
\normboxed{\,
\parder\lapse0
\integrale{\moto_\lapse\di{\bulkout}}{}\elechargeform
+\ointegrale{\partial\bulkout}{}\elecurrflux=0\,,
\,}
\label{fm: chargecons}
\end{equation}
follows from \Stokes\ formula.
\end{proof}
Localising Eq.\eqref{fm: MILopen} and Eq.\eqref{fm: MIL} we get:
\begin{equation}
\setlength{\jot}{8pt}
\left\{
\begin{aligned}
\magwind=\Lieder_{\fourvel}\di\elefluxpot+\elecurrpot\,,
\\
\extder\magwind=\Lieder_{\fourvel}\di\eleflux+\elecurrflux\,.
\end{aligned}
\right.
\label{fm: diffMIL}
\end{equation}
Eq.$\eqref{fm: diffMIL}_1$ may be rewritten in terms of the splitting:
\begin{equation}
\begin{aligned}
\Lieder_{\fourvel}\di\elefluxpot={\Lieder_\timearrow}\di\elefluxpot
+\eleflux\punto\vel
+\extder(\elefluxpot\punto\vel)
\,.
\label{fm: DiffMILsplit}
\end{aligned}
\end{equation}
Taking the exterior derivative of Eq.$\eqref{fm: diffMIL}_2$
we get
\begin{equation}
\extder\elecurrflux=\Lieder_{\fourvel}\di{\extder\eleflux}
\,,
\label{fm: }
\end{equation}
which by Eq.$\eqref{fm: elevect}_3$ and Eq.\eqref{fm: coulombbal}
leads to the vectorial expression:
\begin{equation}
\diverg\di\elecurr+\Lieder_{\fourvel}\di{\rho}+\rho\punto\diverg\di{\vel}=0\,.
\label{fm: }
\end{equation}

\section{Splitting spacetime forms}
\label{sec: SSF}

Spacetime forms, introduced by
Hermann \cite{Minkowski1907},
and further discussed by
Harry \cite{Bateman1910}
elaborating on ideas by \cite{Hargreaves1908},
were later investigated by {\'E}lie \cite{CartanE1924}.
More recent treatments and applications to Electrodynamics 
were contributed in 
\citep{ThorneMacdonald1982,Fecko1997}.

\begin{lemma}[Splitting]\label{lm: fourtothreemoto}
A framing $\,\splitZ\equaldef\timeform\otimes\timearrow\,$
induces a representation
for spacetime $\,\kk$-forms $\,\Exter\,$
in terms of the time-vertical projector 
$\,\splitS=\BI-\splitZ\,$, the time-arrows $\,\timearrow\,$
and the time differential $\,\timeform\,$:
\begin{equation}
\normboxed{\,
\Exter=
\splitS\pull\Exter+\timeform\wedge(\Exter\punto\timearrow)\,.
\,}
\label{fm: splitting}
\end{equation}
\end{lemma}
\begin{proof}
The spatial restriction $\,\splitS\pull\Exter\,$ of a spacetime $\,\kk$-form $\,\Exter\,$
is defined, for
$\,\Ba_{1},\ldots,\Ba_{\kk}\in\VEVE\,$
by:
\begin{equation}
\di{\splitS\pull\Exter}\di{\Ba_{1},\ldots,\Ba_{\kk}}
=\Exter\di{\splitS\Ba_{1},\ldots,\splitS\Ba_{\kk}}\,.
\label{fm: }
\end{equation}
Let us denote spacetime tangent vectors by
\begin{equation}
\BY_{i}:\EVE\mapsto\TEVE\,,\;\,i=1,\ldots,k-1\,.
\label{fm: }
\end{equation}
Adopting the notations:
\begin{equation}
\vcenter{\halign{
\hfil$#$&$#$\hfil&$#$\hfil&$#$\hfil\cr
\BY&\,=\set{\BY_{1},\ldots,\BY_{k-1}}\,,\qquad
\vspace{8pt}\cr
\splitS\BY&\,=\set{\splitS\BY_{1},\ldots,\splitS\BY_{k-1}}\,,
\vspace{8pt}\cr
(\splitS)_{\ii}\BY&\,=\set{\splitS\BY_{1},.,\splitS\BY_{i-1},\splitS\BY_{i+1},.,\splitS\BY_{k}}\,,
\vspace{8pt}\cr
(\splitZ)_{\ii}\BY&\,=\set{\splitS\BY_{1},\ldots,\splitZ\BY_{\ii},\ldots,\splitS\BY_{k-1}}
\,.\cr}}
\label{fm: splittingnotation}
\end{equation}
and taking into account the skew character of forms, 
the result is given by the computation:
\begin{equation}
\vcenter{\halign{
\hfil$#$&$#$\hfil&$#$\hfil&$#$\hfil\cr
&\,\Exter\coppia\BX{\BY}
=\Exter\coppia{\splitS\BX+\splitZ\BX}{\BY}
\vspace{8pt}\cr
&\,=\Exter\coppia{\splitS\BX}{\splitS\BY}
+\Exter\coppia{\splitZ\BX}{\splitS\BY}
\vspace{8pt}\cr
&\,+\sum_{i=1,k-1}\Exter\coppia{\splitS\BX}{(\splitZ)_{\ii}\BY}
\vspace{8pt}\cr
&\,=(\splitS\pull\Exter)\coppia{\BX}{\BY}+\scalar{\timeform}{\BX}\,\Exter\coppia{\timearrow}{\splitS\BY}
\vspace{8pt}\cr
&\,+\sum_{i=1,k-1}(-1)^{i}\scalar{\timeform}{\BY_{i}}\,\Exter\terna{\timearrow}{\splitS\BX}{(\splitS)_{\ii}\BY}
\vspace{8pt}\cr
&\,=\Bigdi{\splitS\pull\Exter+\timeform\wedge(\Exter\punto\timearrow)}\coppia{\BX}{\BY}\,.
\cr}}
\label{fm: comput}
\end{equation}
This concludes the proof.
\end{proof}
An alternative  proof of Lemma \ref{lm: fourtothreemoto}
may be given in terms of components 
by evaluating the spacetime form on a basis 
made of exterior products
of differential of adapted coordinates on $\,\TEVE\,$.

Grouping the ones that do include the differential $\,\dtE\,$ and those that do not,
gives the result 
\citep{Bateman1910,StephenParrott1987,BennTucker1987,Fecko2014}.

This $\,3+1\,$ decomposition of the graded algebra $\,\FORMS\di\TEVE\,$
of spacetime differential forms,
was elegantly revisited in \citep{Fecko1997,Fecko2014} 
by introducing the linear operators:%
\begin{equation}
\setlength{\jot}{8pt}
\left\{
\begin{aligned}
&\immers_\timearrow:\FORMS^\kk\di\TEVE\mapsto\FORMS^{\kk-1}\di\TEVE\quad(\mathrm{contraction})
\,,
\\
&\extens_\timeform:\FORMS^\kk\di\TEVE\mapsto\FORMS^{\kk+1}\di\TEVE \quad(\mathrm{extension})
\,,
\end{aligned}
\right.
\label{fm: name}
\end{equation}
defined by
\begin{equation}
\setlength{\jot}{8pt}
\left\{
\begin{aligned}
&\immers_\timearrow\,\Exter=\Exter\punto\timearrow
\,,
\\
&\extens_\timeform\,\Exter=\timeform\wedge\Exter
\,.
\end{aligned}
\right.
\label{fm: name}
\end{equation}
It is easily verified that the swapped compositions:
\begin{equation}
\setlength{\jot}{8pt}
\left\{
\begin{aligned}
\immers_\timearrow\,\extens_\timeform:\FORMS^\kk\di\TEVE\mapsto\FORMS^\kk\di\TEVE\,,
\\
\extens_\timeform\,\immers_\timearrow:\FORMS^\kk\di\TEVE\mapsto\FORMS^\kk\di\TEVE\,,
\end{aligned}
\right.
\label{fm: name}
\end{equation}
are complementary projectors in $\,\FORMS\di\TEVE\,$.
In fact:
\begin{equation}
\setlength{\jot}{8pt}
\begin{aligned}
\extens_\timeform\,\immers_\timearrow\,\Exter
&\,=\timeform\wedge(\Exter\punto\timearrow)
\,,
\end{aligned}
\label{fm: }
\end{equation}
so that, by graded \Leibniz\ rule (contraction is a derivation of degree $\,-1\,$,
we have:
\begin{equation}
\setlength{\jot}{8pt}
\begin{aligned}
\immers_\timearrow\,\extens_\timeform\,\Exter
&\,=(\timeform\wedge\Exter)\punto\timearrow
\\
&\,=\scalar{\timeform}{\timearrow}\punto\Exter
-\timeform\wedge(\Exter\punto\timearrow)
\,.
\end{aligned}
\label{fm: decompbis}
\end{equation}
Inserting the tuning property Eq.\eqref{fm: tuning} in Eq.\eqref{fm: decompbis}
gives the result.
A comparison with Eq.\eqref{fm: splitting} leads to the conclusion that
the spatial restriction is given by
$\,\splitS\pull=\immers_\timearrow\,\extens_\timeform\,$.

\medskip\goodbreak

In view of the applications of spacetime splitting 
to the theory of electromagnetic induction,
it is convenient to rewrite the result in Eq.\eqref{fm: splitting}
in terms of the spacetime motion velocity field 
$\,\fourvel:\EVE\mapsto\TEVE\,$:
\begin{equation}
\setlength{\jot}{8pt}
\normboxed{\,
\begin{aligned}
\Exter&\,=\splitS\pull\Exter
\\
&\,+\timeform\wedge
\Bigdi{ \splitS\pull(\Exter\punto\fourvel)-(\splitS\pull\Exter)\punto\fourvel }\,.
\end{aligned}
\,}
\label{fm: fourvelsplit}
\end{equation}

To prove the formula in Eq.\eqref{fm: fourvelsplit}, 
the time-arrow is expressed as time-horizontal component of 
the spacetime velocity field
$\,\fourvel:\EVE\mapsto\TEVE\,$ by setting
\begin{equation}
\timearrow=\splitZ\punto\fourvel=\fourvel-\splitS\punto\fourvel\,.
\label{fm: spatimesp}
\end{equation}
The split formula in Eq.\eqref{fm: splitting} may then be rewritten as:%
\begin{equation}
\vcenter{\halign{
\hfil$#$&$#$\hfil&$#$\hfil&$#$\hfil\cr
\Exter&\,=\splitS\pull\Exter+\timeform\wedge\splitS\pull(\Exter\punto\BZ)
\vspace{8pt}\cr
&\,=\splitS\pull\Exter+\timeform\wedge\splitS\pull
\Bigdi{\Exter\punto\bigdi{\fourvel-\splitS\fourvel}}
\,,\cr}}
\label{fm: splittingV}
\end{equation}
and Eq.\eqref{fm: fourvelsplit} follows from the trivial equality
\begin{equation}
\splitS\pull\Bigdi{\Exter\punto(\splitS\fourvel)}
=(\splitS\pull\Exter)\punto\fourvel\,.
\label{fm: equalsimpl}
\end{equation}

The proof exposed in Lemma \ref{lm: fourtothreemoto} and the 
expression Eq.\eqref{fm: fourvelsplit} of the split formula in terms of
the spacetime velocity field were first proposed in \citep{Electro2013}.

\section{Spacetime Electromagnetics}
\label{sec: STElectro}

The above introduced decomposition may be applied
to spacetime electromagnetics forms.

Let us now consider the 
\Faraday\ spacetime two-form $\,\Farad\,$
expressing the \emph{electromagnetic induction field} 
and subsequently in a similar way
the \Ampere\ spacetime two-form $\,\Amper\,$
expressing the \emph{electromagnetic induction flux},
respectively \emph{even} and \emph{odd} forms,
so named in \citep{MisnerThorneWheeler1973}.%
\footnote{\label{fn: expl}
In \cite[p.212]{Sommerfeld1952} the author exclaimed: 
“I wish to create the impression in my readers that 
the true mathematical structure of these entities will appear only now,
as in a mountain landscape when the fog lifts.”
}

\subsection{Faraday spacetime two-form}
\label{sec: FourFaraday}

The electric induction phenomena are
governed by the closed spacetime induction
\Faraday\ two-form $\,\Farad\,$ 
(the \emph{electromagnetic field})
and by its
\emph{potential} one-form $\,\Faradpot\,$
such that
\begin{equation}
\Farad=\extder\Faradpot\,.
\label{fm: }
\end{equation}

From Lemma \ref{lm: fourtothreemoto},
setting $\,\timeform=\dtE\,$,
we infer the following statement.%
\begin{proposition}[Electric induction]
\label{prop: dispmagcur}
Time-vertical restrictions of
spacetime \Faraday\ two-form $\,\Farad\,$ 
and of its \emph{potential} one-form $\,\Faradpot\,$
fulfilling $\,\Farad=\extder\Faradpot\,$,
are the \emph{even} forms: %
\footnote{\label{fn: lenz}
The definition of $\,\eleform\,$
in terms of $\,\fourvel\,$ instead of $\,\timearrow\,$
is innovative and decisive to recover the spatial rule in Prop.\ref{prop: stFarad}.
The minus sign in the expression of $\,\eleform\,$
is motivated by \Lenz\ rule, see Eq.\eqref{fm: electricuno}.}
\begin{equation}
\kern-5pt
\left\{\vcenter{\halign{
\hfil$#$&$#$\hfil&\quad$#$\hfil&$#$\hfil\cr
\magmom&\,=\splitS\pull\Faradpot\,,
&\kern-20pt\textrm{magnetic momentum field}
\vspace{12pt}\cr
\elestatform&\,=\splitS\pull(\Faradpot\punto\fourvel)\,,
&\textrm{electric potential field}
\vspace{12pt}\cr 
\magcurl&\,=\splitS\pull\Farad\,,
&\textrm{magnetic vortex field}
\vspace{12pt}\cr
\eleform&\,=-\splitS\pull(\Farad\punto\fourvel)\,,
&\textrm{electric field}
\cr}}
\right.
\label{fm: eleforms}
\end{equation}
with the representation formulae:
\begin{equation}
\setlength{\jot}{8pt}
\left\{
\begin{aligned}
\Faradpot&\,=
\magmom+\dtE\wedge(\elestatform-\magmom\punto\vel)\,,
\\
\Farad&\,=
\magcurl-\dtE\wedge\Bigdi{\eleform+\magcurl\punto\vel}\,.
\end{aligned}
\right.
\label{fm: Farad}
\end{equation}
where the scalar $\,\elepot\,$ is the electric potential field.
\end{proposition}

According to next proposition,
the spacetime rule of electric induction, expressed in terms of
\Faraday\ spacetime two-form $\,\Farad\,$ amounts to the closedness
property $\,\extder\Farad=\zerovec\,$ which 
by \Volterra\ theorem is equivalent to
existence of a potential 1-form $\,\Faradpot\,$.

\medskip\goodbreak

\begin{proposition}[Gauss-Lenz-Henry-Faraday]
\label{prop: stFarad}
Closedness of \Faraday\ spacetime two-form 
$\,\Farad\,$ is
equi\-valent to the spatial \Gauss\ law for the magnetic vortex
and to \Lenz-\Henry-\Faraday\ spatial induction law:
\begin{equation}
\vcenter{\halign{
$#$\hfil&$#$\hfil&$#$\hfil\cr
\extder\Farad=\zerovec\
&\,\equi
\left\{\vcenter{\halign{
\hfil$#$&$#$\hfil&$#$\hfil&$#$\hfil\cr
&\extder\magcurl=\zerovec\,,
\vspace{8pt}\cr
&\extder\eleform+\Lieder_{\fourvel}\di\magcurl=\zerovec
\,.\cr}}\right.
\vspace{12pt}\cr
&\,\kern-40pt\equi
\left\{\vcenter{\halign{
$#$\hfil&$#$\hfil&$#$\hfil\cr
&\diverg\di\magind=0\,,
\vspace{8pt}\cr
&\rotor\di\elefield+\Lieder_{\fourvel}\di\magind
+{\diverg\di\threevel}\punto\magind=\zerovec
\,.\cr}}\right.
\cr}}
\label{fm: GHFlawdiff}
\end{equation}
\end{proposition}
\begin{proof}
The spacetime extrusion formula Eq.\eqref{fm: extrusion}
with $\,\BOO^\kk=\Farad\,$
gives:
\begin{equation}
\setlength{\jot}{8pt}
\begin{aligned}
\Lieder_{\fourvel}\di{\Farad}
&\,=\di{\extder\Farad}\punto\fourvel
+\extder\di{\Farad\punto\fourvel}\,.
\end{aligned}
\label{fm: electric}
\end{equation}
By applying \Stokes' formula Eq.\eqref{fm: VS}
to the boundary of a spatial compact manifold,
the following commutativity between exterior derivative and spatial projection
may be inferred:
\begin{equation}
\extder\circ\splitS\pull=\splitS\pull\circ\extder\,.
\label{fm: projcommut}
\end{equation}
Here and in the sequel, to simplify the notation, 
the exterior derivative acting on a spacetime form and
the one acting on a spatial form 
will both be denoted by the same symbol $\,\extder\,$.
From definitions Eq.\eqref{fm: eleforms} we get:
\begin{equation}
\kern-5pt
\left\{\vcenter{\halign{
\hfil$#$&$#$\hfil&$#$\hfil&$#$\hfil\cr
\splitS\pull\di{\extder\Farad}
&\,=\extder\di{\splitS\pull\Farad}
=\extder\magcurl\,,
\vspace{12pt}\cr
\splitS\pull(\extder\Farad\punto\fourvel)
&\,=\splitS\pull\Bigdi{\Lieder_{\fourvel}\di\Farad-\extder\di{\Farad\punto\fourvel}}
\vspace{8pt}\cr
&\,\kern-25pt
=\Lieder_{\fourvel}(\splitS\pull\Farad)-\extder\Bigdi{\splitS\pull\di{\Farad\punto\fourvel}}
\vspace{8pt}\cr
&\,\kern-25pt
=\Lieder_{\fourvel}\di\magcurl+\extder\eleform
\,.\cr}}\right.
\label{fm: electricuno}
\end{equation}
The implication $\,\Longrightarrow\,$ in Eq.\eqref{fm: GHFlawdiff} follows.

The converse implication $\,\Longleftarrow\,$ 
is inferred by applying the representation formula Eq.\eqref{fm: fourvelsplit}
to the three-form $\,\extder\Farad\,$:%
\begin{equation}
\setlength{\jot}{8pt}
\begin{aligned}
\extder\Farad
&\,=\splitS\pull(\extder\Farad)
\\
&\,\kern-40pt
+\dtE\,\wedge\Bigdi{\splitS\pull(\extder\Farad\punto\fourvel)
-(\splitS\pull\extder\Farad)\punto\fourvel}\,.
\end{aligned}
\label{fm: electricdue}
\end{equation}
The r.h.s. of Eq.\eqref{fm: GHFlawdiff}
and Eq.\eqref{fm: electricuno} yield:
\begin{equation}
\setlength{\jot}{8pt}
\left\{
\begin{aligned}
&\,\splitS\pull\extder\Farad=\zerovec\,,
\\
&\,\splitS\pull(\extder\Farad\punto\fourvel)=\zerovec\,,
\end{aligned}
\right.
\label{fm: }
\end{equation}
which by Eq.\eqref{fm: electricdue}
imply $\,\extder\Farad=\zerovec\,$.
\end{proof}

Similarly, from the extrusion formula Eq.\eqref{fm: STdiffhomo},
setting $\,\BOO^\kk=\Faradpot\,$ we get:
\begin{equation}
\setlength{\jot}{8pt}
\begin{aligned}
\Lieder_{\fourvel}\di{\Faradpot}
&\,=(\extder\Faradpot)\punto\fourvel
+\extder(\Faradpot\punto\fourvel)\,,
\end{aligned}
\label{fm: name}
\end{equation}
which by the commutativity property
Eq.\eqref{fm: projcommut} and definitions Eq.\eqref{fm: eleforms} gives:
\begin{equation}
\splitS\pull(\extder\Faradpot)
=\extder(\splitS\pull\Faradpot)=\extder\magmom\,,
\label{fm: }
\end{equation}
and
\begin{equation}
\left\{\vcenter{\halign{
\hfil$#$&$#$\hfil&$#$\hfil&$#$\hfil\cr
\splitS\pull(\extder\Faradpot\punto\fourvel)
\vspace{8pt}\cr
&\,\kern-45pt
=\splitS\pull\Bigdi{\Lieder_{\fourvel}\di\Faradpot-\extder\di{\Faradpot\punto\fourvel}}
\vspace{8pt}\cr
&\,\kern-45pt
=\Lieder_{\fourvel}(\splitS\pull\Faradpot)
-\extder\Bigdi{\splitS\pull\di{\Faradpot\punto\fourvel}}
\vspace{8pt}\cr
&\,\kern-45pt
=\Lieder_{\fourvel}\di\magmom+\extder\elestatform
\,.\cr}}\right.
\label{fm: novanta}
\end{equation}
Hence:
\begin{equation}
\vcenter{\halign{
$#$\hfil&$#$\hfil&$#$\hfil\cr
\Farad=\extder\Faradpot\
&\,\equi
\left\{\vcenter{\halign{
$#$\hfil&$#$\hfil&$#$\hfil\cr
&\,\magcurl=\extder\magmom\,,
\vspace{8pt}\cr
&\,\eleform+\Lieder_{\fourvel}\di\magmom=\zerovec
\,.\cr}}\right.
\cr}}
\label{fm: GHFlaw}
\end{equation}
In vector terms the expression becomes:
\begin{equation}
\vcenter{\halign{
$#$\hfil&$#$\hfil&$#$\hfil\cr
\left\{\vcenter{\halign{
$#$\hfil&$#$\hfil&$#$\hfil\cr
&\,\magind=\rotor\di\magvecpot\,,
\vspace{8pt}\cr
&\,\elefield+\Lieder_{\fourvel}\di\magvecpot
+2\,\Eul\di\vel\punto\magvecpot=\zerovec
\,,\cr}}\right.
\cr}}
\label{fm: name}
\end{equation}
with the \Euler\ stretching tensor given by Eq.\eqref{fm: Eul}:
\begin{equation}
\Eul\di\vel\equaldef\inv\metric\punto\unmezzotext\Lieder_\vel\di\metric\,.
\label{fm: eultens}
\end{equation}

\subsection{Amp{\`e}re spacetime two-form}
\label{sec: FourAmpere}

Let us now turn to \Ampere-\Maxwell\ induction law.
Relevant spacetime differential form is
\Ampere\ two-form $\,\Amper\,$ (\emph{electromagnetic induction flux})
which is potential for the \emph{current} three-form $\,\Current\,$.

\goodbreak

From Lemma \ref{lm: fourtothreemoto}
we infer the next statement.%
\begin{proposition}[Magnetic induction]
\label{prop: dispmagcur}
The time vertical restrictions of
the spacetime \Ampere\ two-form $\,\Amper\,$ 
and of the four-current\ three-form: 
\begin{equation}
\Current=\extder\Amper
\label{fm: }
\end{equation}
are given by:%
\footnote{\label{fn: defvel}
The definition of $\,\magwind\,$ and $\,\elecurrflux\,$
in terms of $\,\fourvel\,$ instead of $\,\timearrow\,$
is decisive to recover the 
spatial rule in Prop.\ref{prop: stAmper}.
}
\begin{equation}
\left\{\vcenter{\halign{
\hfil$#$&$#$\hfil&\quad$#$\hfil&$#$\hfil\cr
\eleflux&\,=\splitS\pull\Amper\,,
&\kern-10pt\textrm{electric displacement flux}
\vspace{12pt}\cr
\magwind&\,=\splitS\pull(\Amper\punto\fourvel)\,,
&\kern-5pt\textrm{magnetic induction flux}
\vspace{12pt}\cr 
\elechargeform&\,=\splitS\pull\Current\,,
&\textrm{electric charge}
\vspace{12pt}\cr
\elecurrflux&\,=\splitS\pull(\Current\punto\fourvel)\,,
&\textrm{electric current flux}
\cr}}
\right.
\label{fm: magforms}
\end{equation}
with the representation formulae:
\begin{equation}
\vcenter{\halign{
\hfil$#$&$#$\hfil&$#$\hfil&$#$\hfil\cr
&\,\Amper=
\eleflux+\dtE\wedge(\magwind-\eleflux\punto\vel)\,,
\vspace{8pt}\cr
&\,\Current=
\elechargeform+\dtE\wedge(\elecurrflux-\elechargeform\punto\vel)
\,.\cr}}
\label{fm: elerepr}
\end{equation}
\end{proposition}

An evaluation analogous to the one in Prop.\ref{prop: stFarad}
yields the next result.
\begin{proposition}[Coulomb,Amp{\`e}re,Maxwell]
\label{prop: stAmper}
Equality between the current three-form $\,\Current\,$
and the exterior derivative of 
\Ampere-\Maxwell\ two-form $\,\Amper\,$
is equivalent to \Coulomb's balance law for electric charge
and to the magnetic induction law:
\begin{equation}
\vcenter{\halign{
$#$\hfil&$#$\hfil&$#$\hfil\cr
&\,\extder\Amper=\Current
\equi
\left\{\vcenter{\halign{
$#$\hfil&$#$\hfil&$#$\hfil\cr
&\extder\eleflux=\elechargeform\,,
\vspace{8pt}\cr
&\extder\magwind=\Lieder_{\fourvel}\di\eleflux+\elecurrflux
\,,\cr}}\right.
\vspace{12pt}\cr
&\kern-5pt\equi
\left\{\vcenter{\halign{
$#$\hfil&$#$\hfil&$#$\hfil\cr
&\diverg\di\eledisp=\rho\,,
\vspace{8pt}\cr
&\rotor\di\magfield
=\Lieder_{\fourvel}\di\eledisp+\diverg\di\vel\punto\eledisp+\elecurr
\,.\cr}}\right.
\cr}}
\label{fm: Amp}
\end{equation}
\end{proposition}

The following property states a really awesome equivalence.

\begin{proposition}[Equivalence between spacetime and spatial formulations]
\label{prop: stspazequiv}
The pair of closedness properties:
\begin{equation}
\setlength{\jot}{8pt}
\left\{
\begin{aligned}
&\,\extder\Farad=\zerovec\,,
\\
&\,\extder\Current=\zerovec\,,
\end{aligned}
\right.
\label{fm: name}
\end{equation}
of the spacetime forms $\,\Farad\,$
and $\,\Current\,$,
are equivalent to the spatial electromagnetic rules respectively named after
\Gauss-\Lenz-\Henry-\Faraday\ for electric induction
and after
\Coulomb-\Oersted-\Ampere-\Maxwell\ for magnetic induction.
\end{proposition}
\begin{proof}
The equivalence follows directly from the computations 
in Prop.\ref{prop: stFarad} and
Prop.\ref{prop: stAmper}.
This equivalence holds true
in the general case of deforming continuous bodies.
\end{proof}

\section{Electromagnetic power}
\label{sec: EMPow}

The power locally expended by the electromagnetic fields 
is the sum of electric and magnetic powers:
\begin{equation}
\normboxed{\,
\setlength{\jot}{8pt}
\begin{aligned}
\powerEH\equaldef
&\,\eleform\wedge\Bigl(\Lieder_\fourvel\di\eleflux+\elecurrflux\Bigr)
\\
+&\,\magwind\wedge\Lieder_\fourvel\di\magcurl\,.
\end{aligned}
\,}
\label{fm: powerEH}
\end{equation}

\begin{lemma}[Umov electromagnetic power]
In any outer oriented, bounded, compact and connected spatial domain $\,\controltout\,$
the global electromagnetic power depends only on the boundary values
of electric and magnetic fields through the incoming flux:
\footnote{{\ }%
The vector field $\,\elefield\times\magfield:\EVE\mapsto\VEVE\,$
was first introduced in \citep{Umov1874}
and later reproduced
in \citep{Poynting1884} and \citep{Heaviside1885}.
}
\begin{equation}
\normboxed{\,
\integrale{\controltout}{}\powerEH=-\integrale{\partial\controltout}{}\powerflux\,,
\,}
\label{fm: umovlaw}
\end{equation}
of Nikolay \Umov\ odd two-form:%
\begin{equation}
\normboxed{\,
\powerflux\equaldef\eleform\wedge\magwind\in\FORMS^2\di\VEVE\,.
\,}
\label{fm: umovdef}
\end{equation}
\end{lemma}
\begin{proof}
The induction laws Eq.$\eqref{fm: diffMIL}_2$ and Eq.\eqref{fm: Diffvort}
give to the local electromagnetic power the expression:
\begin{equation}
\normboxed{\,
\setlength{\jot}{8pt}
\begin{aligned}
\powerEH\equaldef
&\,\eleform\wedge\extder\magwind-\magwind\wedge\extder\eleform
\\
=&\,-\extder\di{\eleform\wedge\magwind}
\,,
\end{aligned}
\,}
\label{fm: powerEHsplit}
\end{equation}
so that \Stokes\ formula yields the result.
\end{proof}

The vector formalism usually adopted in literature
can be recovered by observing that:
\begin{equation}
\eleform\wedge\magwind
=\volformg\punto\di{\elefield\times\magfield}\,,
\label{fm: UmovVect}
\end{equation}
so that:
\begin{equation}
\extder\di{\eleform\wedge\magwind}=\diverg\di{\elefield\times\magfield}\punto\volformg\,.
\label{fm: }
\end{equation}

\section{Changes of Frame}
\label{sec: FCcha}

A natural axiomatic statement is that tensor fields 
on the spacetime manifold $\,\EVE\,$ transform by push
under the action of a spacetime automorphism 
$\,\map:\EVE\mapsto\EVE\,$
describing a smooth frame-change.

The \Faraday\ and \Ampere-\Maxwell\ two-forms
$\,\Farad\,$ and $\,\Amper\,$
will accordingly transform as:
\begin{equation}
\setlength{\jot}{8pt}
\left\{
\begin{aligned}
(\Farad)_\map&\,=\map\push\Farad\,,
\\
(\Amper)_\map&\,=\map\push\Amper\,.
\end{aligned}
\right.
\label{fm: FAM}
\end{equation}

All other electromagnetic fields
also transform according to the natural rule, by invariance 
of their scalar value under push of the involved arguments.

This conclusion can be deduced observing that:
\begin{equation}
\setlength{\jot}{8pt}
\left\{
\begin{aligned}
(\tE)_\map&\,=\map\push\tE=\tE\circ\inv\map\,,
\\
\extder(\tE)_\map&\,=\extder\di{\map\push\tE}=\map\push\dtE\,,
\\
\timearrow_\map&\,=\map\push\timearrow\,,
\\
\BI_\map&\,=\map\push\BI\,,
\\
\splitZ_\map&\,=\extder(\tE)_\map\otimes\timearrow_\map
=\map\push\di{\dtE\otimes\timearrow}=\map\push\splitZ\,,
\\
\splitS_\map&\,=\BI_\map-\splitZ_\map
=\map\push\BI-\map\push\splitZ=\map\push\splitS\,.
\end{aligned}
\right.
\label{fm: name}
\end{equation}
The second and the next to last rules follow from commutativity
of exterior derivative and push by the diffeomorphic frame-change map.

\section{Frame covariance}
\label{sec: Framecov}


In literature it is usually affirmed that
\Maxwell\ equations are not form-invariant under
\Euclid\ frame changes,
but are such under \Lorentz\ transformations.

In our view, the mathematically unspecified notion of
form-invariance must be replaced by the
natural requirement of covariance
under a change of frame.

\begin{definition}[Covariance of a rule]
The transformed fields,
got by pushing the involved tensor fields along the frame-change mapping,
are required to fulfil the transformed rule when the original tensor fields obey the original rule.
\end{definition}

Covariance of the electromagnetic  induction rules 
is based on the following preliminary result.
\begin{lemma}[Covariance of spacetime velocity]\label{lem: covstvel}
The spacetime velocity is covariant
under any transformation $\,\map\,$ in the group of automorphisms in $\,\EVE\,$.
\end{lemma}
\begin{proof}
The expression of the pushed spacetime motion:
\begin{equation}
(\map\push\moto)_\lapse=\map\circ\moto_\lapse\circ\inv\map\,,
\label{fm: }
\end{equation}
taking the derivative $\,\parder\lapse0\,$ yields:
\begin{equation}
\fourvelmap=\TT\map\punto\fourvel\circ\inv\map
=\map\push\fourvel\,.
\label{fm: stvelcovar}
\end{equation}
which is the pertinent transformation rule.
\end{proof}

When a full spacetime formulation is adopted, 
it is readily verified that  the following fundamental result
holds.
\begin{proposition}[Covariance of induction laws]
\label{prop: pushLie}
The integral formulation Eq.\eqref{fm: EIL},
or the equivalent full differential expression 
Eq.\eqref{fm: DiffEIL},
and similarly Eq.\eqref{fm: MIL} or Eq.\eqref{fm: diffMIL},
are covariant under any spacetime change of frame.
\end{proposition}
\begin{proof}
The validity of the result
relies on basic properties of \Lie\ 
and exterior derivatives under the action of automorphic 
spacetime frame change.
Indeed,
spacetime tensor fields 
$\,\stress:\EVE\mapsto\TENS\di\TEVE\,$ 
and spacetime exterior forms 
$\,\Boo:\EVE\mapsto\FORMS\di\TEVE\,$,
fulfil the natural transformation
and commutativity property:%
\begin{equation}
\setlength{\jot}{8pt}
\left\{
\begin{aligned}
&\,\map\push\Bigdi{{\Lieder_{\fourvel}}\di{\stress}}
={\Lieder_{(\map\push\fourvel)}}\di{\map\push\stress}\,,
\\
&\,\extder\circ(\map\push\Boo)=\map\push\circ\extder\Boo
\,.
\end{aligned}
\right.
\label{fm: pushLie}
\end{equation}
Accordingly, under a frame-change $\,\map:\EVE\mapsto\EVE\,$,
Eq.\eqref{fm: DiffEIL} does transform into:
\begin{equation}
-\map\push\eleform
=\map\push\Bigdi{\Lieder_{\fourvel}\di\magmom}
=\Lieder_{(\map\push\fourvel)}\di{\map\push\magmom}\,.
\label{fm: pushEIL}
\end{equation}
The covariance property of the spacetime velocity 
expressed by Eq.\eqref{fm: stvelcovar} gives:
\begin{equation}
\Lieder_{(\map\push\fourvel)}\di{\map\push\magmom}
=\Lieder_{(\fourvelmap)}\di{\map\push\magmom}\,.
\label{fm: pushEILtrue}
\end{equation}
Covariance of the electric or magnetic induction rules thus follows.
\end{proof}

Let us note that:
\begin{equation}
\setlength{\jot}{8pt}
\left\{
\begin{aligned}
&\,\fourvelmap=\splitS\fourvelmap+\timearrow\,,
\\
&\,\map\push\fourvel
=\map\push\di{\splitS\fourvel+\timearrow}
=\map\push\di{\splitS\fourvel}+\map\push\timearrow
\,.
\end{aligned}
\right.
\label{fm: name}
\end{equation}
Accordingly, in the framing $\,\splitZ\equaldef\dtE\otimes\timearrow\,$
the transformation rule of the time-vertical component of the spacetime velocity
is given by:%
\begin{equation}
\setlength{\jot}{8pt}
\begin{aligned}
\splitS\fourvelmap&\,=\map\push\di{\splitS\fourvel}+\fourvelrel
\,,
\end{aligned}
\label{fm: transfrelvel}
\end{equation}
with the relative spacetime velocity between framings
defined by:
\begin{equation}
\fourvelrel\equaldef\map\push\timearrow-\timearrow\,.
\label{fm: sparelvel}
\end{equation}

Under \Newton\ frame changes 
clock rates are preserved,
i.e. 
$\,\map\push\dtE=\dtE\,$
so that the relative spacetime velocity 
$\,\fourvelrel\,$ between framings  is time-vertical.

From Eq.\eqref{fm: transfrelvel} and Eq.\eqref{fm: sparelvel}
we may infer that
covariance of the spatial component of spacetime velocity
holds only with respect to the 
subgroup of frame transformations
inducing no relative velocity (a trivial  case).

Here lies the mathematical reason why covariance is lost,
even under \Galilei\ changes of frame 
when, in place of adopting the correct expression in
Eq.\eqref{fm: DiffEIL},
the following improperly incomplete induction rule is adopted,
with the addend
$\,\extder(\magmom\punto\vel)\,$ dropped off
the split expression Eq.\eqref{fm: DiffEILsplit}:%
\begin{equation}
\normboxed{\,
\begin{aligned}
-\eleform={\Lieder_\timearrow}\di\magmom
+\magcurl\punto\vel
\,.
\label{fm: DiffEILsplitsimpl}
\end{aligned}
\,}
\end{equation}

For a direct comparison with the formulations in literature
we observe that, 
the definition in
Eq.$\eqref{fm: elemag}_2$ and
Eq.\eqref{fm: magcurl}:
\begin{equation}
\setlength{\jot}{8pt}
\left\{
\begin{aligned}
\magcurl&=\volformg\punto\magind\,,
\\
\magmom&=\metric\punto\magvecpot\,,
\end{aligned}
\right.
\label{fm: magcurlmom}
\end{equation}
the definition of vector product in Eq.\eqref{fm: vectprod}
and the time-independence of the metric in Eq.\eqref{fm: notimemetric},
lead to a vectorial expression of Eq.\eqref{fm: DiffEILsplitsimpl} given by: %
\footnote{\label{fn: incompl}
This incomplete expression, adopted by
\cite{Heaviside1885,Heaviside1892}, \cite{Hertz1892} and \cite{Lorentz1892}
was reproduced in all subsequent treatments in literature,
e.g. 
\citep[Eq.(85)]{Deschamps1981} and
\citep[Eq.(17.2)]{LandauLifshits1987}.
}
\begin{equation}
\normboxed{\,
-\,\elefield={\Lieder_\timearrow}\di\magvecpot
+\magind\times\vel\,.
\label{fm: DiffEILvectsplitsimpl}
\,}
\end{equation}

\section{Observers point of view}
\label{sec: Obs}

We may now deduce in a straightforward
way the transformation rules for spacetime fields,
due to the action of a spacetime frame-change
$\,\map:\EVE\mapsto\EVE\,$
as described
by a given framing $\,\splitZ\,$, see Eq.\eqref{fm: framing}.

More precisely we shall compute, 
for each spatial electromagnetic field or flux,
the expression of the components
of the transformed field or flux
in a coordinate system adapted to the original framing.

This definition is adopted in degree that the push
by the transformation map
be substituted to the improper requirement of 
\emph{form-invariance}. 

The latter notion is in fact not susceptible of a mathematical definition 
and is therefore misleading, as witnessed
by the manifest contradiction between
the suggested procedure and the conclusion drawn in
\citep[Part II, {\S}4]{Einstein1905a}.

\section{Special relativity}
\label{sec: SR}

Let us consider a spacetime frame
$\,\set{\fourax_0,\fourax_1,\fourax_2,\fourax_3}\,$
\emph{adapted} to a given framing $\,\splitZ=\dtE\otimes\timearrow\,$,
with the first vector given by 
$\,\fourax_0=\timearrow/\lightvel,$
and the remaining time-vertical.

All basis vectors are dimensionless.

To a  \Lorentz\ boost
$\,\mapL:\EVE\mapsto\EVE\,$
with velocity $\,\traslvel\,$ in the $\,\fourax_1\,$ direction:
\begin{equation}
\traslvel=\traslpar\,\fourax_1\,,
\label{fm: }
\end{equation}
dropping the invariant basis vectors $\,\set{\fourax_2,\fourax_3}\,$,
there corresponds the tangent transformation:
\begin{equation}
\TT\mapL:\TEVE\mapsto\TEVE\,,
\label{fm: }
\end{equation}
given by:%
\footnote{\label{fn: VL}
These transformations where introduced
and named after \Lorentz, 
by \cite{Poincare1905},
who provided a partial amendment of the ones proposed 
by \cite{VoigtWoldemar1887a,VoigtWoldemar1887b}
and by \cite{Lorentz1904}.
}
\begin{equation}
\renewcommand{\arraystretch}{1.5}
\begin{bmatrix}
\mapL\push\timearrow\\
\mapL\push\fourax_{1}
\end{bmatrix}
=\relfactor\,
\begin{bmatrix}
1&\traslpar\;
\\
\traslpar/\lightvel^2&1\;
\end{bmatrix}
\punto
\begin{bmatrix}
\timearrow\\
\fourax_{1}
\end{bmatrix}\,.
\label{fm: VLtangbasis}
\end{equation}

\goodbreak

In terms of the adimensional speed 
$\,\beta\di\traslpar\equaldef\traslpar/\lightvel\,$ 
of the boost
(ratio between boost speed $\,\traslpar\,$
and light speed \emph{in vacuo} $\,\lightvel\,$), 
the relativistic factor has the expression:
\begin{equation}
\relfactor\equaldef (1-\traslpar^2/\lightvel^2)^{-1/2}
=1/\sqrt{1-\beta\di\traslpar^2}\,.
\label{fm: relfactor}
\end{equation}
Transformations $\,\mapL:\EVE\mapsto\EVE\,$ of the group defined by
Eq.\eqref{fm: VLtangbasis}-\eqref{fm: relfactor}
are designed to get invariance of the 
nonsingular spacetime metric tensor 
$\,\metric_\BM:\TEVE\mapsto\TEVEs\,$
\citep{Minkowski1908,Weyl1922}:%
\begin{equation}
\metric_\BM=\splitS\pull\metric-\lightvel^2\,\di{\dtE\otimes\dtE}
\,.
\label{fm: gMINK}
\end{equation}
Here $\,\metric:\VEVE\mapsto\dual{(\VEVE)}\,$
is the positive definite \emph{spatial metric}.
Invariance under the boost $\,\mapL:\EVE\mapsto\EVE\,$ means:%
\begin{equation}
\metric_\BM=\mapL\pull\metric_\BM
\,.
\label{fm: invarianceMINK}
\end{equation}
The inverse boost 
is got by replacing $\,\traslpar\,$ with $\,-\traslpar\,$, 
so that the relativistic factor $\,\relfactor\,$ is unchanged.
\par\noindent
In addition to Eq.\eqref{fm: VLtangbasis} we have for $\,\lapse=2,3\,$:
\begin{equation}
\mapL\push\fourax_\lapse=\fourax_\lapse\,.
\label{fm: trasvbasis}
\end{equation}
Explicitly we write:
\begin{equation}
\left\{\vcenter{\halign{
\hfil$#$&$#$\hfil&$#$\hfil&$#$\hfil\cr
\mapL\push\timearrow&\,=\relfactor\,(\timearrow+\traslpar\,\fourax_1)\,,
\vspace{8pt}\cr
\mapL\push\fourax_{1}&\,=\relfactor\,\Bigdi{(\traslpar/\lightvel^2)\,\timearrow+\fourax_1}
\,,\cr
}}\right.
\label{fm: transfbasis}
\end{equation}
with inverse ($\,-\traslpar\,$ in place of $\,\traslpar\,$)
given by:
\begin{equation}
\left\{\vcenter{\halign{
\hfil$#$&$#$\hfil&$#$\hfil&$#$\hfil\cr
\mapL\pull\timearrow&\,=\relfactor\,(\timearrow-\traslpar\,\fourax_1)\,,
\vspace{8pt}\cr
\mapL\pull\fourax_{1}&\,=\relfactor\,\Bigdi{-(\traslpar/\lightvel^2)\,\timearrow+\fourax_1}
\,.\cr
}}\right.
\label{fm: transvectors}
\end{equation}

A vector $\,\Vect\in\TEVE\,$
has components transformed by the matrix
inverse-transpose of the one in Eq.\eqref{fm: VLtangbasis}:
\begin{equation}
\renewcommand{\arraystretch}{1.5}
\begin{bmatrix}
\VV_{\mapL\push\timearrow}
\\
\VV_{\mapL\push\fourax_{1}}
\end{bmatrix}
=\relfactor\,
\begin{bmatrix}
1&-\traslpar/\lightvel^2\;
\\
-\traslpar&1\;
\end{bmatrix}
\punto
\begin{bmatrix}
\VV_\timearrow
\\
\VV_{\fourax_{1}}
\end{bmatrix}\,.
\label{fm: VLcomp}
\end{equation}

In the limit $\,\traslpar/\lightvel\to0\,$
we get:
\begin{equation}
\setlength{\jot}{8pt}
\left\{
\begin{aligned}
&\,\relfactor\to1\,,
\\
&\,\traslpar/\lightvel^2\to0\,.
\end{aligned}
\right.
\label{fm: name}
\end{equation}

The \Lorentz\ tangent map
Eq.\eqref{fm: VLtangbasis}
reduces then to
the \Galilei\ transformation rule, for a relative translational speed $\,\traslpar\,$
in direction of $\,\fourax_{1}\,$:
\begin{equation}
\renewcommand{\arraystretch}{1.5}
\begin{bmatrix}
\mapG\push\timearrow\\
\mapG\push\fourax_{1}
\end{bmatrix}
=
\begin{bmatrix}
\!\!1&\traslpar\;
\\
\;\;0\phantom{ro}&1\;
\end{bmatrix}
\punto
\begin{bmatrix}
\timearrow\\
\fourax_{1}
\end{bmatrix}\,.
\label{fm: tangbasis}
\end{equation}

 \section{Frame changes in special relativity}
\label{sec: framechange}

A spacetime tensor field $\,\stress\,$ is transformed
by the action of a \Lorentz\ automorphism
$\,\mapL:\EVE\mapsto\EVE\,$
into the pushed field $\,\mapL\push\stress\,$.
A framing $\,\splitZ\,$ is likewise sent into the pushed framing $\,\mapL\push\splitZ\,$.

A spacetime frame
$\,\set{\fourax_0,\fourax_1,\fourax_2,\fourax_3}\,$
\emph{adapted} to $\,\splitZ\,$
is pushed to a spacetime frame \emph{adapted} to
the pushed framing.

A comparison of a tensor field with its pushed counterpart
will be made in the sequel by evaluating the longitudinal and transversal components
of both fields in the \emph{adapted} spacetime frame
$\,\set{\fourax_0,\fourax_1,\fourax_2,\fourax_3}\,$.

These evaluations are carried out to contrast method and conclusions
exposed in literature.

To be honest, there is no real need of this stuff because 
in electromagnetism all involved fields are fully covariant under any spacetime frame-change
and all rules of electromagnetic induction are also fully covariant,
when properly expressed in geometric terms,
thanks to the commutativity property 
between push transformation and exterior derivative
and of naturality of \Lie\ derivatives with respect to push,
as expressed in Eq.\eqref{fm: pushLie}.

These properties are direct consequences of basic mathematical
notions concerning the relation in Eq.\eqref{fm: VS}
between integrals over compact manifolds
and over their boundaries,
and the transformation of integrals
under the action of diffeomorphic maps,
the fields acted upon by the integrals being exterior forms
of maximal degree on the relevant domains.


The transformation of electromagnetic fields and fluxes
under spacetime changes of frame has a central role in many
treatments and is therefore certainly worth to be explicitly investigated,
a task performed in the next subsections.

\subsection{Electric induction}
\label{sec: eleind}

\subsubsection{Electric field}
\label{sec: elefield}

The electric spacetime time-vertical one-form $\,\eleform\,$
is transformed by the \Lorentz\ change of observer into the one-form:
\begin{equation}
\mapL\push\eleform\,,
\label{fm: }
\end{equation}
Its longitudinal component, 
along the direction $\,\fourax_1\,$ of the boost,
is given by:
\begin{equation}
\vcenter{\halign{
\hfil$#$&$#$\hfil&$#$\hfil&$#$\hfil\cr
\di{\mapL\push\eleform}\punto\fourax_1
&\,=\mapL\push\Bigdi{\eleform\punto\di{\mapL\pull\fourax_1}}
\vspace{8pt}\cr&\,
\kern-40pt
=\relfactor\punto\mapL\push\Bigdi{\eleform\punto
\Bigdi{\fourax_1-\xcancel{(\traslpar/\lightvel^2)\,\timearrow} }}
\vspace{12pt}\cr&\,
\kern-40pt
=\di{\relfactor\punto\eleform\punto\fourax_1}\circ\inv\mapL
\,.\cr}}
\label{fm: Ele1}
\end{equation}
Cancellation is due to time-verticality of $\,\eleform\,$.

The frame-transformation for the spacetime
transversal component along directions
$\,\fourax_\lapse\,$, with $\,\lapse=2,3\,$, are:%
\begin{equation}
\vcenter{\halign{
\hfil$#$&$#$\hfil&$#$\hfil&$#$\hfil\cr
\di{\mapL\push\eleform}\di{\fourax_\lapse}
&\,=\mapL\push\Bigdi{\eleform\punto\di{\mapL\pull\fourax_\lapse}}
\vspace{8pt}\cr
&\,=\mapL\push\Bigdi{\eleform\punto\fourax_\lapse}
\vspace{8pt}\cr
&\,=\di{\eleform\punto\fourax_1}\circ\inv\mapL
\,.\cr}}
\label{fm: Ele2}
\end{equation}

The transversal components of the electric field $\,\eleform\,$
along $\,\fourax_\lapse\,$, with $\,\lapse=2,3\,$, are 
then invariant:
\begin{equation}
\eleform\punto{\fourax_\lapse}\to\di{\eleform\punto\fourax_\lapse}\circ\inv\mapL\,.
\label{fm: eleort}
\end{equation}

\subsubsection{Magnetic vortex}
\label{sec: magvor}

The frame-transformation formula for the components
of the magnetic vortex $\,\magcurl\,$ in the longitudinal planes 
$\,\set{\threeax_1,\threeax_\lapse}\,$, with $\,\lapse=2,3\,$, writes:
\begin{equation}
\setlength{\jot}{8pt}
\begin{aligned}
&\,(\mapL\push\magcurl)\di{\fourax_1,\fourax_\lapse}
=\mapL\push\Bigdi{\magcurl\punto\di{\mapL\pull\fourax_1}\punto\fourax_\lapse}
\\
&\,=\mapL\push\Bigdi{
\relfactor\,
\Magcurl\Bigdi{
\fourax_1-\xcancel{(\traslpar/\lightvel^2)\,\timearrow},\fourax_\lapse}}
\,.
\end{aligned}
\label{fm: Mag1}
\end{equation}
Cancellation is due to time-verticality of $\,\magcurl\,$.
The components of magnetic vortex in longitudinal planes
are amplified by the relativistic factor:
\begin{equation}
\magcurl\di{\fourax_1,\fourax_\lapse}\to
\relfactor\punto\mapL\push\Bigdi{\magcurl\di{\fourax_1,\fourax_\lapse}}\,.
\label{fm: }
\end{equation}
On the other hand, the component of $\,\magcurl\,$ in the transversal plane 
$\,\set{\fourax_2,\fourax_3}\,$ is invariant:
\begin{equation}
\setlength{\jot}{8pt}
\begin{aligned}
(\mapL\push\magcurl)\di{\fourax_2,\fourax_3}
&\,=\mapL\push\Bigdi{\magcurl\di{\fourax_2,\fourax_3}}\,.
\end{aligned}
\end{equation}

\subsubsection{Vectorial notation}
\label{sec: vecnota}

In terms of spatial vector fields, we have, for $\,\lapse=2,3\,$:
\begin{equation}
\setlength{\jot}{8pt}
\begin{aligned}
&\,\eleform\di{\fourax_1}=\metric\di{\elefield^\parallel,\fourax_1}\,,
\\
&\,\eleform\di{\fourax_\lapse}=\metric\di{\elefield^\perp,\fourax_\lapse}\,,
\\
&\,\magcurl\di{\fourax_2,\fourax_3}
=\volform\di{\magind,\fourax_2,\fourax_3}
=\metric\di{\magind^\parallel,\fourax_1}\,,
\\
&\,\magcurl\di{\fourax_1,\fourax_\lapse}
=\volform\di{\magind,\fourax_1,\fourax_\lapse}
=\metric\di{\magind^\perp,\fourax_\lapse}\,.
\end{aligned}
\label{fm: parorth}
\end{equation}
Here above $\,\parallel\,$ and $\,\perp\,$ denote
the components parallel and orthogonal to the boost direction
$\,\fourax_1\,$.

Eq.$(18.42)$ and $(18.43)$
in \citep[p.330]{PanofskiPhillips1962}
and Table $(26.3)$ in \citep[26.3]{Feynman1964}
contain the currently adopted transformation rules 
for electric and magnetic spatial vector fields
under a \Lorentz\ boost.

In {\S}\ref{sec: comparison} 
a Synoptic Table offers a comparison of these rules, 
labeled as \emph{old},
versus the ones contributed here,
labeled as \emph{new}.

Agreement holds only for the parallel magnetic induction vector field
$\,\magind^\parallel\,$
(orthogonal to the transversal plane).

On the contrary, all other
\emph{old} transformation rules exposed in literature,
pertaining to transversal magnetic induction vector field
$\,\magind^\perp$
(parallel to the transversal plane)
and to electric vector field $\,\elefield\,$,
are not in agreement with the \emph{new} ones.
No entanglements are found as outcome of the new analysis.

\subsection{Magnetic induction}
\label{sec: magind}

The electric displacement two-form $\,\eleflux\,$,
magnetic winding one-form $\,\magwind\,$,
electric charge three-form $\,\elechargeform\,$,
and current two-form $\,\elecurrflux\,$,
all time-vertical and \emph{odd}.

The transformation rules of their components
are interpreted in the original framing as:
\begin{equation}
\setlength{\jot}{8pt}
\kern-10pt
\left\{
\begin{aligned}
\di{\mapL\push\eleflux}\punto\fourax_1\punto\fourax_\lapse
&\,=\mapL\push\Bigdi{\eleflux\punto\di{\mapL\pull\fourax_1}\punto\fourax_\lapse}
\\
&\,
=\mapL\push\Bigdi{\relfactor\,\eleflux\punto\fourax_1\punto\fourax_\lapse
\\
&\,
-\relfactor\,(\traslpar/\lightvel^2)\,\xcancel{\eleflux\punto\timearrow\punto\fourax_\lapse}}\,,
\end{aligned}
\right.
\label{fm: Mag0}
\end{equation}
\begin{equation}
\setlength{\jot}{8pt}
\kern-10pt
\left\{
\begin{aligned}
\di{\mapL\push\magwind}\punto\fourax_1
&\,=\mapL\push\Bigdi{\magwind\punto\di{\map\pull\fourax_1}}
\\
&\,\kern-40pt
=\mapL\push\Bigdi{\relfactor\,\magwind\punto\bigdi{
\fourax_1-\xcancel{(\traslpar/\lightvel^2)\,\timearrow}} }\,,
\end{aligned}
\right.
\label{fm: Mag1}
\end{equation}
\begin{equation}
\setlength{\jot}{0pt}
\kern-10pt
\left\{
\begin{aligned}
\di{\mapL\push\elecurrflux}\punto\fourax_1\punto\fourax_\lapse
&\,=\mapL\push\Bigdi{\elecurrflux\punto\di{\map\pull\fourax_1}\punto\fourax_\lapse}
\\
&\,
=\mapL\push\Bigdi{\relfactor\,\elecurrflux\punto\fourax_1\punto\fourax_\lapse
\\
&\,
-\relfactor\,(\traslpar/\lightvel^2)\,\xcancel{\elecurrflux\punto\timearrow\punto\fourax_\lapse}}\,,
\end{aligned}
\right.
\label{fm: Mag2}
\end{equation}
\begin{equation}
\setlength{\jot}{8pt}
\kern-10pt
\left\{
\begin{aligned}
\di{\mapL\push\elechargeform}\punto\fourax_1\punto\fourax_{23}
&\,=\mapL\push\Bigdi{\elechargeform\punto\di{\map\pull\fourax_1}\punto\fourax_{23}}
\\ 
&\,\kern-20pt
=\mapL\push\Bigdi{\relfactor\,\elechargeform\punto\fourax_1\punto\fourax_{23}}
\\ 
&\,\kern-20pt
-\mapL\push\Bigdi{
\relfactor\,(\traslpar/\lightvel^2)\,\xcancel{\elechargeform\punto\timearrow\punto\fourax_{23}}
}\,,
\end{aligned}
\right.
\label{fm: Mag3}
\end{equation}
with the shorthand $\,\fourax_{23}=\fourax_{2}\punto\fourax_{3}\,$. 

\medskip\goodbreak

The \emph{new} transformation rules exposed below
in the Synoptic Table
provide also an \emph{errata corrige} to the rule in
\citep{Electro2013}, where the electric-magnetic entanglement,
although vanishing in the classical limit,
was still present due to a trivial lack of cancellation by spatiality.

We may conclude that,
between electrodynamical fields, 
transformed by the action of the \Lorentz\ group
and interpreted in the original frame,
relativistic entanglements do not occur.

\subsection{General transformation rule}
\label{sec: genrule}

A direct inspection of the proofs in 
{\S}\ref{sec: eleind}
and
{\S}\ref{sec: magind}
reveals that resulting transformation rule
for spacetime (electromagnetic) differential forms
under the action of a \Lorentz\ frame-change,
depends only on the list of basis vectors relevant 
to the involved components
and not on the spacetime differential forms themselves.

Precisely, the components transformation rules depend on
whether the list of basis vector arguments does include the boost direction or does not. 
\begin{itemize}
\item
In the first case the transformation is an amplification 
by the relativistic factor.
\item
In the second case the transformation is just by invariance.
\end{itemize}
Any way, no entanglement does occur.

\section{Comparisons}
\label{sec: comparison}

The Synoptic Table below provides a comparison
between the \emph{new} transformation rules for electric and magnetic
spatial vector fields and the \emph{old} rules.

The manifest outcome is that entanglements involved in the \emph{old} rules
do not occur in the \emph{new} ones.

\medskip

\begin{table}[h!]
\kern-10pt
\small
\renewcommand{\arraystretch}{2.5}
\noindent
\begin{tabular}{| r  l | l |}
\hline
\multicolumn{3}{| c |}{\bf{Synoptic Table}}
\\
\hline
\multicolumn{2}{|c|}{\textbf{\emph{\small new}}}
&\multicolumn{1}{|c|}
{\textbf{\emph{\small old}}}
\\
\hline
$\coppia{\elefield^\parallel}{\elefield^\perp}\to\kern-10pt$
&$\coppia{\relfactor\,\elefield^\parallel}{\elefield^\perp}$
&$\coppia{\elefield^\parallel}{\relfactor\,(\elefield^\perp+\traslvel\times\magind)}$
\\
$\coppia{\magind^\parallel}{\magind^\perp}\to\kern-10pt$
&$\coppia{\magind^\parallel}{\relfactor\,\magind^\perp}$
&$\;\coppia{\magind^\parallel}{\relfactor\,(\magind^\perp-\traslvel\times\elefield)}$
\\
\hline
$\coppia{\magfield^\parallel}{\magfield^\perp}\to\kern-10pt$
&$\coppia{\relfactor\,\magfield^\parallel}{\magfield^\perp}$
&$\coppia{\magfield^\parallel}{\relfactor\,(\magfield^\perp+\traslvel\times\eledisp)}$
\\
$\coppia{\eledisp^\parallel}{\eledisp^\perp}\to\kern-10pt$
&$\coppia{\eledisp^\parallel}{\relfactor\,(\eledisp^\perp}$
&$\coppia{\eledisp^\parallel}{\relfactor\,(\eledisp^\perp+\traslvel\times\magfield)}$
\\
\hline
$\coppia{\elecurr^\parallel}{\elecurr^\perp}\to\kern-10pt$
&$\coppia{\elecurr^\parallel}{\relfactor\,\elecurr^\perp}$
&$\coppia{\elecurr^\parallel}{\relfactor\,(\elecurr^\perp+\traslvel\times\magind)}$
\\
\hline
\end{tabular}
\label{tb: transfnovel}
\end{table}

\goodbreak

The term $\,\traslvel\times\magind\,$, in the \emph{old} expression
for the transversal component of electric field,
is responsible for the invalid relativistic support of \Lorentz\ force.

The geometric analysis provides similarly the
transformation rules for all other  time-vertical forms,
as interpreted in the original framing.

\medskip\goodbreak

Under a \Lorentz\ automorphism,
the integral, over a body configuration $\,\cP\,$,
transform in such a way that:%
\begin{equation}
\left\{\vcenter{\halign{
$#$\hfil&$#$\hfil&$#$\hfil&$\quad#$\quad\hfil\cr
&\,\integrale{\cP}{}\mapL\push\elechargeform
&\,=\,\relfactor\punto\integrale{\cP}{}\elechargeform\,,&\textrm{electric charge}
\vspace{8pt}\cr
&\,\integrale{\cP}{}\mapL\push\volformg
&\,=\,\relfactor\punto\integrale{\cP}{}\volformg\,,&\textrm{metric volume}
\vspace{8pt}\cr
&\,\integrale{\cP}{}\mapL\push\massform
&\,=\,\relfactor\punto\integrale{\cP}{}\massform\,.&\textrm{material mass}
\cr}}\right.
\label{fm: transfvarie}
\end{equation}
As a consequence:%
\footnote{\label{fn: our}
Contrary to our evaluation,
the transformation $\,\rho\to\relfactor\,\rho\,$
is attributed to the charge density $\,\rho\,$ per unit volume,
in standard treatments \citep{Weyl1922,JefimenkoOleg1999}.
}
\begin{equation}
\kern-10pt
\left\{\vcenter{\halign{
$#$\hfil&$#$\hfil&$\;#$\quad\hfil\cr
&\,\rho\to\,\rho\,,&\textrm{electric charge per unit volume}
\vspace{6pt}\cr
&\,\rho_{m}\to\,\rho_{m}\,.&\textrm{mass per unit volume}
\cr}}\right.
\label{fm: densit}
\end{equation}

The charge density $\,\rho\,$  per unit volume,
is an \emph{even} scalar field on the trajectory manifold $\,\TjE\subset\EVE\,$.

Invariance follows from the rules
Eq.$\eqref{fm: transfvarie}_{1,2}$
and from the definition Eq.$\eqref{fm: elevect}_4$:
\begin{equation}
\elechargeform=\rho\punto\volformg\,.
\label{fm: }
\end{equation}

An analogous reasoning shows invariance of 
the mass per unit volume, 
an \emph{even} scalar field $\,\rho_{m}\,$ on the trajectory manifold $\,\TjE\subset\EVE\,$,
defined by:%
\begin{equation}
\massform=\rho_{m}\punto\volformg\,.
\label{fm: massdensity}
\end{equation}

A perfect analogy holds between the transformation rules
in Eq.$\eqref{fm: transfvarie}_{1,2,3}$ for electric charge,
metric volume and mass, since they 
are all integrals of spatial forms of maximal degree
and the rule in \msec\ref{sec: genrule} does apply.

The formula in
Eq.$\eqref{fm: transfvarie}_3$ describes in a simple
direct way the estimate of the transformed \emph{mass}
as detected by an observer in terms of the 
\emph{mass-form}
transformed according to a \Lorentz\ frame-change.

Let us prove explicitly the formula in Eq.$\eqref{fm: transfvarie}_2$.

To this end, for sake of simplicity, the time-vertical tangent vector fields 
$\,\set{\fourax_1,\fourax_2,\fourax_3}\,$ are taken mutually $\,\metric$-orthogonal.
The metric on the spatial bundle
$\,\metric:\VEVE\mapsto\dual{(\VEVE)}\,$
can be extended by means of the projection pull-back
to a spacetime singular metric: 
\begin{equation}
\splitS\pull\metric:\TEVE\mapsto\dual{(\TEVE)}\,.
\label{fm: }
\end{equation}
The singular metric so got is time-vertical, i.e.
vanishing when the time-arrow $\,\timearrow\,$ is an argument.

Recalling Eq.\eqref{fm: transvectors},
the longitudinal component is transformed as:
\begin{equation}
\setlength{\jot}{8pt}
\kern-15pt
\begin{aligned}
\Bigdi{\mapL\push\splitS\pull\metric}\di{\BX_{1},\BX_{1}}
&\,=\mapL\push\Bigdi{\bigdi{\splitS\pull\metric}\di{\mapL\pull\BX_{1},\mapL\pull\BX_{1}}}
\\
&\,=\mapL\push\Bigdi{
\metric\di{\relfactor\punto\BX_{1},\relfactor\punto\BX_{1}}
}
\\
&\,=\relfactor^2\punto\mapL\push\Bigdi{
\metric\di{\BX_{1},\BX_{1}}
}\,.
\end{aligned}
\label{fm: metriclong}
\end{equation}
The remaining components on the diagonal are left invariant.
The metric volume form is given by:
\begin{equation}
\Bigdi{\volformg\di{\BX_1,\BX_2,\BX_3}}^2
=\det\Bigdi{\BG\di{\BX_1,\BX_2,\BX_3}}\,.
\label{fm: volformGram}
\end{equation}
With $\,\BG\di{\BX_1,\BX_2,\BX_3}\,$ \Gram\ matrix of the metric $\,\metric\,$.
Substituting Eq.\eqref{fm: metriclong}
into Eq.\eqref{fm: volformGram} we get:
\begin{equation}
\volformg\di{\BX_1,\BX_2,\BX_3}\to\relfactor\punto\volformg\di{\BX_1,\BX_2,\BX_3}\,.
\label{fm: voltransf}
\end{equation}

According to Eq.\eqref{fm: metriclong},
the transformed spatial length in the longitudinal direction,
evaluated by the original observer, is amplified by the relativistic factor,
as first deduced in
\citep{TimeLength2014}:
\begin{equation}
\metric\di{\BX_1,\BX_1}^{\unmezzotext}\to\relfactor\punto\metric\di{\BX_1,\BX_1}^{\unmezzotext}\,.
\label{fm: lengthampl}
\end{equation}

Also, according to Eq.$\eqref{fm: transfvarie}_1$
the transformed electric charge,
evaluated by the original observer, is amplified by the relativistic factor.

The same result applies to the metric volume 
provided by Eq.\eqref{fm: voltransf} and reported in Eq.$\eqref{fm: transfvarie}_2$.
An analogous proof applies also to the material mass,
as reported in Eq.$\eqref{fm: transfvarie}_3$
leading to the following consideration.

The \Maclaurin\ series formula for the relativistic factor
in terms of the adimensional speed 
$\,\beta\di\traslpar\equaldef\traslpar/\lightvel\,$,
truncated at the third term:
\begin{equation}
\relfactor=(1-\traslpar^2/\lightvel^2)^{-1/2}
\simeq1+0+\unmezzotext\,\traslpar^2/\lightvel^2\,,
\label{fm: Mclaurin}
\end{equation}
substituted in Eq.$\eqref{fm: transfvarie}_3$ multiplied by $\,\lightvel^2\,$ yields:
\begin{equation}
\relfactor\punto\integrale{\cP}{}\massform\punto\lightvel^2
\simeq
\integrale{\cP}{}\massform\punto\lightvel^2
+\integrale{\cP}{}\unmezzotext\punto\massform\punto\traslpar^2\,.
\label{fm: relenergy}
\end{equation}

The expression in Eq.\eqref{fm: relenergy} is named \emph{total energy}.
The former of the terms:
\begin{equation}
\integrale{\cP}{}\massform\punto\lightvel^2
\,,\quad
\integrale{\cP}{}\unmezzotext\punto\massform\punto\traslpar^2
\,.
\label{fm: }
\end{equation}
is the \emph{rest energy} of the body $\,\cP\,$,
while the latter is similar to the expression for the \emph{kinetic energy},
but with boost speed $\traslpar$ in place of the velocity field of the body.%
\footnote{\label{fn: spread}
Confusion between the two is surprisingly spread all over
standard physics literature on relativity dealing with particle kinematics.
}

An analogous transformation law applies to \emph{kinetic momentum}.
The continuum definition is provided by the variational expression:
\begin{equation}
\integrale{\cP}{}\delta\Boo^1\wedge(\massform\punto\vel)
=\integrale{\cP}{}\scalar{\delta\Boo^1}{\vel}\punto\massform\,,
\label{fm: kinmom}
\end{equation}
and, according to Eq.$\eqref{fm: transfvarie}_3$ is transformed into:
\begin{equation}
\relfactor\punto\integrale{\cP}{}\scalar{\delta\Boo^1}{\vel}\punto\massform\,.
\label{fm: relatkinmom}
\end{equation}

For a continuous body, the 
mass is a maximal material form 
to be integrated on the current configuration $\,\cP\,$.
The previous analysis reveals that in this context
the transformation rules dictated by special relativity
are deduced in a simplest way by means of Eq.\eqref{fm: relatkinmom}.

These evaluations should be compared with
the involved argument exposed in
physically biased tracts on special relativity
where the transformed quantities appear as tentative 
definitions to be confirmed by thought collision experiment
\citep[Ch.V]{Rindler1989},
\citep[\msec(7.1.1)]{ForshawSmith2009},
\citep{Fernflores2019}.

For what concerns the clock $\,\dtE\in\dual{(\TEVE)}\,$,
we observe that the rate of the pushed clock,
when evaluated in the original framing,
is \emph{faster} by the relativistic factor:%
\begin{equation}
\setlength{\jot}{8pt}
\begin{aligned}
\scalar{\mapL\push\dtE}{\timearrow}
&\,=\mapL\push\scalar{\dtE}{\mapL\pull\timearrow}
\\
&\,=\relfactor\punto\mapL\push\scalar{\dtE}{\timearrow-\xcancel{\traslpar\,\fourax_1}}
\\&\,=
\relfactor\ge\,1
\,,
\end{aligned}
\label{fm: clockfast}
\end{equation}
since $\scalar{\dtE}{\timearrow}=1$.
Cancellation is due to time-verticality of $\fourax_1$.


In the classical limit $\,\traslpar/\lightvel\to0\,$,
from Eq.\eqref{fm: relfactor} we get $\,\relfactor=1\,$
and the result reduces to the one concerning \Newton\ frame-changes 
\begin{equation}
\map_{N}:\EVE\mapsto\EVE\,,
\label{fm: }
\end{equation}
characterised by invariance of the clock rate:
\begin{equation}
\map_{N}\push\dtE=\dtE\,.
\label{fm: clockinv}
\end{equation}

By virtue of this property, spatial vector fields
are transformed by \Newton\ frame-changes 
into vector fields that are still spatial
in the same framing.

\section{Conclusions}
\label{sec: conclusion}

Three main contributions have been brought to theoretical
Electrodynamics and Relativity.

The first contribution concerns the development of a spacetime formulation
of electric induction law in terms of electric field and magnetic potential,
(\emph{á la} \Maxwell\ and \ThomsonJJ) 
in place of the seemingly convenient but eventually misleading
reduced formulation (\emph{á la} \Heaviside-\Hertz-\Lorentz).
in terms of exterior derivatives.

In the reduced formulations a velocity dependent exact differential term
is zeroed by the action of taking the exterior derivative,
but this fact
destroys frame-covariance of the induction laws.

The integral formulation ought to be made in terms of one-forms 
integrated along $1$D paths in spacetime.

It has been shown \citep{Electro2014}
that this new formulation solves all
troubles concerning the \emph{flux-rule}
exposed in \citep[II.17-2]{Feynman1964}.

Indeed the \emph{flux-rule} (or better the \emph{vorticity-rule})
was there applied out of its range of validity 
which is limited to boundary paths undergoing
regular motions, as assessed here in Prop.\ref{prop: fluxrule}.

Moreover, the formulation in terms of differential forms puts into
evidence a  new term, depending on the stretching of deformable bodies,
completely overlooked by standard vectorial formulations.

The second contribution provides a special expression referring
to the case of translational motions in a field of 
constant magnetic momentum and uniform magnetic vorticity.

It is thus possible to apply
a simple formula for the electric field in terms of the 
magnetic vortex, which is exactly one-half of 
what is usually labeled as \Lorentz\ force rule.

The third contribution is concerned with the
spacetime formulation of Electromagnetism 
in terms of the electromagnetic
\Faraday\ and \Ampere\ two-forms
and the detection of the transformation rules
for electric and magnetic fields and fluxes
under change of frame and in particular
under \Lorentz\ transformations
of special relativity.

The electromagnetic induction laws,
expressed in terms of \Lie\ and exterior derivatives,
are covariant under any change of frame.

This means they simply transform by push according to the
diffeomorphic transformation map,
since both these derivatives and all involved geometrical entities 
transform by push, in a natural way.

This statement amends the claim that
\Maxwell\ laws are not invariant under \Galilei\ group of frame transformations
\cite[p.21, Eq.(2.6)]{ChoquetBruhat2009}.

In this respect I observe that
invariance is not the proper qualification to be asked for,
since tensor fields of degree greater than zero (i.e. other than scalar fields)
are involved in the laws of electromagnetic induction
and therefore covariance, that is variance by push,
should rather be invoked.

The transformation rule by covariance, 
when applied to \Lorentz\ frame changes,
and interpreted in the original frame,
reveals that relativistic transformations previously considered in literature
ought to be thoroughly modified.

Indeed, in amendment of standard statements,
the conclusion of the new investigation is that
electromagnetic entanglements between electric and magnetic spatial fields
are completely absent,
as clearly depicted in the Synoptic Table of {\S}\ref{sec: comparison}.

In the classical limit $\,\traslpar/\lightvel\to0\,$ and
the relativistic factor tends to unity $\,\relfactor\to1\,$.

Moreover,
$\,\traslpar/\lightvel^2\to0\,$ in Eq.\eqref{fm: VLtangbasis},
so that invariance of electric and magnetic
forms under spatial \Galilei\ transformation is recovered in the limit,
as expected on physical and mathematical ground,
due to continuous dependence of the transformation
on the light speed.

In particular, the 
amended relativistic transformation rule for the electric field
deprives the \Lorentz\ force rule of any relativistic support.

Last but not least, 
according to the present treatment,
all spatial differential forms of maximal degree,
such as metric volume form, mass form and charge form,
are transformed in the same way,
by amplification according to the relativistic factor,
as given by Eq.\eqref{fm: transfvarie}.

Length in direction of boost and clock rates is likewise modified
by the same amplification.
These results are of special relevance and 
suggest a revision of physical interpretations in special relativity.

Since early contributions \citep{Poincare1905}
electric charge conservation
under \Lorentz\ transformations was assumed
in deducing transformation rules of electromagnetic fields.

\goodbreak

Quite the other way, in relativistic dynamics
mass is not assumed to be conserved
but rather to be amplified according to the relativistic factor. 

According to the geometrical analysis exposed in the present contribution,
mass, electric charge and metric volume,
which all are spatial differential forms of the maximal degree,
have all an identical behaviour under relativistic frame transformations.

The spacetime treatment, performed in terms of differential forms
and \Lie\ derivatives along the motion,
brings to conceptual and methodological decisive improvements 
over the still presently ubiquitously adopted
standard vectorial expressions.

Theoretical discussions are moreover significantly simplified 
and clarified by the adopted geometric framework.

This is especially evident in discussing questions 
about frame covariance of induction laws 
and in evaluating the transformations induced by frame changes
when interpreted in the original framing.

\goodbreak

As a matter of fact, 
all modern treatments of electrodynamics still include 
inappropriate entanglements
borrowed from the analysis in 
\citep{Lorentz1904,Poincare1905,Einstein1905a}
which were based
on the incomplete interpretation of 
the original formulation 
exposed in \citep{Maxwell1861,Maxwell1865}
and on the consequent misleading simplification brought
by \cite{Heaviside1885,Heaviside1892,Hertz1892,Lorentz1892}.

Unfortunately,
the clarification contributed by \cite{ThomsonJJ1893} 
about the original formulation of electromagnetic induction laws 
by his master James \Clerk-\Maxwell\
was completely overlooked in the pertinent literature
of the XX century. 

The geometric analysis first carried out in \citep{InductionLaws2012},
and revised and further developed in the present paper,
provides
an independent confirmation of these clarifications
bringing them again under the spotlight
after more than a century of oblivion.

The mathematical theory of differential forms and integration on manifolds, 
largely due to {\'E}lie \cite{CartanE1899,CartanE1923,CartanE1924,CartanE1945},
Georges \cite{deRham1955} and Hassler \cite{Whitney1957},
as well illustrated in \citep{MarsdenRatiuAbraham2003,HitchinNigel2003,Fecko2006},
is presently self-proposing as the
suitable tool for spacetime formulation of
electromagnetic induction laws.

\goodbreak

This comment refers especially to
treatments involving deformable bodies in motion,
and for the description of spacetime transformations of 
electromagnetic fields in special relativity.

A sound evidence of merits of clarity and conciseness
of the exterior differential machinery with respect to 
the standard vectorial one, emerges by comparing the sharp and general reasoning
in Prop.\ref{prop: pushLie} to the involved treatment in  \citep{JefimenkoOleg1999}
relying on questionable electromagnetic entanglements and on form invariance.

Thanks to this powerful theory, 
a direct recourse to the relevant notions and properties
permits to get rid of the alleged assumption
of \emph{form-invariance} of electromagnetic induction laws
and of \emph{conservation} of electric charge under \Lorentz\ frame-changes,
and to state natural and consistent rules of transformation for physical fields
represented by differential forms in spacetime.%

\section{Some hints for collateral reading}
\label{sec: suggestions}

I would draw attention
of readers interested in historical and attributional issues 
in differential geometry, to two nice brief papers
that could easily escape to a first search.

One is by \cite{Samelson2001} where evidence
about the birth of exterior derivatives and of their powerful properties
first investigated by 
\cite{Volterra1889a,Volterra1889b}
are given.

The other one is by \cite{Trautman2008} 
about the \emph{naissance} of \Lie-derivatives theory.

Historical notes
on the development of the
laws of electromagnetic induction
were recently contributed by
Ovidio Mario \cite{Bucci2014}.

For mathematical biased scholars we also suggest the
treatments in
\citep{Marmo2005,Marmo2006,DeNicolaTulczyjew2009}
concerned in particular with the notion of orientations in spacetime.

A comprehensive exposition of differential forms, 
integration on manifolds and orientation,
is offered in  \citep{Fecko1997,Fecko2006}
\citep{MarsdenRatiuAbraham2003} and \citep{HitchinNigel2003}.

The role of differential forms in Electrodynamics 
was well outlined in 
\citep{Deschamps1970,Deschamps1981}
and is effectively described in
\citep{Fecko2014}.

\goodbreak

Discrete topological formulations of Electromagnetics
and relevant computational aspects
are discussed in 
\citep{Tonti1995,Tonti2002},
\citep{Bossavit1991,Bossavit1998,Bossavit2004,Bossavit2005}, 
\citep{GrossKotiuga2004,KurzAuchmann2012},
\citep{Stern2015}
and references therein.

\setlength{\bibsep}{0pt plus 0.3ex}


\newpage

\end{document}